\documentclass[fleqn,usenatbib]{mnras}

\usepackage{times}
\usepackage{newtxtext,newtxmath}

\usepackage[T1]{fontenc}
\usepackage{ae,aecompl}

\usepackage{graphicx}	
\usepackage{amsmath}	
\usepackage{amssymb}	
\usepackage{wasysym}

\usepackage{tabularx}
\newcolumntype{Y}{>{\centering\arraybackslash}X}
\newcolumntype{P}[1]{>{\centering\arraybackslash}p{#1}}
\usepackage{pdflscape}
\usepackage{hyperref}
\title[IllustrisTNG ETG density profiles]{Early-type galaxy density profiles from IllustrisTNG: \\I. Galaxy correlations and the impact of baryons}

\author[Y. Wang et al.]{Yunchong Wang$^{1,2}$\thanks{E-mail: \url{ycwang15@mit.edu}},
Mark Vogelsberger$^{1}$,
Dandan Xu$^{3,2}$,
Shude Mao$^{2,4}$,
\newauthor
Volker Springel$^{5}$,
Hui Li$^{1}$,
David Barnes$^{1}$,
Lars Hernquist$^{6}$,
Annalisa Pillepich$^{7}$,
\newauthor
Federico Marinacci$^{6}$,
R\"{u}ediger Pakmor$^{8}$,
Rainer Weinberger$^{6}$ and
Paul Torrey$^{9}$
\\
$^{1}$Kavli Institute for Astrophysics and Space Research, Department of Physics, MIT,  Cambridge, MA 02139, USA\\
$^{2}$Tsinghua Center for Astrophysics, Department of Physics, Tsinghua University, Beijing, 100084, China\\
$^{3}$Institute for Advanced Studies, Tsinghua University, Beijing, 100084, China\\
$^{4}$National Astronomical Observatories, Chinese Academy of Sciences, Beijing, 100012, China\\
$^{5}$Max-Planck-Institut f\"{u}r Astrophysik, Karl-Schwarzschild-Str. 1, D-85748, Garching, Germany\\
$^{6}$Harvard-Smithsonian Center for Astrophysics, 60 Garden Street, Cambridge, MA 02138\\
$^{7}$Max-Planck-Institut f\"{u}r Astronomie, K\"{o}nigstuhl 17, 69117 Heidelberg, Germany\\
$^{8}$Heidelberg Institute for Theoretical Studies, Schloss-Wolfsbrunnenweg 35, D-69118 Heidelberg, Germany\\
$^{9}$Department of Astronomy, University of Florida, 211 Bryant Space Sciences Center, Gainesville, FL 32611, USA
}

\date{Accepted ***. Received ***; in original form ***}

\pubyear{2019}

\begin{document}
\label{firstpage}
\pagerange{\pageref{firstpage}--\pageref{lastpage}}
\maketitle

\begin{abstract}
We explore the isothermal total density profiles of early-type galaxies (ETGs) in the IllustrisTNG simulation. For the selected 559 ETGs at $z = 0$ with stellar mass $10^{10.7}\mathrm{M}_{\astrosun} \leqslant M_{\ast} \leqslant 10^{11.9}\mathrm{M}_{\astrosun}$, the total power-law slope has a mean of $\langle\gamma^{\prime}\rangle = 2.011 \pm 0.007$ and a scatter of $\sigma_{\gamma^{\prime}} = 0.171$ over the radial range 0.4 to 4 times the stellar half mass radius. Several correlations between $\gamma^{\prime}$ and galactic properties including stellar mass, effective radius, stellar surface density, central velocity dispersion, central dark matter fraction and in-situ-formed stellar mass ratio are compared to observations and other simulations, revealing that IllustrisTNG reproduces many correlation trends, and in particular, $\gamma^{\prime}$ is almost constant with redshift below $z = 2$. Through analyzing IllustrisTNG model variations we show that black hole kinetic winds are crucial to lowering $\gamma^{\prime}$ and matching observed galaxy correlations. The effects of stellar winds on $\gamma^{\prime}$ are subdominant compared to AGN feedback, and differ due to the presence of AGN feedback from previous works. The density profiles of the ETG dark matter halos are well-described by steeper-than-NFW profiles, and they are steeper in the full physics (FP) run than their counterparts in the dark matter only (DMO) run. Their inner density slopes anti-correlates (remain constant) with the halo mass in the FP (DMO) run, and anti-correlates with the halo concentration parameter $c_{200}$ in both types of runs. The dark matter halos of low-mass ETGs are contracted whereas high-mass ETGs are expanded, suggesting that variations in the total density profile occur through the different halo responses to baryons.
\end{abstract}




\section{Introduction}
\label{sec:1}

The general structure formation scenario of the Universe in the $\Lambda$-Cold Dark Matter (CDM) cosmology model consists of a `bottom up' assembly of dark matter halos and subsequent gas cooling and star formation leading to the formation of the galaxies we see today~(e.g. \citealt{1978MNRAS.183..341W,1984Natur.311..517B,1994MNRAS.271..781C}). This scenario has been tested robustly by semi-analytical modeling~\citep{1993MNRAS.264..201K,1999ASPC..163..227B,2000MNRAS.319..168C}, N-body numerical simulations~\citep{1985ApJ...292..371D,2005Natur.435..629S} and cosmological hydrodynamic simulations~\citep{2014MNRAS.444.1518V,2014Natur.509..177V,2014MNRAS.444.1453D,2015MNRAS.446..521S} over the past few decades. 

Early-type galaxies (hereafter, ETGs) are massive elliptical and lenticular galaxies with little gas and old stellar populations. As final products of galaxy mergers and secular formation processes, they provide crucial insights for testing and constraining the theory of structure formation within their large-scale cosmological environment. Over the past decade, the joint efforts of many galaxy surveys have resulted in a significant sample of observed ETGs, including strong lensing surveys such as the Lensing Structure and Dynamics Survey~\citep[LSD]{2002ApJ...575...87T,2003ApJ...583..606K,2004ApJ...611..739T}, the Sloan Lens ACS Survey~\citep[SLACS]{2006ApJ...638..703B,2006ApJ...640..662T,2009ApJ...705.1099A}, the SLACS for the Masses Survey~\citep[S4TM]{2017ApJ...851...48S}, the Strong Lensing Legacy Survey~\citep[SL2S]{2007A&A...461..813C,2012ApJ...761..170G}, the BOSS Emission-Line Lens Survey~\citep[BELLS]{2012ApJ...757...82B,2012ApJ...744...41B}, and surveys targeted at stellar kinematics and dynamics such as SPIDER~\citep{2010MNRAS.408.1313L}, $\mathrm{ATLAS^{3D}}$~\citep{2011MNRAS.413..813C}, the SLUGGS Survey~\citep{2014ApJ...796...52B,2016MNRAS.458L..44F}, and the MASSIVE Survey~\citep{2014ApJ...795..158M}. Through detailed stellar dynamics and strong lensing modeling, the masses, sizes, stellar and total density distributions and other dynamical features of ETGs are constrained to different levels. 
 
Interestingly, the average total power-law density slope of observed ETGs has been found to be close to isothermal within a few effective radii, i.e., $\rho(r) \propto r^{-\gamma^{\prime}}$, where  $\gamma^{\prime} = 2$, which describes a sphere of collisional ideal gas in equilibrium between thermal pressure and self gravity. This coincidence sometimes is also referred to as the `bulge-halo conspiracy': while neither the stellar- (baryonic-) component nor the dark matter halo exhibits an isothermal density distribution, the sum of the two appears to follow such a profile with little intrinsic scatter (e.g. analytical two-component stellar-dark matter models, see \citealt{2009MNRAS.393..491C}). The observational evidence for the presence of near-isothermal density profiles in ETGs is prevalent and convincing, from dynamically modeled ETGs at $z \approx 0 $~\citep{2014MNRAS.445..115T,2016MNRAS.460.1382S,2017MNRAS.467.1397P,2018MNRAS.476.4543B}, strong lensing ETGs which trace back to $z = 1$~\citep{2006ApJ...649..599K,2009ApJ...703L..51K,2009MNRAS.399...21B,2011MNRAS.415.2215B,2010ApJ...724..511A,2011ApJ...727...96R,2013ApJ...777...98S,2018MNRAS.480..431L,2018MNRAS.475.2403L}, and X-ray observations of ETGs~\citep{2006ApJ...646..899H,2010MNRAS.403.2143H}. Apart from the near-isothermal behavior, the total density profiles of observed ETGs also show clear correlations with galaxy parameters, such as the total stellar mass, effective radius, stellar surface density, central velocity dispersion and central dark matter fraction~\citep{2010ApJ...724..511A,2013ApJ...777...98S,2015ApJ...814...26N,2015ApJ...803...71S,2017MNRAS.467.1397P}. A mild shallowing trend of the total density profile with increasing redshift is also observed for strong lenses~\citep{2006ApJ...649..599K,2010ApJ...724..511A,2011MNRAS.415.2215B,2011ApJ...727...96R,2013ApJ...777...98S}. 

Dedicated theoretical studies including semi-analytical modeling and simulations, aiming at explaining the `bulge-halo conspiracy', have converged on a scenario in which ETGs form in a two-stage fashion: dissipative gas cooling first triggers active star formation and dark matter halo contraction which steepens the density profile, then non-dissipative mergers and accretion follow and make the density profile shallower~\citep{2007ApJ...658..710N,2009ApJ...703.1531N,2009ApJ...706L..86N,2009ApJ...691.1168H,2009ApJS..182..216K,2012ApJ...754..115J,2013MNRAS.433.3297D,2013ApJ...766...71R,2016MNRAS.458.2371R}. This is in line with the constraints from the more profound ETG scaling laws, such as the fundamental plane relations~\citep{1987nngp.proc..175F,1996MNRAS.280..167J,2006MNRAS.366.1126C,2006ApJ...641...21R} and the $M_{\mathrm{BH}}-\sigma_{\mathrm{v}}$ relation~\citep{2001ApJ...552L..13C,2003ApJ...596..903P,2006ApJ...641...90R,2008ApJ...680..143G,2013ApJ...764..184M}, which also indicate the important role of dissipative processes. Although ETG evolution is dominated by `dry' mergers in the latter phase of the `two-phase' scenario~\citep{2012ApJ...754..115J,2013ApJ...766...71R,2016MNRAS.458.2371R,2017MNRAS.464.3742R}, \citet{2014ApJ...786...89S} claimed dissipative `wet' mergers are indispensable in addition to `dry' mergers during the formation of ETGs in order to establish the observed redshift evolution of the ETG density profile. AGN feedback is also found to evolve the total and dark matter density profile of ETGs shallower with time through zoom-in and cosmological simulations~\citep {2013MNRAS.433.3297D,2017MNRAS.472.2153P,2019MNRAS.483.4615P}. These efforts were crucial steps towards a self-consistent ETG formation and evolution scenario, bridging the gap between theoretical assumptions and observational uncertainties.

Furthermore, since the `bulge-halo conspiracy' emphasizes the interplay between baryonic and dark matter components that comprise the ETGs, it is of critical importance to study the individual effects of baryons and dark matter. Conventionally, the dark matter halo profile is believed to be universally well-described by an NFW profile~\citep{1997ApJ...490..493N}. However, recent studies from observations~\citep{2010ApJ...721L.163A,2010ApJ...709.1195T} and simulations~\citep{2010MNRAS.402...21N,2015ApJ...811....2H,2017MNRAS.472.4343C} have revealed that the dark matter profile is in fact non-universal, and the influence of baryonic processes on dark matter halos varies with halo size, shape, baryon fraction and environment~\citep{2010Natur.463..203G,2013ApJ...765...25N,2016MNRAS.461.2658D,2018arXiv180907255C}. The observed dark matter halo contraction~\citep{2012ApJ...752..163S,2012ApJ...747L..15G,2014MNRAS.439.2494O,2015ApJ...814...26N,2016MNRAS.456..870B} is in agreement with modified adiabatic contraction models~\citep{2004ApJ...616...16G,2010MNRAS.407..435A}, ruling out the standard adiabatic contraction~\citep{1986ApJ...301...27B}, and is consistent with the weak dissipative processes predicted by the theoretical studies noted above in shaping ETG density profiles. This has been shown to be true for IllustrisTNG elliptical galaxies at $z = 0$ compared to observations, and some tension in the contraction level still exists at $z = 2$~\citep{2018MNRAS.481.1950L}.

With the advent of a new generation of cosmological hydrodynamic simulations, i.e. the Illustris simulations\footnote{\url{http://www.illustris-project.org/}}~\citep{2014MNRAS.444.1518V, 2014Natur.509..177V, 2014MNRAS.445..175G, 2015MNRAS.452..575S,2015A&C....13...12N}, the EAGLE Project\footnote{\url{http://icc.dur.ac.uk/Eagle/}}~\citep{2015MNRAS.446..521S,2015MNRAS.450.1937C}, the Horizon-AGN Simulation\footnote{\url{https://www.horizon-simulation.org/}}~\citep{2016MNRAS.463.3948D,2017MNRAS.467.4739K}, and the Magneticum Pathfinder\footnote{\url{http://www.magneticum.org/}} simulations (Dolag et al. to be submitted), large statistical samples of simulated ETGs have become available that reproduce the observed density profiles and correlation trends (e.g. see \citealt{2017MNRAS.464.3742R,2017MNRAS.469.1824X,2018MNRAS.479.5448B} and references therein). In this paper, we use the IllustrisTNG project\footnote{\url{http://www.tng-project.org}}~\citep{2018MNRAS.480.5113M,2018MNRAS.477.1206N,2018MNRAS.475..624N,2018MNRAS.475..648P,2018MNRAS.475..676S}, an updated set of simulations with a new physical model extending the original Illustris project, and select a realistic sample of simulated ETGs to study the statistical properties and correlations with global galactic properties of their density profiles. We will: \textbf{a)} investigate the distribution of the total density slopes in different radial ranges; \textbf{b)} compare the correlations between the simulated total density slopes with a number of galaxy properties as well as the redshift evolution of the slopes with a diverse dataset from observations and other simulations to explore the outcome and systematic biases of our IllustrisTNG ETG sample; \textbf{c)} explore these correlations in the TNG methodology box with different feedback model variations and reveal the individual effects of different parts of the feedback physics on the total density profile; \textbf{d)} compare between full physics and dark matter only simulations to elucidate the impact of baryons in shaping density profiles of the total and dark matter components. 

This paper is organized as follows: in Section \ref{sec:2} we briefly introduce the IllustrisTNG simulations, our sample selection criteria and the post-processing tools that we use for our analysis; in Section \ref{sec:3} we present the measured total power-law density slopes of the IllustrisTNG ETG samples and their correlations with a number of global galactic properties as well their redshift evolution; in Section \ref{sec:4} we present the analysis of the various galaxy correlations of the total density profile under different simulation model variations, and discuss in detail the interpretation of the effects of AGN and stellar feedback on the total density profile with a comparison to previous literature; in Section \ref{sec:5} we describe the difference between the full physics simulation and the dark matter only simulation to identify the separate contributions of baryonic and dark matter in the origin of the density profiles; in Section \ref{sec:6} we summarize the general properties we obtained for the IllustrisTNG ETGs in this work and discuss further questions that still need to be answered in the future. Throughout this paper, we adopt the Planck $\Lambda$CDM cosmology~\citep{2016A&A...594A..13P}, which is also used in the IllustrisTNG simulations; i.e., $h = 0.6774$, $\Omega_{\mathrm{m}} = 0.3089$, $\Omega_{\mathrm{\Lambda}} = 0.6911$, $\Omega_{\mathrm{b}} = 0.0486$, and $\sigma_{\mathrm{8}} = 0.8159$.

The full dataset of the IllustrisTNG ETGs that we present in this work will be available on the IllustrisTNG website (\url{http://www.tng-project.org/}).  

\section{Methodology}
\label{sec:2}
\subsection{The simulation}
\label{sec:2.1}

The IllustrisTNG project (IllustrisTNG hereafter, see  \citealt{2018MNRAS.480.5113M,2018MNRAS.477.1206N,2018MNRAS.475..624N,2018MNRAS.475..648P,2018MNRAS.475..676S}) is a suite of state-of-the-art magneto-hydrodynamic cosmological simulations. Evolved using the moving mesh hydrodynamics code \textsc{arepo}~\citep{2010MNRAS.401..791S}, the simulations were built upon the many successes of the original Illustris project~\citep{2014MNRAS.444.1518V, 2014Natur.509..177V, 2014MNRAS.445..175G, 2015MNRAS.452..575S,2015A&C....13...12N} and the Illustris models~\citep{2013MNRAS.436.3031V,2014MNRAS.438.1985T}, but with improved prescriptions for both stellar and AGN feedback~\citep{2017MNRAS.465.3291W,2018MNRAS.473.4077P}. The full physics IllustrisTNG simulation suite reproduces many key relations in observed galaxies, including the galaxy-color bimodality in the Sloan Digital Sky Survey~\citep{2018MNRAS.475..624N}, the evolution of the mass-metallicity relation~\citep{2017arXiv171105261T,2018MNRAS.477L..16T}, the galaxy size-mass relation evolution~\citep{2018MNRAS.474.3976G}, the fraction of dark matter within galaxies at $z=0$~\citep{2018MNRAS.481.1950L}, the intra-cluster metal distribution in galaxy clusters~\citep{2018MNRAS.474.2073V}, and the cool-core structure in galaxy clusters~\citep{2018MNRAS.tmp.1999B}. These verifications of the IllustrisTNG physics model stand as strong confirmation of the plausibility of the simulated IllustrisTNG galaxy and galaxy cluster populations. 

In this work we make use of the highest resolution version of the TNG100 simulation, which employs $2\times 1820^{3}$ resolution elements in a ($75/h \approx 110.7$ Mpc)$^3$ box. The baryonic and dark matter mass resolutions are $m_{\mathrm{baryon}} = 1.4\times10^{6} \mathrm{M}_{\mathrm{\astrosun}}$ and $m_{\mathrm{DM}} = 7.5\times 10^{6}\mathrm{M}_{\mathrm{\astrosun}}$, respectively. A softening length of $\epsilon = 0.74\,\mathrm{kpc}$ (below $z = 1$) is adopted for the dark matter and stellar components, while the gravitational softening of the gas cells is fully adaptive (minimum 0.19 comoving kpc). We also make use of the TNG100-dark matter only (DMO) simulation, which has the same initial conditions, total mass and softening length as the TNG100-full physics run. The dark matter resolution in the DMO run is $m_{\mathrm{DM}} = 8.9\times 10^{6} \mathrm{M}_{\mathrm{\astrosun}}$.

\subsection{Galaxy classification and sample selection}
\label{sec:2.2}

\begin{figure}
\includegraphics[width=\columnwidth]{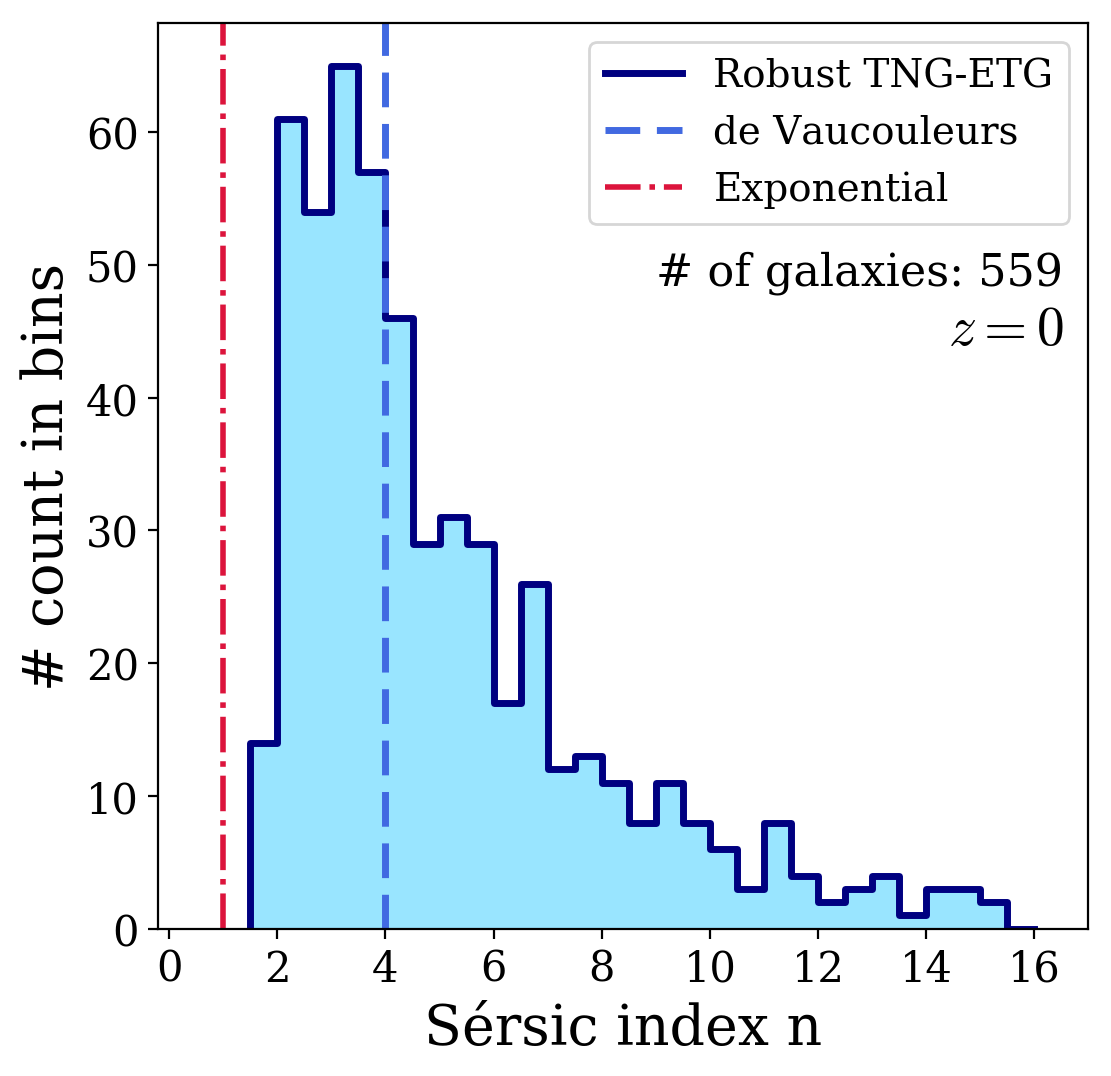}
\caption{The distribution of the S$\mathrm{\acute{e}}$rsic index of the single S$\mathrm{\acute{e}}$rsic model fitted to our final sample of 559 $z=0$ IllustrisTNG ETGs. The blue solid histogram indicates the number distribution of our selected IlustrisTNG ETGs, and they typically have S$\mathrm{\acute{e}}$rsic index $n \geqslant 2$, which well represents the luminosity profiles of an ETG sample.}
\label{fig:sersic_index}
\end{figure} 

Galaxy type classification is achieved using the same procedure as in \citet{2017MNRAS.469.1824X}. Here we only briefly summarize the key steps and features involved. 

Galaxies are identified as gravitationally bound structures of dark matter particles, stellar particles and gas elements, using the \textsc{subfind} algorithm \citep{2001MNRAS.328..726S,2009MNRAS.399..497D}. To derive the optical light of galaxies, each stellar particle is assigned a brightness magnitude in a given observational filter band based on its star formation age and metallicity using \citet{2003MNRAS.344.1000B} stellar population synthesis (SPS) model \textsc{galaxev}. A simple projection-dependent dust attenuation model is adopted in a post-processing fashion in order to take into account dust absorption and scattering effects.   

Using the SDSS $r$-band rest-frame luminosity, we fit both a single de Vaucouleurs profile~\citep{1948AnAp...11..247D}
and a single exponential profile to a galaxy's radial surface brightness distribution in a given projection, for example X projection along the simulation box. As our first criterion, if a galaxy can be better fitted by the former than the latter (i.e., with a smaller $\chi^2$), then it is classified as an ETG.

In addition, we also fit a combined de Vaucouleurs profile plus a S$\mathrm{\acute{e}}$rsic profile~\citep{1963BAAA....6...41S} 
to the total radial light profile in order to describe a combined light distribution from both the galactic bulge and the disk components. As our second criterion, if the integrated luminosity fraction of the de Vaucouleurs component is larger than half, then the galaxy is classified as an ETG.

In our final ETG sample, we only include galaxies whose radial light profiles in all three independent projections (along X, Y, and Z axes of the simulation box) satisfy both criteria above. In order to compare with the results from observations and other simulations (see Section \ref{sec:3.2} for details), we also select only central galaxies (the largest substructure identified by the \textsc{subfind} algorithm in each FoF group) with total stellar mass $5\times 10^{10} \mathrm{M}_{\astrosun} \leqslant M_{\ast} \leqslant 8\times 10^{11} \mathrm{M}_{\astrosun}$ (about $3.6\times10^4$ to $5.8\times10^5$ stellar particles). Our selection criteria results in 559 well-resolved early-type galaxies at redshift $z = 0$, and 720 at $z=0.2$, which are used to compare with local and higher-redshift (median redshift $z\approx0.23$) ETG samples from both observations and other simulations. We note that by visually checking the mock SDSS $g,r,i-$band composite images of these galaxies, most of them are bulge-dominated red galaxies that typically resemble observed ETG samples. A small fraction ($\approx 10\%$) of galaxies that appear as `red-spirals' or that seem to have gone through recent star formation are included in our IllustrisTNG ETG sample based on their luminosity profile classification. Although we do not show, we have checked for consistency that with or without these sub-samples, the results in the remaining of this paper does not change. 

To demonstrate the robustness of our final ETG sample, we present Fig.~\ref{fig:sersic_index}, which shows the histogram of the S$\mathrm{\acute{e}}$rsic index measured for 559 $z = 0$ galaxies (in their $X$-projection) in our final IllustrisTNG ETG sample. As shown in the figure, our selected IllustrisTNG ETGs typically have S$\mathrm{\acute{e}}$rsic index $n \geqslant 2$, representing the luminosity profiles of observed ETG samples. 

\subsection{Analysis}
\label{sec:2.3}

An exact power-law model $\rho(r) \propto r^{-\gamma^{\prime}}$ provides a reasonable approximation to the galaxy total radial density profile, i.e. the combined radial matter density profile of all dark matter, stellar, and gas particles of a given galaxy. In practice, whether in observation or simulation, the slope index $\gamma^{\prime}$ of an approximate power-law is always measured within a given radial range $(r_{\mathrm{1}}, r_{\mathrm{2}})$, usually from about a tenth the galaxy effective radius out to a few effective radii (see Section \ref{sec:3.2} for details). For each galaxy in our IllustrisTNG ETG sample, we adopt the position of the particle with the minimum gravitational potential in its host halo as the center of the galaxy. Assuming spherical symmetry, we calculate the total radial density distribution in 100 radial bins, equally divided in logarithmic scale. We then perform a linear fit (with equal radial weighting) to $\log\rho(r)-\log r$ within a given radial interval $(r_{\mathrm{1}}, r_{\mathrm{2}})$ and define the best linear fit slope as the power-law density slope $\gamma^{\prime}$ for the total radial density distribution. This best-fit power-law slope of the total matter radial density profile is a 3D measure of the radial distribution of both baryonic and dark matter within the galaxy, so we choose to scale ($r_{1}$, $r_{2}$) using the 3D stellar half mass radius $R_{1/2}$ (which includes all stellar particles assigned to this galaxy by \textsc{subfind}) throughout this paper. Although the observational modeling techniques implemented for deriving the power-law slopes involve various assumptions, the quantitative analysis of their systematic biases is beyond the scope of this paper. Therefore, we make direct comparison of the `true' 3D power-law slope of the total mass density profile of the IllustrisTNG ETGs with the observational power-law slopes; i.e. we do not attempt to mock the observational procedures to derive $\gamma^{\prime}$.

We further study the correlations between the power-law slope $\gamma^{\prime}$ and the following galaxy properties: total stellar mass, effective radius, stellar surface density, central velocity dispersion, stellar orbital anisotropy, central dark matter fraction and in-situ-formed stellar mass ratio. \textbf{i)} The total stellar mass $M_{\ast}$ of an ETG is defined as the sum of the mass of all the stellar particles assigned to the ETG by \textsc{subfind}. \textbf{ii)}  We adopt the projected 2D stellar half mass radius (including all stellar particles assigned to the galaxy identified by \textsc{subfind}) along the X-axis of the simulation box as a measure of the effective radius $R_{\mathrm{eff}}$ of IllustrisTNG ETGs. The 2D projected half light radius of the IllustrisTNG quenched galaxies has been shown to agree with observations within $1\,\sigma$ error bars, assuming 0.25 dex observational uncertainty in the measurement of stellar mass~\citep{2018MNRAS.474.3976G}. However,  we point out that different assumed stellar mass uncertainties, luminosity fitting methods, dust attenuation models, aperture size and shape, and projection effects all add up to the systematic biases in the size measurement. The 2D $R_{\mathrm{eff}}$ is overestimated by $\approx 0.1\,$dex in the stellar mass range of our IllustrisTNG ETG sample as shown by~\citet{2018MNRAS.474.3976G}. \textbf{iii)} The stellar surface density is defined as $\Sigma_{\ast} = M_{\ast}/2\pi R_{\mathrm{eff}}^{2}$. \textbf{iv)} The central velocity dispersion $\sigma_{\mathrm{e/2}}$ is calculated for all the stellar particles projected (along the $X$-projection of the simulation box) within the central 1/2 of $R_{\mathrm{eff}}$, with each stellar particle weighted by its (rest-frame) SDSS $r$-band luminosity. In addition, we also calculate the stellar orbital anisotropy parameter $\beta$, which is defined for all the stellar particles within the central 3D sphere with radius of $R_{\mathrm{eff}}$, each stellar particle weighted by its mass. \textbf{v)} The central dark matter fraction $f_{\mathrm{DM}}$ is calculated as the 3D fraction of dark matter mass over the total mass of all dark matter, stellar and gas particles enclosed within the central 3D sphere with radius of $R_{\mathrm{eff}}$. \textbf{vi)} The in-situ-formed stellar mass ratio $f_{\mathrm{in-situ}}$ is the sum of all the stellar particles formed within the main progenitor branch of the ETG versus the total stellar mass of the ETG, using the stellar assembly catalogs derived for the galaxy version of the \textsc{sublink} merger tree~\citep{2015MNRAS.449...49R}.

\section{Statistical properties}
\label{sec:3}

In this section, we present the results for the total power-law density slope $\gamma^{\prime}$ of our selected ETG sample. It is found that the profiles are indeed close to isothermal with little intrinsic scatter. Further comparison to observations and other simulations reveals that the IllustrisTNG ETGs also show consistently tight correlations between global galactic properties and the total power-law density slope. The redshift evolution of $\gamma^{\prime}$ of the IllustrisTNG ETGs is almost constant below $z = 1$ and increases with increasing redshift above $z = 1$.

\subsection{IllustrisTNG ETG total density slopes at $z = 0$}
\label{sec:3.1}

\begin{figure}
\includegraphics[width=\columnwidth]{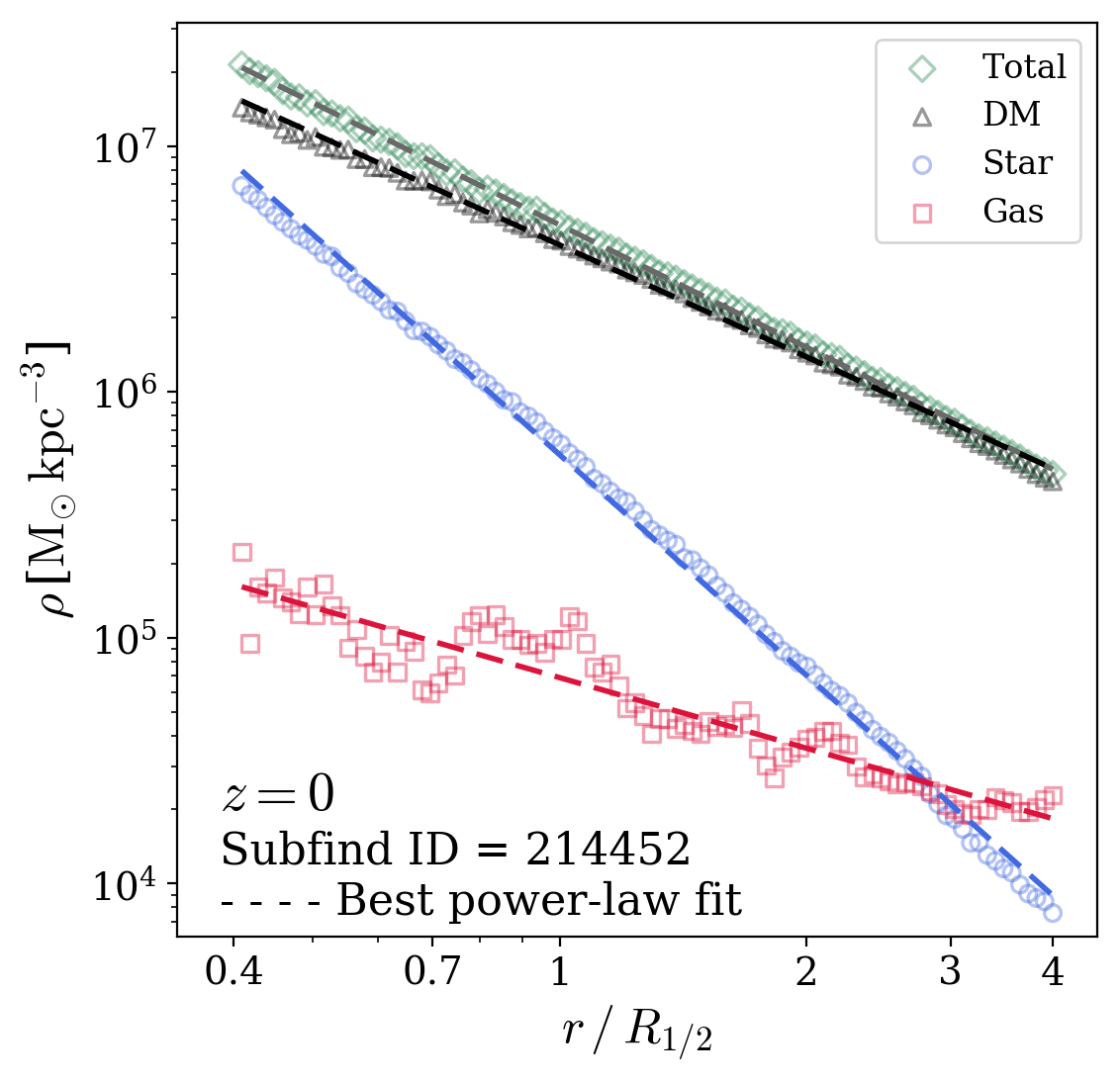}
\caption{An illustration of a selected IllustrisTNG ETG 3D radial density profile in the radial range of [0.4$\,R_{\mathrm{1/2}}$, 4.0$\,R_{\mathrm{1/2}}$], where $R_{\mathrm{1/2}}$ is the 3D stellar half mass radius. The green, black, blue, and red open markers represent the 3D density of the total, dark matter, star and gas components, respectively. The best linear fit to the density components are depicted by the dashed lines with the same color. The slopes of all the components vary little over the range of [$R_{\mathrm{1/2}}$, $4\,R_{\mathrm{1/2}}$], and the dominance of dark matter increases with the increase of the outer radial range, which accounts for the weak evolution of $\langle\gamma^{\prime}\rangle$ and the decrease in $\sigma_{\gamma^{\prime}}$ across this radial range.}
\label{fig:slope_eg}
\end{figure} 

Fig.~\ref{fig:slope_eg} presents the radial density distributions of the total (green) as well as dark matter (black), star (blue) and gas (red) of an example IllustrisTNG ETG. The best power-law fits to their radial density profiles are given by the dashed lines of the same color. Obviously, the total, stellar and dark matter components are well-described by power-law models in the radial range [$0.4\,R_{\mathrm{1/2}}$, $4\,R_{\mathrm{1/2}}$]. The stellar component profile has a steeper slope than the total density profile, while the dark matter component profile has a shallower slope than the total density profile. The stellar component dominates over the gas component, which is characteristic of gas-poor early-type galaxies.

The presence and prevalence of isothermal density profiles in the inner region of ETGs has already been demonstrated in the original Illustris simulation~\citep{2017MNRAS.469.1824X}. For the IllustrisTNG ETG sample, we follow the analysis method described in Section \ref{sec:2.3} to calculate each galaxy's total power-law density slope $\gamma^{\prime}$ within different radial ranges. We fix the inner radial limit to be $0.4\,R_{\mathrm{1/2}}$ so that the inner radius for the ETG with the smallest $R_{1/2}$ is larger than the simulation softening length ($\epsilon = 0.74\,\mathrm{kpc}$ below $z = 1$) to avoid fiducial core features in the density profiles. The minimum value for the inner radius $0.4\,R_{1/2}$ in our IllustrisTNG ETG sample is $1.17\,\mathrm{kpc}$ ($0.93\,\mathrm{kpc}$) at $z = 0$ ($z = 0.2$). We select four different radii as the outer radial limits, namely, $R_{\mathrm{1/2}}$, $2\,R_{\mathrm{1/2}}$, $3\,R_{\mathrm{1/2}}$, and $4\,R_{\mathrm{1/2}}$. This set of radial ranges is consistent with most of the observational samples in comparison, which usually measures the total density profile within a few times the effective radius (see Section \ref{sec:3.2.1} and Appendix \ref{sec:AA}). The median value for $R_{1/2}$ in our IllustrisTNG ETG sample is $7.25\,\mathrm{kpc}$ ($8.41\,\mathrm{kpc}$) at $z = 0$ ($z = 0.2$).

\begin{figure}
\includegraphics[width=\columnwidth]{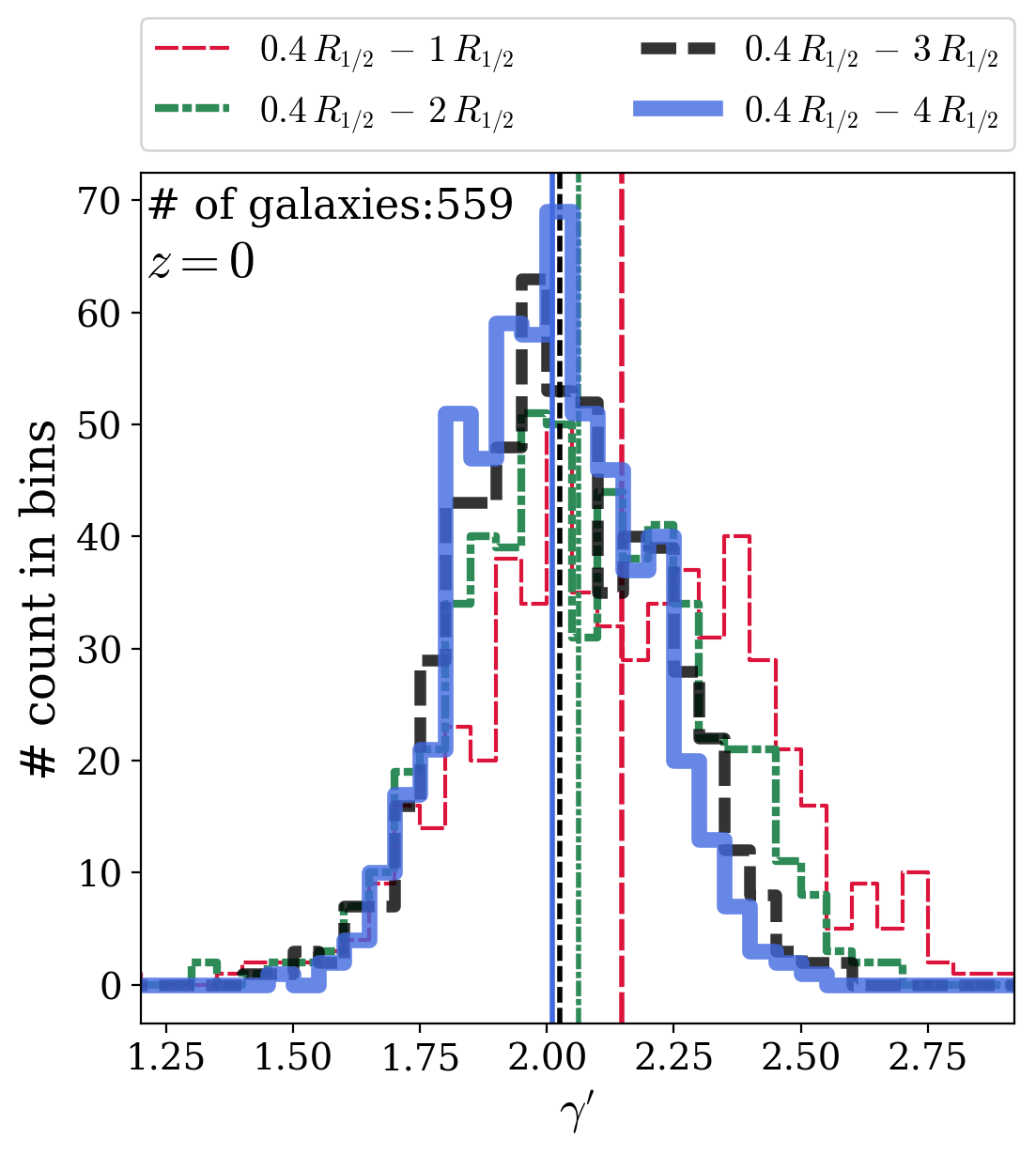}
\caption{The total power-law density slope distribution of all 559 ETGs selected from IllustrisTNG at $z=0$. Different colors represent the distribution of the total density slope in the radial range indicated in the legend. Each histogram with a certain color represents the number distribution of the total density slope measured over a certain radial range. The dashed lines correspond to the mean of the total density slope in each radial range, with the same color representing the same range as the histogram.}
\label{fig:denprof}
\end{figure}

The number count distributions of $\gamma^{\prime}$ measured within different radial ranges of the 559 ETGs in our IllustrisTNG sample at $z = 0$ are shown in Fig.~\ref{fig:denprof}. The mean $\langle\gamma^{\prime}\rangle$ of each distribution is indicated by the dashed line. 

As it can be seen, the mean distributions of $\gamma^{\prime}$ that are measured within different radial ranges are in general close to 2 and have smaller mean and scatter as the outer radius increases. The decrease in the mean is due to the increasing dominance of dark matter with increasing outer radius which leads to a shallower slope (Fig.~\ref{fig:slope_eg}), and the slight decrease in the scatter is due to the reduced Poisson noise with the increase of the total number of particles included for deriving $\gamma^{\prime}$ in wider overall radial ranges. A summary of the mean and scatter of $\gamma^{\prime}$ for the ETG sample $z = 0$ is given in Table~\ref{tab:denprof}.

\begin{table}
		\begin{center}
		\begin{tabular}{lccc}
			\hline
			Radial range & $\langle\gamma^{\prime}\rangle$ & $\sigma_{\gamma^{\prime}}$\\
			\hline
			$0.4\,R_{\mathrm{1/2}}-1\,R_{\mathrm{1/2}}$ & $2.148 \pm 0.012$ & $0.279$ \\
			$0.4\,R_{\mathrm{1/2}}-2\,R_{\mathrm{1/2}}$ & $2.063 \pm 0.010$ & $0.231$ \\
			$0.4\,R_{\mathrm{1/2}}-3\,R_{\mathrm{1/2}}$ & $2.025 \pm 0.008$ & $0.196$ \\
			$0.4\,R_{\mathrm{1/2}}-4\,R_{\mathrm{1/2}}$ & $2.011 \pm 0.007$ & $0.171$ \\
			\hline
		\end{tabular}
        \end{center}
		\caption{The mean and scatter of the total power-law density slope $\gamma^{\prime}$ of the four radial ranges over which we measured the slope for the 559 IllustrisTNG ETGs we selected. The mean $\langle\gamma^{\prime}\rangle$ is shown along with its $1\,\sigma$ error and does $not$ take into account any weighting of the global galactic properties, while the scatter $\sigma_{\gamma^{\prime}}$ shows the standard deviation of $\gamma^{\prime}$. It is obvious that the total density slope is close to isothermal  ($\gamma^{\prime} \approx 2$) throughout [$R_{\mathrm{1/2}}$, $4\,R_{\mathrm{1/2}}$] with little evolution of the mean and scatter, although it is noticeable that the mean and the scatter become smaller with increasing outer radial range.}
		\label{tab:denprof}
\end{table}

\subsection{Correlations between the total power-law slope and other galaxy properties and their comparisons to observations}
\label{sec:3.2}

As mentioned in the introduction, the presence of isothermal density profiles has been observed to be unanimous among ETGs with little intrinsic scatter~\citep{2006ApJ...649..599K,2006ApJ...646..899H,2009ApJ...703L..51K,2009MNRAS.399...21B,2011MNRAS.415.2215B,2010ApJ...724..511A,2010MNRAS.403.2143H,2011ApJ...727...96R,2013ApJ...777...98S,2014MNRAS.445..115T,2016MNRAS.460.1382S,2017MNRAS.467.1397P,2018MNRAS.476.4543B,2018MNRAS.475.2403L}. This is also true for the IllustrisTNG ETGs that we have selected as demonstrated above. The formation of such an isothermality can not be observed directly. However, studies from both observations and simulations on correlations of the total density slope with the global galactic properties could shed light on the physics during the formation and evolution process of the total density profile. 

As shown in Section \ref{sec:3.1}, $\gamma^{\prime}$ is remarkably close to $2$ even measured in large radial ranges out to $4\,R_{1/2}$ in the dark matter dominated regions of galaxies. This is in line with observations (e.g. \citealt{2010ApJ...724..511A,2013ApJ...777...98S,2015ApJ...804L..21C}) and other simulations (e.g. \citealt{2012ApJ...754..115J,2017MNRAS.469.1824X, 2017MNRAS.464.3742R}) which suggest tight constraints on galaxy formation models. Thus, we show the correlations between $\gamma^{\prime}$ that is measured within $0.4\,R_{\mathrm{1/2}}$ to $4\,R_{\mathrm{1/2}}$ and the total stellar mass $M_{\ast}$, the effective radius $R_{\mathrm{eff}}$, the stellar surface density $\Sigma_{\ast}$, the central velocity dispersion $\sigma_{\mathrm{e/2}}$, the stellar orbital anisotropy parameter $\beta$, the central dark matter fraction $f_{\rm DM}$ and the in-situ-formed stellar mass ratio $f_{\mathrm{in-situ}}$ of the IllustrisTNG ETGs, along with comparisons to existing literature datasets in this section. In the following, we first give a detailed account of the adopted comparison datasets.  

\subsubsection{Comparison datasets}
\label{sec:3.2.1}

The adopted datasets are divided into three subsets categorizing (1) local ETGs through stellar dynamic modeling, (2) higher-redshift ETGs from strong lensing surveys and (3) other numerical simulations.

The first category consists of early-type galaxies from the SPIDER~\citep{2010MNRAS.408.1313L}, $\mathrm{ATLAS^{3D}}$~\citep{2011MNRAS.413..813C, 2013MNRAS.432.1709C} and SLUGGS~\citep{2014ApJ...796...52B} surveys as well as from the observations of the Coma cluster~\citep{2007MNRAS.382..657T}. In particular, \citet{2014MNRAS.445..115T} measured the total density slopes of SPIDER and $\mathrm{ATLAS^{3D}}$ galaxies by fitting observed central kinematics ($\sigma_{\mathrm{e}}$) with two-component dynamical mass modeling using spherical Jeans equation. Similarly, \citet{2015ApJ...804L..21C} measured the mass profile of fast-rotator early-type galaxies from SLUGGS and $\mathrm{ATLAS^{3D}}$ surveys based on 2D stellar kinematics, and found near-isothermal total density profiles from $0.1\,R_{\mathrm{eff}}$ to $4\,R_{\mathrm{eff}}$. \citet{2016MNRAS.460.1382S} extended the circular velocity out to $16\,R_{\mathrm{eff}}$ with a median of $6\,R_{\mathrm{eff}}$ using \textsc{Hi} circular velocity, the total density slopes were derived using the Jeans Anisotropy Modeling (JAM) method. \citet{2017MNRAS.467.1397P} utilized a more exhaustive dataset of $\mathrm{ATLAS^{3D}}$ for central 2D kinematics modeling and derived total density slopes $\gamma^{\prime}$ and central dark matter fraction $f_{\rm DM}$ within $R_{\mathrm{eff}}$. \citet{2018MNRAS.476.4543B} measured the total density slopes of the SLUGGS galaxies from $0.1\,R_{\mathrm{eff}}$ to $4\,R_{\mathrm{eff}}$ using the JAM modeling method (same as model III in \citealt{2017MNRAS.467.1397P}). Their comparison to EAGLE and Magneticum simulated total density slopes confirmed the consistency of the simulation data with observations. We note that for all the datasets mentioned above, the quoted stellar masses are obtained by assuming a Chabrier initial mass function~\citep{2003PASP..115..763C}. For comparisons to this dataset, we use the IllustrisTNG ETG sample at $z=0$.

In the second category, the adopted strong lensing surveys contain early-type lensing galaxies up to redshift $z \approx 1$. Among these, the Lenses Structure and Dynamics (LSD) Survey~\citep{2002ApJ...575...87T,2003ApJ...583..606K,2004ApJ...611..739T} provided 5 ETG lenses at $z \approx 0.5 - 1.0$ with total slopes shallower than isothermal. \citet{2010ApJ...724..511A} reported the effective radii, central velocity dispersions, stellar masses and central dark matter fractions of galaxies from the Sloan Lens ACS (SLACS) Survey~\citep{2006ApJ...638..703B, 2008ApJ...682..964B, 2009ApJ...705.1099A}. The total power-law density slopes were derived by combining strong lensing and  stellar kinematic measurements, as practiced in \citet{2006ApJ...649..599K}.  
Also included are measurements from \citet{2011MNRAS.415.2215B}, who combined strong lensing and 2D stellar kinematic modeling techniques and applied it to the SLACS ETG sample. The CFHTLS-Strong Lensing Legacy Survey~\citep{2007A&A...461..813C, 2011ApJ...727...96R} (SL2S) offered a deeper and wider sky coverage compared to SLACS. The total power-law density slopes and general galaxy properties of SL2S galaxies mentioned above were derived in \citet{2013ApJ...777...98S}. The central dark matter fraction values for the \citet{2013ApJ...777...98S} ETG sample are taken from~\citet{2015ApJ...800...94S}. For consistency of the comparison, we convert the stellar mass measured assuming a Salpeter Initial Mass Function (IMF) in \citet{2011ApJ...727...96R} and \citet{2013ApJ...777...98S} to a Chabrier IMF using the conversion formula $M_{\ast}^{\mathrm{Chab}} = 0.61M_{\ast}^{\mathrm{Salp}}$~\citep{2014ARA&A..52..415M}. For comparisons to this dataset, we use the IllustrisTNG ETG sample at $z=0.2$, which is the median redshift of the above-mentioned strong lensing dataset. 

The third subset consists of other numerical simulation data from both large volume and zoom-in simulations. We include the $z = 0$ data of Magneticum, Oser and Wind simulations from \citet{2017MNRAS.464.3742R} for comparison with the IllustrisTNG ETG sample at $z=0$. The Magneticum Pathfinder simulations (Dolag et al. to be submitted) is a set of hydrodynamic cosmological simulations evolved using the SPH (smooth particle hydrodynamics) code \textsc{gadget3} with an updated SPH formulation. It implements AGN feedback, weak kinetic feedback from galactic winds and metal line cooling~\citep{2014MNRAS.442.2304H,2015ApJ...812...29T}. The zoom-in simulations of Oser and Wind have relatively higher resolution compared to Magneticum but exclude AGN feedback. The Oser simulation includes star formation and self-regulated SN feedback, with primordial gas cooling and without galactic winds~\citep{2010ApJ...725.2312O,2012ApJ...744...63O}. The Wind simulations include metal enrichment and strong galactic winds, producing consistent SFR and baryon conversion efficiency in low mass halos, but overestimate SFR in high mass halos in the absence of AGN feedback~\citep{2013MNRAS.436.2929H,2015MNRAS.449..528H}.

We only include data of the ETGs with stellar mass $\mathrm{log}\,(M_{\ast}/M_{\astrosun})\in[10.7, 11.9]$ from the above-mentioned literature, which is the same stellar mass range for our IllustrisTNG ETG sample. We refer the reader to Appendix \ref{sec:AA} for more details on the three comparison datasets included in this section.

\subsubsection{The correlation with the total stellar mass $M_{\mathrm{\ast}}$}
\label{sec:3.2.2}

\begin{figure}
\includegraphics[width=\columnwidth]{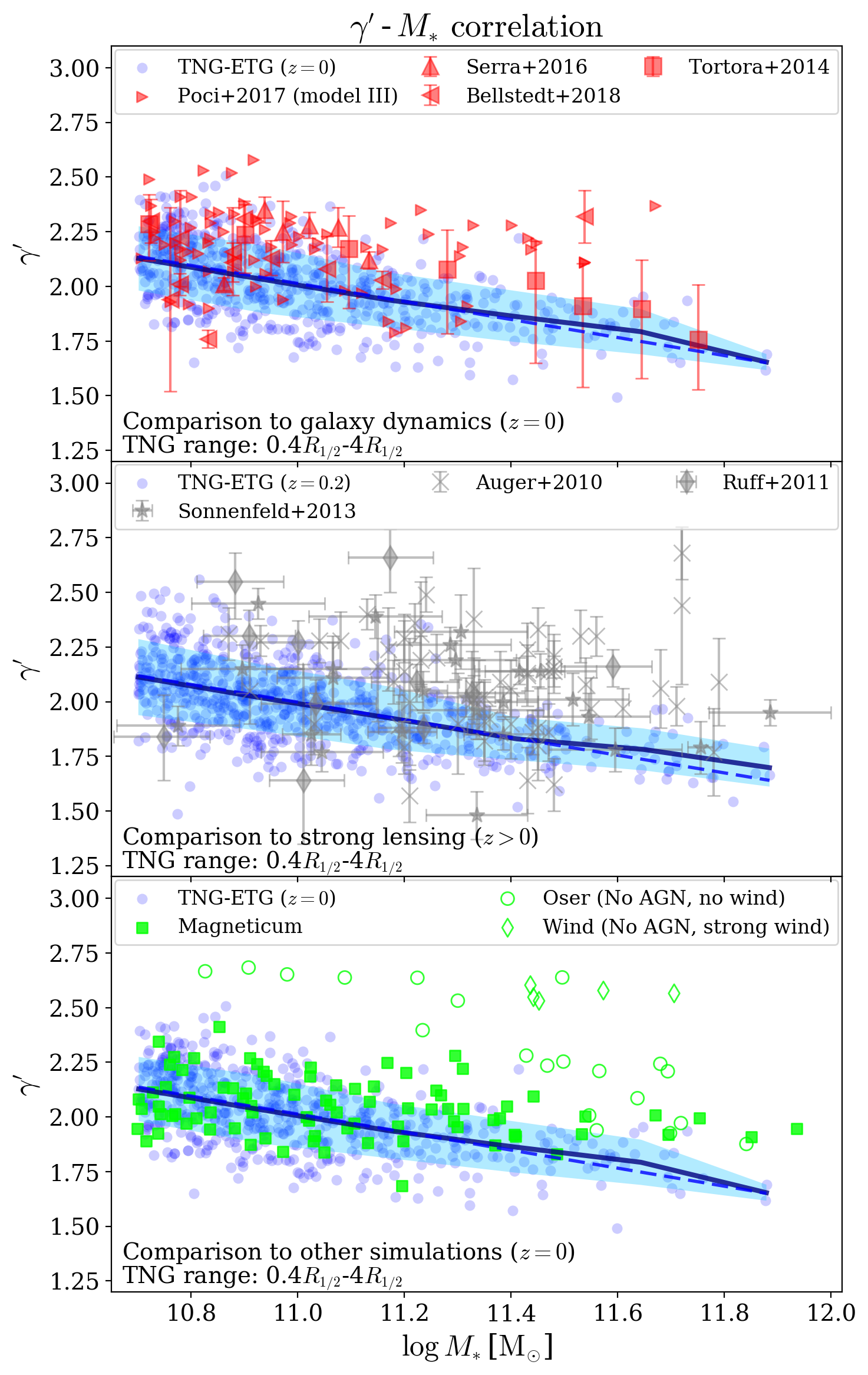}
\caption{The correlation of the total power-law density slope $\gamma^{\prime}$ and the total stellar mass $M_{\mathrm{\ast}}$. The IllustrisTNG ETGs are shown by the blue scattered dots. The solid blue curve gives the mean of the IllustrisTNG ETG slopes, and the blue shaded region shows the standard deviation of the slope distribution. The dashed blue line is the best linear fit to the IllustrisTNG ETG data points, with $\partial\gamma^{\prime}/\partial \mathrm{log} M_{\mathrm{\ast}} = -0.41 \pm 0.03$ and a Pearson correlation coefficient $r_{\mathrm{p}} = -0.58$ for $z=0$, and $\partial\gamma^{\prime}/\partial \mathrm{log} M_{\mathrm{\ast}} = -0.40 \pm 0.02$ and a Pearson correlation coefficient $r_{\mathrm{p}} = -0.57$ for $z=0.2$. The comparison datasets of dynamic modeling (red), strong lensing (grey) and other simulations (green) are shown in the subplots from top to bottom, respectively. The IllustrisTNG ETG slopes are calculated over the radial range [$0.4\,R_{\mathrm{1/2}}$, $4\,R_{\mathrm{1/2}}$].}
\label{fig:gam_mstar}
\end{figure}

The total stellar mass of the IllustrisTNG ETGs are taken as the sum of all the stellar particles assigned to its host subhalo by \textsc{subfind}. The stellar mass range is selected to be in the range of $10^{10.7}\mathrm{M}_{\mathrm{\astrosun}} \leqslant M_{\mathrm{\ast}} \leqslant  10^{11.9}\mathrm{M}_{\mathrm{\astrosun}}$ in order to compare to observations (we also use this stellar mass cut for the correlations with other global galactic properties below). Since the outer radial range for the observed total power-law density slope $\gamma^{\prime}$ spans a wide range from $R_{\mathrm{eff}}$ to $16\,R_{\mathrm{eff}}$ and $\gamma^{\prime}$ does not vary significantly from using $2\,R_{\mathrm{1/2}}$ to $4\,R_{\mathrm{1/2}}$ as the outer radial range, we use the total density slope calculated within the radial range of [0.4$\,R_{\mathrm{1/2}}$, 4$\,R_{\mathrm{1/2}}$] for the comparisons to observations and other simulation data. We also use $\gamma^{\prime}$ measured within this radial range for correlations with other global galactic properties studied in this work.

The correlation of the total power-law density slope and total stellar mass is shown in Fig.~\ref{fig:gam_mstar}. The IllustrisTNG ETGs are shown by the blue scattered dots. The solid blue curve gives the mean of the IllustrisTNG ETG slopes, and the blue shaded region shows the standard deviation of the slope distribution. Comparisons between the IllustrisTNG data (at two redshifts) and the three datasets, namely, stellar kinematic data of local ETGs (red), strong lensing data of higher-redshift galaxies (grey) and other simulation data (green), are shown in the three subplots from top to bottom. 

The dashed blue line is the best linear fit to the IllustrisTNG ETG data points, with $\partial\gamma^{\prime}/\partial \mathrm{log} M_{\mathrm{\ast}} = -0.43 \pm 0.03$ and a Pearson correlation coefficient $r_{\mathrm{p}} = -0.58$ for $z=0$, and $\partial\gamma^{\prime}/\partial \mathrm{log} M_{\mathrm{\ast}} = -0.40 \pm 0.02$ and a Pearson correlation coefficient $r_{\mathrm{p}} = -0.57$ for $z=0.2$. The power-law slope $\gamma^{\prime}$ decreases mildly as stellar mass $M_{\mathrm{\ast}}$ increases. It can be seen that the IllustrisTNG ETG data are generally in good agreement with both observation datasets and the Magneticum Simulation data which includes AGN feedback and weak galactic winds for the baryonic physical prescription. However, we also notice that IllustrisTNG ETGs follow a $\gamma^{\prime}-M_{\ast}$ correlation that has smaller scatter than the observational datasets, and the trend is also steeper compared to the Magneticum ETGs.

It is also visible from the third subplot of Fig.~\ref{fig:gam_mstar} that the Oser and Wind simulations overestimate $\gamma^{\prime}$ by $\sim 0.5$. This is related to the fact that these two zoom-in simulations omit AGN feedback, which enhances accreted stellar population and randomizes stellar orbits~\citep{2013MNRAS.433.3297D}. With the absence of AGN feedback, the influence of metal cooling and stellar winds can increase the ratio of in-situ formed stellar populations~\citep{2013MNRAS.436.2929H}, thereby increasing $\gamma^{\prime}$ through enhancing the steeper stellar density profile in the total density profile of Oser and Wind simulated galaxies. This leads to much higher $\gamma^{\prime}$ values compared to IllustrisTNG ETGs and Magneticum ETGs, and as \citet{2017MNRAS.464.3742R} mentioned, demonstrates the importance of AGN feedback for preventing over-cooling as well as expanding the galaxy eventually leading to shallower total density profiles. Also, as \citet{2018MNRAS.476.4543B} stated, the almost flat $\gamma^{\prime}-M_{\ast}$ correlation across a large dynamic range of stellar masses demands for 
inclusion of smaller mass galaxies to further constrain the simulation models. We will discuss the importance of AGN feedback in producing realistic $\gamma^{\prime}$ values and how it affects the steepness of the $\gamma^{\prime}-M_{\ast}$ correlation in detail in Section~\ref{sec:4}.

\subsubsection{The correlation with the effective radius $R_{\mathrm{eff}}$}
\label{sec:3.2.3}

\begin{figure}
\includegraphics[width=\columnwidth]{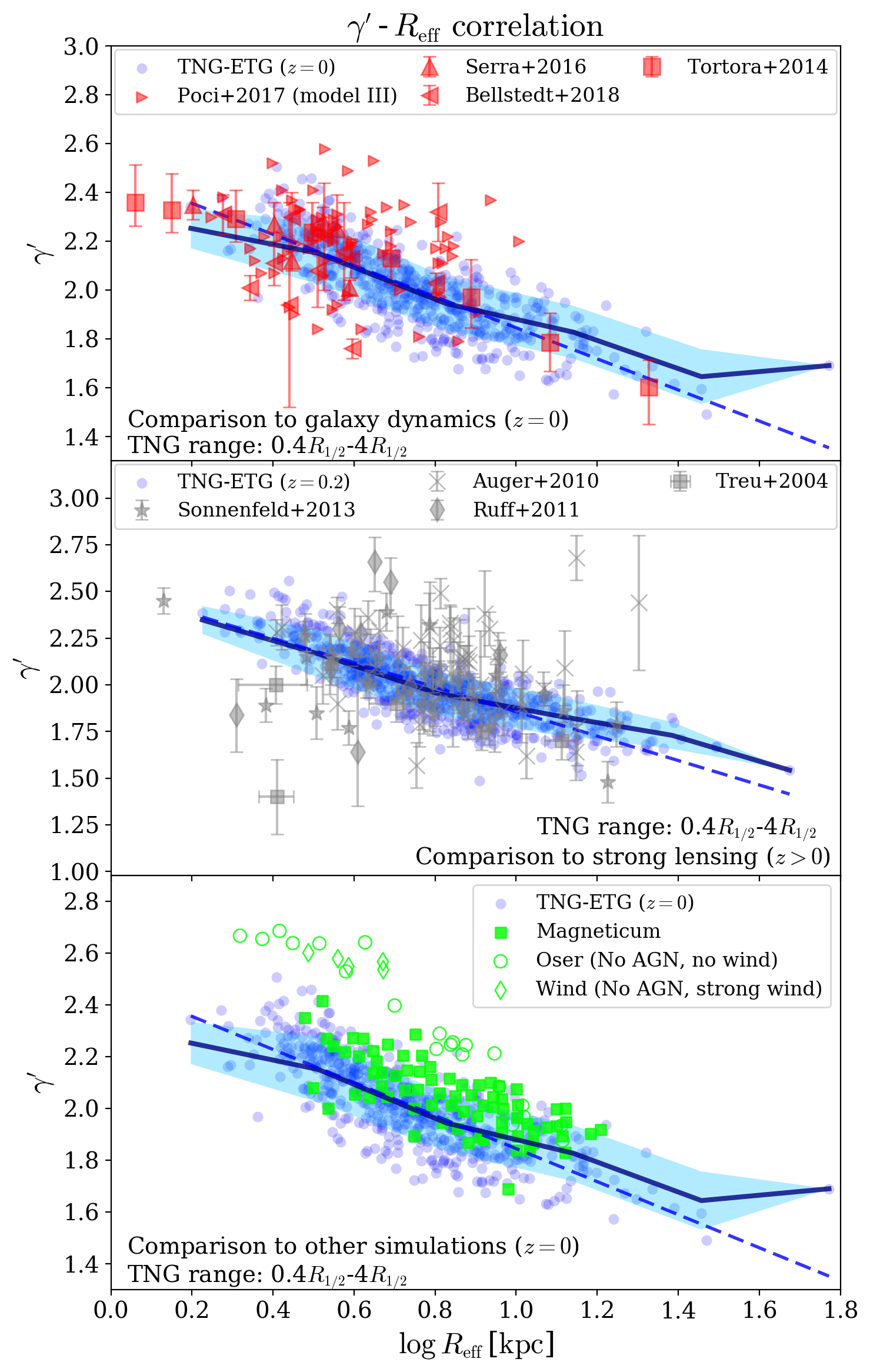}
\caption{The correlation of the total power-law density slope $\gamma^{\prime}$ and the effective radius $R_{\mathrm{eff}}$. The solid blue curve gives the mean of the IllustrisTNG ETG slopes, and the blue shaded region shows the standard deviation of the slope distribution. The dashed blue line is the best linear fit to the IllustrisTNG ETG data points, with $\partial\gamma^{\prime}/\partial \mathrm{log} R_{\mathrm{eff}} = -0.64 \pm 0.02$ and a Pearson correlation coefficient $r_{\mathrm{p}} = -0.80$ for $z=0$, and $\partial\gamma^{\prime}/\partial \mathrm{log} R_{\mathrm{eff}} = -0.65 \pm 0.02$ and a Pearson correlation coefficient $r_{\mathrm{p}} = -0.81$ for $z=0.2$. The comparison datasets of dynamic modeling (red), strong lensing (grey) and other simulations (green) are shown in the subplots from top to bottom, respectively. The IllustrisTNG ETG slopes are calculated over the radial range [0.4$\,R_{\mathrm{1/2}}$, 4$\,R_{\mathrm{1/2}}$].}
\label{fig:gam_reff}
\end{figure}

The correlation of the total power-law density slope and the effective radius $R_{\mathrm{eff}}$ is shown in Fig.~\ref{fig:gam_reff}. The best linear fit to the IllustrisTNG ETG data points gives $\partial\gamma^{\prime}/\partial \mathrm{log} R_{\mathrm{eff}} = -0.64 \pm 0.02$ and a Pearson correlation coefficient $r_{\mathrm{p}} = -0.80$ for $z=0$, and $\partial\gamma^{\prime}/\partial \mathrm{log} R_{\mathrm{eff}} = -0.65 \pm 0.02$ and a Pearson correlation coefficient $r_{\mathrm{p}} = -0.81$ for $z=0.2$. The anti-correlation between $\gamma^{\prime}$ and $R_{\mathrm{1/2}}$ is more established than that of the $\gamma^{\prime}$-$M_{\ast}$ relation.

As it can be seen from Fig.~\ref{fig:gam_reff}, the IllustrisTNG ETGs produce similar trends of the $\gamma^{\prime}-R_{\mathrm{eff}}$ correlation than the comparison datasets, but have larger $R_{\mathrm{eff}}$ in the large size end, given that the stellar mass range of the comparison dataset samples is selected to match the IllustrisTNG ETG sample. \citet{2018MNRAS.474.3976G} found that the $R_{\mathrm{eff}}$ of the IllustrisTNG quenched galaxies are larger than the observed 2D effective radius ~\citep{2003MNRAS.343..978S,2012arXiv1211.6122B,2014ApJ...788...28V} by 0.1 dex (about 2 times) in the stellar mass range of $10^{10.7}\mathrm{M}_{\astrosun} \leqslant M_{\ast} \leqslant 10^{11.9}\mathrm{M}_{\astrosun}$ (see Fig. 2b in \citealt{2018MNRAS.474.3976G}). The agreement at higher redshift ($z=0.2$) with the strong lensing dataset is marginally better than the agreement with the dynamical modeling dataset and the simulations (especially IllustrisTNG compared with Magneticum) at $z = 0$ in the large size end. The $\gamma^{\prime}$ values of IllustrisTNG ETGs are much shallower than Oser and Wind simulations that do not include AGN feedback, which agrees better with observations and reflects the necessity of AGN feedback in the lower galaxy size (mass) range. Please see Section~\ref{sec:4} for a more detailed discussion of AGN feedback effects on the correlation between $\gamma^{\prime}$ and galaxy sizes.

\subsubsection{The correlation with stellar surface density $\Sigma_{\ast}$}
\label{sec:3.2.4}

\begin{figure}
\includegraphics[width=\columnwidth]{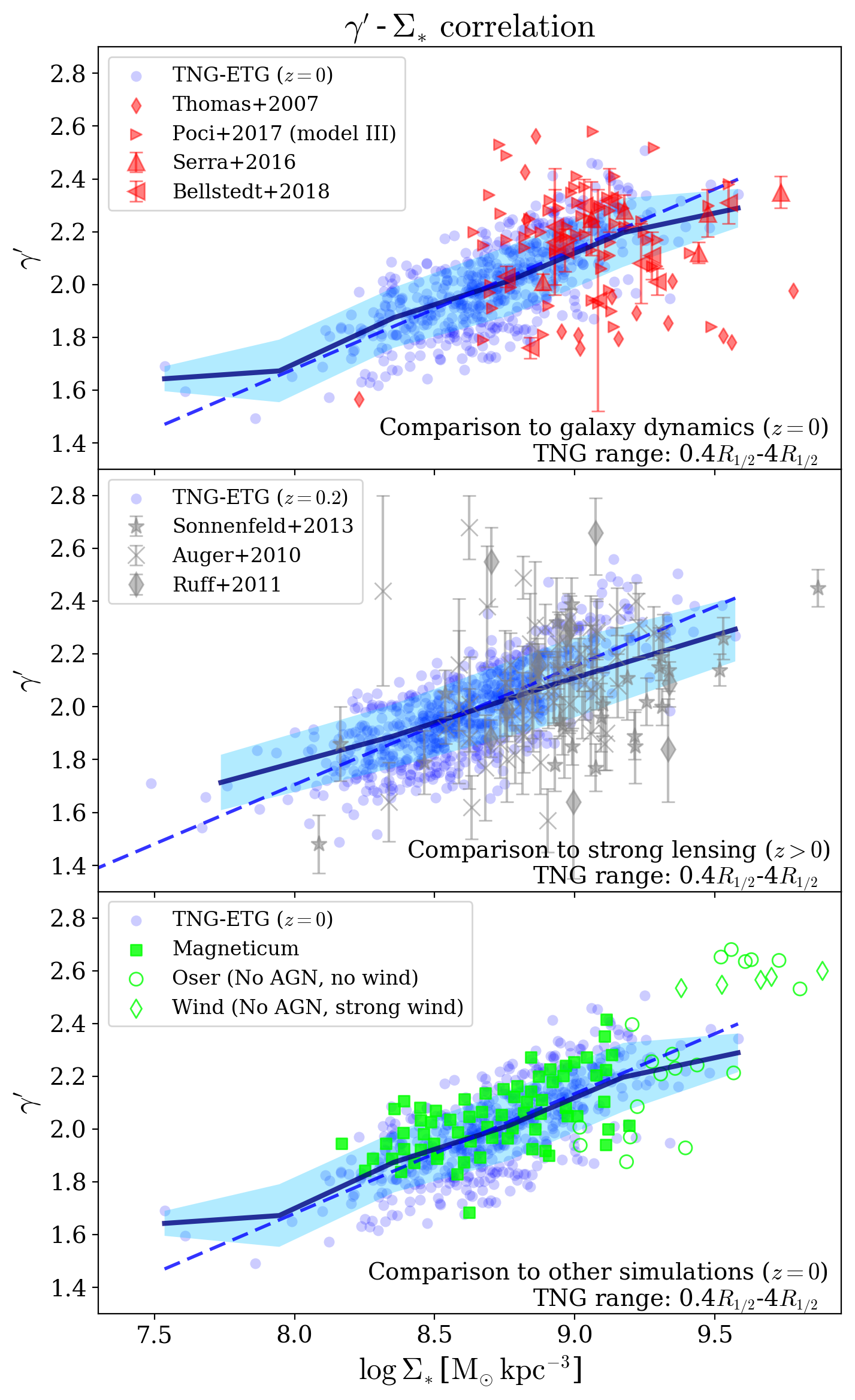}
\caption{The correlation of the total power-law density slope $\gamma^{\prime}$ and the stellar surface density $\Sigma_{\ast}$. The IllustrisTNG ETGs are shown by the blue scattered dots. The solid blue curve gives the mean of the IllustrisTNG ETG slopes, and the blue shaded region shows the standard deviation of the slope distribution. The dashed blue line is the best linear fit to the IllustrisTNG ETG data points, with $\partial\gamma^{\prime}/\partial \mathrm{log} \Sigma_{\ast} = 0.45 \pm 0.02$ and a Pearson correlation coefficient $r_{\mathrm{p}} = 0.73$ for $z=0$, and $\partial\gamma^{\prime}/\partial \mathrm{log} \Sigma_{\ast} = 0.45 \pm 0.02$ and a Pearson correlation coefficient $r_{\mathrm{p}} = 0.75$ for $z=0.2$. The comparison datasets of dynamic modeling (red), strong lensing (grey) and other simulations (green) are shown in the subplots from top to bottom, respectively. The IllustrisTNG ETG slopes are calculated over the radial range [0.4$\,R_{\mathrm{1/2}}$, 4$\,R_{\mathrm{1/2}}$].}
\label{fig:gam_sstar}
\end{figure}

We follow \citet{2013ApJ...777...98S} and define the stellar surface density as $\Sigma_{\ast} = M_{\ast}/2\pi R_{\mathrm{eff}}^{2}$. The correlation of the total power-law density slope and $\Sigma_{\ast}$ is shown in Fig.~\ref{fig:gam_sstar}. The best linear fit to the IllustrisTNG ETG data points gives $\partial\gamma^{\prime}/\partial \mathrm{log} \Sigma_{\ast} = 0.45 \pm 0.02$ and a Pearson correlation coefficient $r_{\mathrm{p}} = 0.73$ for $z=0$, and $\partial\gamma^{\prime}/\partial \mathrm{log} \Sigma_{\ast} = 0.45 \pm 0.02$ and a Pearson correlation coefficient $r_{\mathrm{p}} = 0.75$ for $z=0.2$. The total power-law density slope increases as the stellar surface density increases. As pointed out by \citet{2010ApJ...724..511A} and \citet{2014MNRAS.438.3594D}, such a positive correlation is expected since higher stellar surface density implies a higher central baryon concentration, leading to a steeper density slope. 

We note that the positive correlations in the compared observation and simulation datasets vary significantly, with \citet{2010ApJ...724..511A} giving $\partial \gamma^{\prime}/\partial \Sigma_{\ast} = 0.85 \pm 0.19$ for the SLACS sample, \citet{2017MNRAS.467.1397P} giving $\partial \gamma^{\prime}/\partial \Sigma_{\ast} = 0.174 \pm 0.045$ for the $\mathrm{ATLAS^{3D}}$ sample, and \citet{2017MNRAS.464.3742R} giving $\partial \gamma^{\prime}/\partial \Sigma_{\ast} = 0.38$ and $0.57$ for Magneticum and Oser at $z = 0$, respectively. Within the uncertainties of the comparison datasets, the IllustrisTNG $\gamma^{\prime} - \Sigma_{\ast}$ correlation is in general agreement with observations and Magneticum. However, there exists underestimation of the IllustrisTNG stellar surface density in the lower $\Sigma_{\ast}$, which is a consequence of the combination between the slightly larger sizes of the IllustrisTNG ETGs in the higher $R_{\mathrm{eff}}$ end discussed in Section \ref{sec:3.2.3}, and the rather flat $\gamma^{\prime}-M_{\ast}$ correlation shown in Section~\ref{sec:3.2.2}.

\subsubsection{The correlation with stellar kinematic properties} 
\label{sec:3.2.5}

\begin{figure}
\includegraphics[width=\columnwidth]{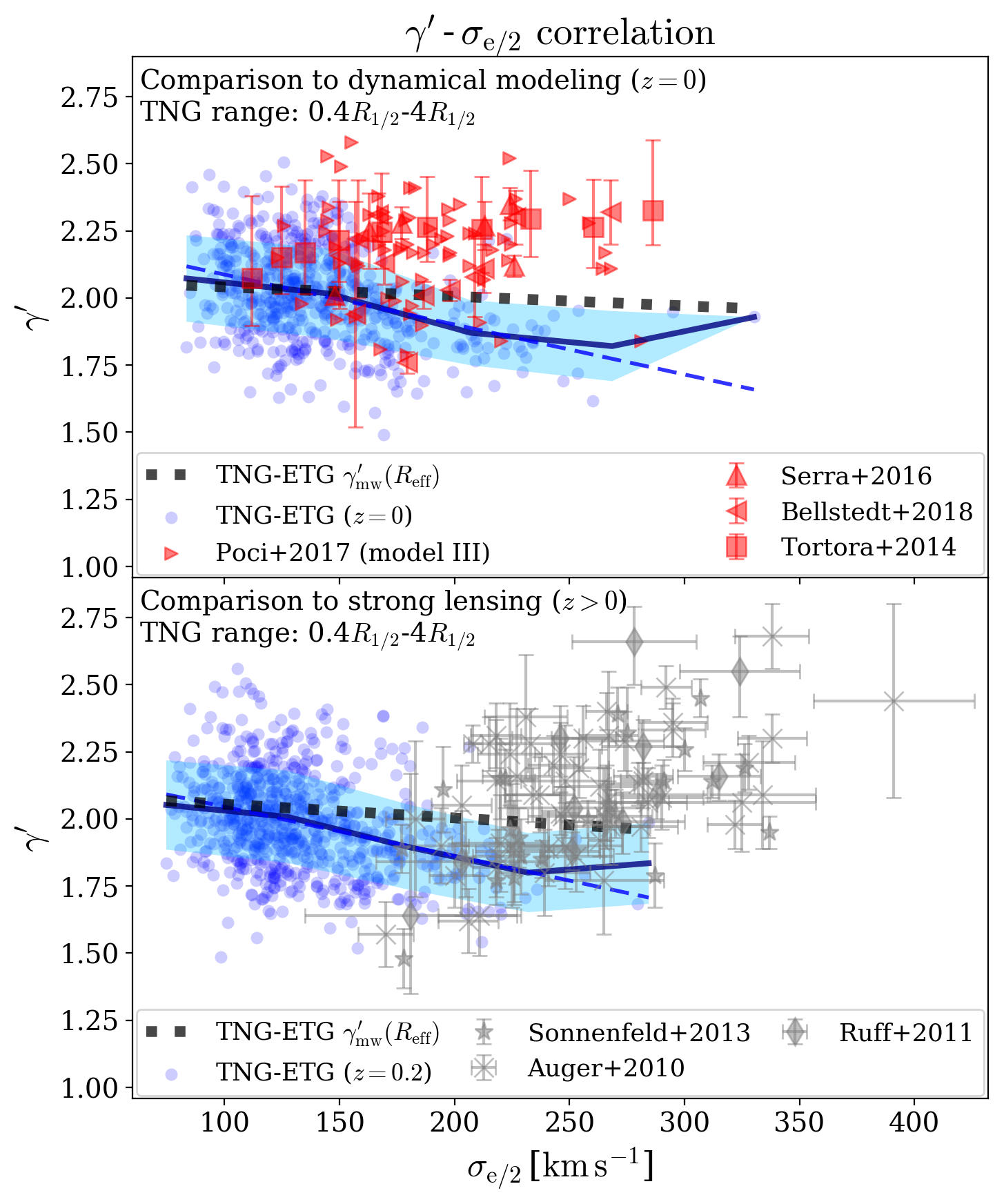}
\caption{The correlation of the total power-law density slope $\gamma^{\prime}$ and the central velocity dispersion $\sigma_{\mathrm{e/2}}$. The IllustrisTNG ETGs are shown by the blue scattered dots. The solid blue curve gives the mean of the IllustrisTNG ETG slopes, and the blue shaded region shows the standard deviation of the slope distribution. The dashed blue line is the best linear fit to the IllustrisTNG ETG data points, with $\partial\gamma^{\prime}/\partial \sigma_{\mathrm{e/2}} = -0.0019 \pm 0.0002$ and a Pearson correlation coefficient $r_{\mathrm{p}} = -0.37$ for $z=0$, and $\partial\gamma^{\prime}/\partial \sigma_{\mathrm{e/2}} = -0.0018 \pm 0.0002$ and a Pearson correlation coefficient $r_{\mathrm{p}} = -0.33$ for $z=0.2$. The best linear fit to the $\gamma_{\mathrm{mw}}^{\prime}(R_{\mathrm{eff}})-\sigma_{\mathrm{e/2}}$ correlation for IllustrisTNG ETGs are shown by the black dotted curve in the two panels. The comparison datasets of dynamic modeling (red) and strong lensing (grey) are drawn by scattered dots with error bars in the top and bottom subplots, respectively. The IllustrisTNG ETG slopes are calculated over the radial range [$0.4\,R_{\mathrm{1/2}}$, $4\,R_{\mathrm{1/2}}$].}
\label{fig:gam_sv}
\end{figure}

The central velocity dispersion $\sigma_{\mathrm{e/2}}$ is measured as the projected stellar-luminosity-weighted line-of-sight velocity dispersion within a 2D aperture of $0.5\,R_{\mathrm{eff}}$ for the IllustrisTNG ETGs. The correlation of the total power-law density slope and the central velocity dispersion is shown in Fig.~\ref{fig:gam_sv}. The best linear fit to the IllustrisTNG ETG data points gives $\partial\gamma^{\prime}/\partial \sigma_{\mathrm{e/2}} = -0.0019 \pm 0.0002$ and a Pearson correlation coefficient $r_{\mathrm{p}} = -0.37$ for $z=0$, and $\partial\gamma^{\prime}/\partial \sigma_{\mathrm{e/2}} = -0.0018 \pm 0.0002$ and a Pearson correlation coefficient $r_{\mathrm{p}} = -0.33$ for $z=0.2$. 

We notice that the velocity dispersions of the IllustrisTNG ETGs are systematically lower than their observational counterparts, which span a similar range in stellar masses. As it can be seen in Section \ref{sec:5.3} (Fig.~\ref{fig:gam_m200_fp}), the total galaxy and halo mass ($M_{200}$) of the simulated galaxies are also markedly smaller than those derived for observed galaxies. Both systematic inconsistencies, combined with the excess of central dark matter fractions as found in \citet{2018MNRAS.481.1950L}, indicate potential problems with the baryonic models of the simulation, which could result in a different mix of baryons and dark matter in central regions of galaxies as well as an overestimation of baryonic sizes of galaxies. These could result in not only the underestimation of the velocity dispersion, but also alter the overall trend of the $\gamma^{\prime} - \sigma_{\mathrm{e/2}}$ correlation.

Furthermore, a problematic morphology-size relation has been found for IllustrisTNG compared with Pan-STARRS observations~\citep{2019MNRAS.483.4140R}. Bulge-dominated galaxies have larger sizes than their disk-dominated counterparts with similar stellar masses in IllustrisTNG, which is opposite to the Pan-STARRS observations. This trend reversal could lead to high-velocity dispersion galaxies (higher bulge dominance) in IllustrisTNG having larger sizes ($R_{\mathrm{eff}}$) and hence lower $\gamma^{\prime}$ (see Fig.~\ref{fig:gam_reff}) compared to the low-velocity dispersion galaxies (lower bulge dominance). Therefore, the morphology-size relation mismatch with observations in IllustrisTNG could be a major factor to account for the apparent $\gamma^{\prime}-\sigma_{\mathrm{e/2}}$ correlation mismatch shown in Fig.~\ref{fig:gam_sv}.

Apart from simulation limitations, we have to beware that observational studies of the $\gamma^{\prime}-\sigma_{\mathrm{e/2}}$ correlation is prone to systematic biases. The observed trends of the two comparison datasets from dynamical modeling and strong gravitational lensing show marked differences at $\sigma_{\mathrm{e/2}}>200$km/s. In fact, \citet{2010ApJ...724..511A} measured $\partial\gamma^{\prime}/\partial\sigma_{\mathrm{e/2}} = 0.07 \pm 0.08$ for the SLACS sample of higher-redshift lensing galaxies, and \citet{2017MNRAS.467.1397P} obtained $\partial\gamma^{\prime}/\partial\sigma_{\mathrm{e}} = 0.442 \pm 0.081$ through 2D kinematics modeling for the local ETG sample. The visible systematic difference in the $\sigma_{\mathrm{e/2}}$ distribution between the two observational datasets can be explained by the fact that within the same stellar-mass range, galaxies with larger central velocity dispersion will have higher probabilities to act as gravitational lenses. Thus, strong lens galaxies in general have high stellar surface densities which leads to steeper slopes and adds an additional source of systematic bias to the total density slopes. 

Also, mass-weighted slopes obtained in dynamic modeling of ETGs (e.g. \citealt{2014MNRAS.445..115T}) could also alter the $\gamma^{\prime}$ values. These assumptions could switch the $\gamma^{\prime}-\sigma_{\mathrm{e/2}}$ correlation from negative to positive as shown for Illustris ETGs in \citet{2017MNRAS.469.1824X} (see their Fig. 17), and contribute to the $\gamma^{\prime}-\sigma_{\mathrm{e/2}}$ trend discrepancy as displayed in Fig.~\ref{fig:gam_sv}. The mass weighted slope at $\gamma^{\prime}_{\mathrm{mw}}\,(R_{\mathrm{eff}})$ is defined as~\citep{2014MNRAS.438.3594D}:
\begin{equation}
\label{equ:gmw}
\gamma^{\prime}_{\mathrm{mw}}(R_{\mathrm{eff}}) = \frac{-1}{M(r)}\int_{0}^{R_{\mathrm{eff}}} \gamma^{\prime}_{\mathrm{l}}(r) 4\pi r^{2} \rho(r) dr = 3 - \frac{4\pi R_{\mathrm{eff}}^{3}\rho(R_{\mathrm{eff}})}{M(R_{\mathrm{eff}})}\,,
\end{equation}
where $M(r)$ is the total mass enclosed within $r$, $\rho(r)$ is the local matter density at $r$, and the local total power-law density slope is defined as $\gamma^{\prime}_{\mathrm{l}}(r) = -d\,\mathrm{log} \,\rho(r)/d\,\mathrm{log}\,r$. The mass-weighted slope $\gamma^{\prime}_{\mathrm{mw}}$ will be identical to the best-fit total power-law density slope $\gamma^{\prime}$ defined in Section~\ref{sec:2.3} when the total radial density profile is in an exact power-law form.

In Fig.~\ref{fig:gam_sv}, we show the best linear fit to the $\gamma_{\mathrm{mw}}^{\prime}(R_{\mathrm{eff}})-\sigma_{\mathrm{e/2}}$ correlation for IllustrisTNG ETGs. The best linear fit gives $\partial\gamma_{\mathrm{mw}}^{\prime}(R_{\mathrm{eff}}) = -0.004 \pm 0.0002$ and a Pearson correlation coefficient $r_{\mathrm{p}}=-0.0879$ at $z=0$, while $\partial\gamma_{\mathrm{mw}}^{\prime}(R_{\mathrm{eff}}) = -0.005 \pm 0.0002$ and a Pearson correlation coefficient $r_{\mathrm{p}}=-0.1161$ at $z=0.2$. As shown in Fig.~\ref{fig:gam_sv}, even if we adopt the mass-weight slope definition, the discrepancy of the $\gamma^{\prime}-\sigma_{\mathrm{e/2}}$ correlation between IllustrisTNG and the dynamic modeling dataset are only mitigated but not fully reconciled, with a mildly negative $\gamma_{\mathrm{mw}}^{\prime}(R_{\mathrm{eff}})-\sigma_{\mathrm{e/2}}$ correlation at both $z=0$ and $z=0.2$. This again suggests limitations in the baryonic models of the simulation, and that different density slope definitions alone could not fully reconcile the discrepancy between simulated results and observed ones apart from the observational modeling biases. 

To further investigate the possible effect from stellar orbital anisotropy, we follow the same practice and adopt the definition of the 3D anisotropy parameter under spherical symmetry~\citep{2008gady.book.....B}:
\begin{equation}
\beta = 1 - \frac{\sigma_{\mathrm{\phi}}^{2} + \sigma_{\mathrm{\theta}}^{2}}{2\sigma_{\mathrm{r}}^{2}}\,,
\end{equation}
where $\sigma_{\mathrm{r}}$, $\sigma_{\mathrm{\phi}}$ and $\sigma_{\mathrm{\theta}}$ are the velocity dispersion within the stellar half mass radius in the radial, azimuthal and polar directions, respectively. $\beta = 0$ corresponds to the isotropic case, $\beta > 0$ stands for radially biased orbits and $\beta < 0$ stands for tangentially biased orbits. In practice, we calculate $\beta$ for all the stellar particles enclosed within a 3D aperture with radius of $R_{\mathrm{eff}}$, each stellar particle weighted by its mass.

\begin{figure}
\includegraphics[width=\columnwidth]{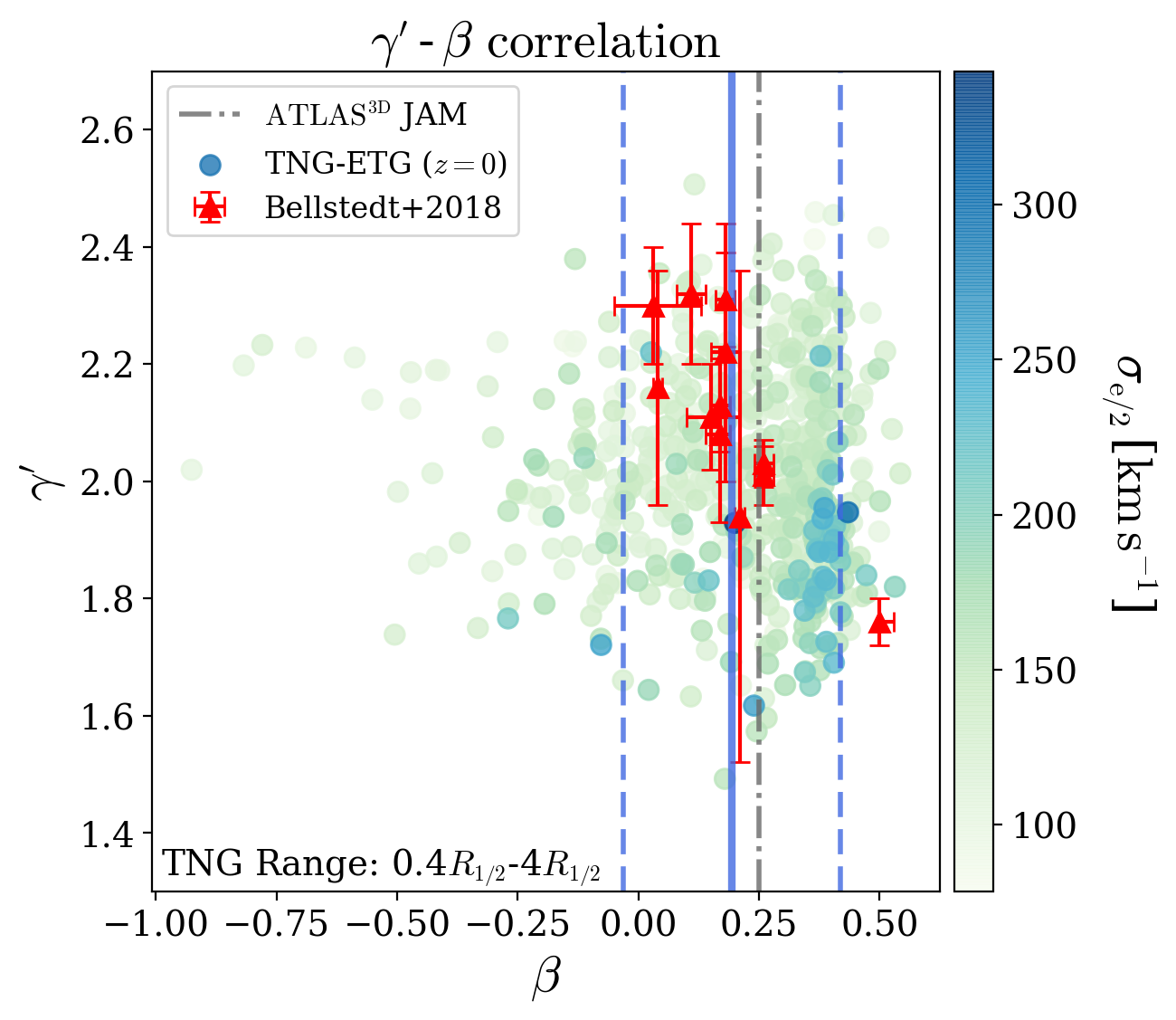}
\caption{The correlation of the total power-law density slope $\gamma^{\prime}$ and the stellar orbital anisotropy parameter $\beta$. The IllustrisTNG ETGs are shown by the scattered dots, and the color index indicates the central velocity dispersion $\sigma_{\mathrm{e/2}}$. The mean of the IllustrisTNG ETG anisotropy parameter $\langle\beta\rangle = 0.226 \pm 0.008$ is shown by the blue line, whereas the standard deviation of the $\beta$ distribution $\sigma_{\beta} = 0.194$ is shown by the dashed lines. $\beta = 0.25$ assumed by $\mathrm{ATLAS^{3D}}$~\citep{2013MNRAS.432.1862C} is indicated by the grey dashed dotted line. Red triangles with error bars are JAM modeling $\beta$ values from \citet{2018MNRAS.476.4543B}. There is no clear correlation between $\gamma^{\prime}$ and $\beta$ for the IllustrisTNG ETGs. Also, stellar orbits are more radially (tangentially) biased in ETGs with higher (lower) central velocity dispersion.}
\label{fig:gam_beta}
\end{figure}

In Fig.~\ref{fig:gam_beta}, we show the $\gamma^{\prime}-\beta$ distribution for the IllustrisTNG ETGs (solid circles) at $z=0$, color-coded with their central velocity dispersion $\sigma_{\mathrm{e/2}}$. Over-plotted are observational data (red triangles with error bars) from \citet{2018MNRAS.476.4543B}, where $\beta$ came from applying the JAM modeling technique to the stellar kinematic data. We adopt the $\gamma^{\prime}$ and $\beta$ values of their model II where the dark matter halo is modeled by a gNFW profile and the stellar mass distribution is parametrized from the observed luminosity. The $\mathrm{ATLAS^{3D}}$ Survey employs an MGE (JAM) modeling method to obtain the inferred stellar mass from IFU spectroscopy~\citep{2013MNRAS.432.1862C}. They chose a fixed anisotropy parameter $\beta = 0.25$ (grey dotted dashed line) to infer the bulge fractions from 2D stellar kinematics, which represents the typical anisotropy parameter for early type galaxies~\citep{2008MNRAS.390...71C} . 

The IllustrisTNG ETGs have a mean anisotropy $\langle\beta\rangle = 0.226 \pm 0.008$ (blue solid line), and a scatter of $\sigma_{\beta} = 0.194$ (blue dashed lines), possessing radially biased stellar orbits typical for early type galaxies, also in good agreement with observations. The total power-law density slope $\gamma^{\prime}$ does not show significant correlation with $\beta$. The most massive galaxies with higher central velocity dispersions tend to have radially anisotropic and shallower density profiles, which corresponds to giant elliptical slow rotators. The opposite corresponds to lenticular fast rotators~\citep{2017ApJ...838...77L}. 

In practice, adopted modeling approaches could also add to the total density slope systematics apart from the lens selection bias. In particular, the total power-law density slopes for the strong-lensing sample are routinely derived under the assumption of an isotropic velocity dispersion. As shown by, e.g., \citet{2006ApJ...649..599K, 2009ApJ...703L..51K, 2017MNRAS.469.1824X}, the true total density slope of a galaxy can be overestimated (underestimated) with the isotropic assumption, if its stellar kinematics is radially (tangentially) anisotropic. Hence, if isotropy ($\beta=0$) is assumed for all ETGs in our sample just as the slope measurement process for strong lenses, overestimation of $\gamma^{\prime}$ in the ETGs with positive $\beta$ and large $\sigma_{\mathrm{e/2}}$ (underestimation of $\gamma^{\prime}$ in ETGs vice versa) could change the trend of the $\gamma^{\prime}-\sigma_{\mathrm{e/2}}$ correlation, which accounts for the discrepancy between IllustrisTNG and the strong lensing dataset in Fig.~\ref{fig:gam_sv}. However, quantifying the amount of systematic bias in modeling methods, strong lens sample selection bias, and limitations of the simulation models are beyond the scope of this paper and will be discussed in future work.

\subsubsection{The correlation with the central dark matter fraction $f_{\mathrm{DM}}$} 
\label{sec:3.2.6}

\begin{figure}
\includegraphics[width=\columnwidth]{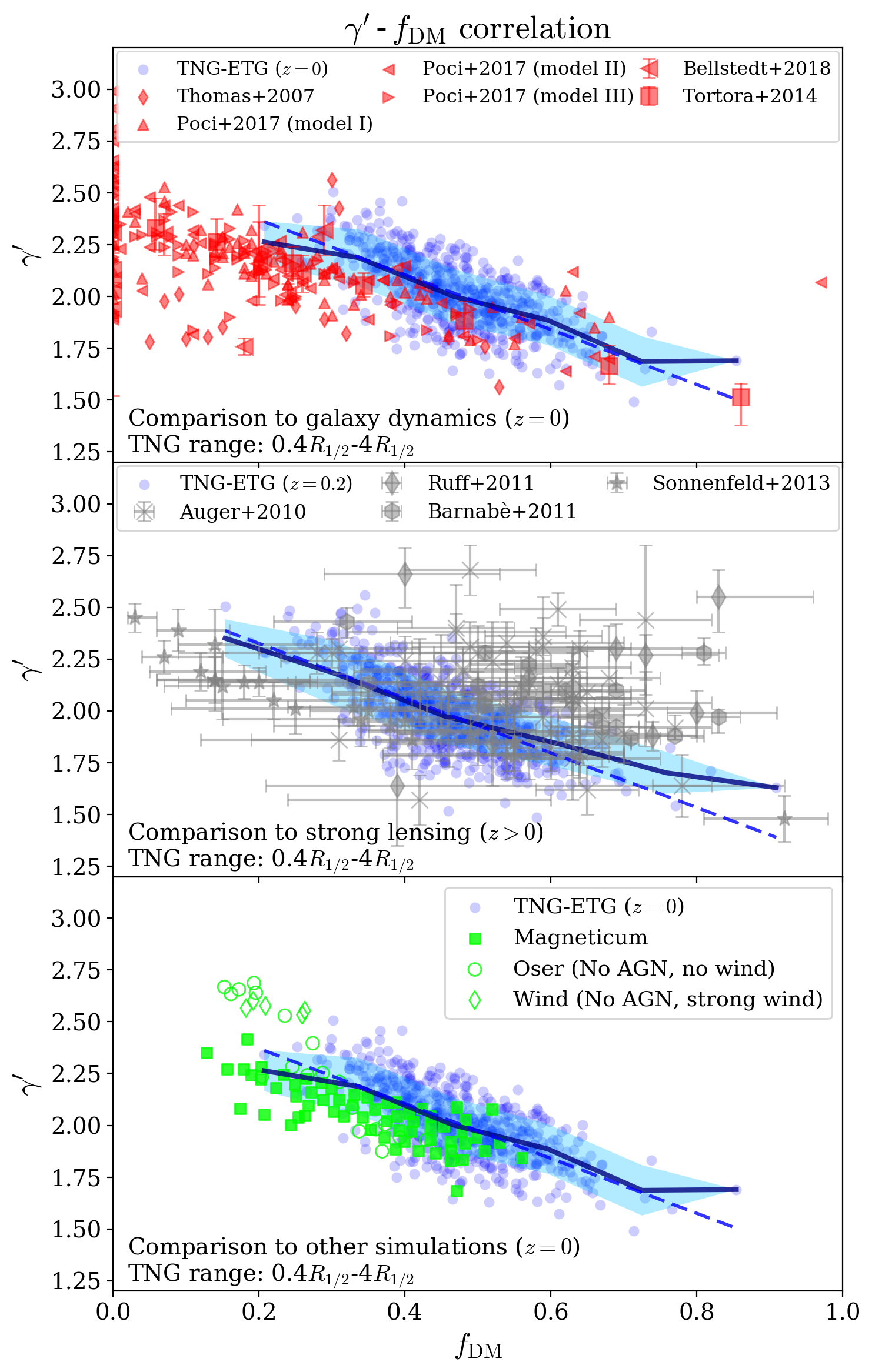}
\caption{The correlation of the total power-law density slope $\gamma^{\prime}$ and the central dark matter fraction $f_{\mathrm{DM}}$. The IllustrisTNG ETGs are shown by the blue scattered dots. The solid blue curve gives the mean of the IllustrisTNG ETG slopes, and the blue shaded region shows the standard deviation of the slope distribution. The dashed blue line is the best linear fit to the IllustrisTNG ETG data points, with $\partial\gamma^{\prime}/\partial f_{\mathrm{DM}} = -1.33 \pm 0.06$ and a Pearson correlation coefficient $r_{\mathrm{p}} = -0.70$ for $z=0$, and $\partial\gamma^{\prime}/\partial f_{\mathrm{DM}} = -1.32 \pm 0.05$ and a Pearson correlation coefficient $r_{\mathrm{p}} = -0.70$ for $z=0.2$. The comparison datasets of dynamic modeling (red), strong lensing (grey) and other simulations (green) are shown in the subplots from top to bottom, respectively. The IllustrisTNG ETG slopes are calculated over the radial range [$0.4\,R_{\mathrm{1/2}}$, $4\,R_{\mathrm{1/2}}$].}
\label{fig:gam_fdm}
\end{figure}

The central dark matter fraction $f_{\mathrm{DM}}$ is defined as the mass ratio of dark matter over the total mass of all simulation particles enclosed within a sphere of a given radius for the IllustrisTNG ETGs. We measure the central dark matter fraction of the IllustrisTNG ETGs within a 3D aperture with radius of $R_{\mathrm{eff}}$, to match the adopted aperture size used in most of the observational data analysis (see Appendix \ref{sec:AA}).

The correlation of the total power-law density slope and the central dark matter fraction is shown in Fig.~\ref{fig:gam_fdm}. We note that the stellar masses of all comparison datasets have been converted to those assuming a Chabrier IMF for the consistency of the total stellar mass. A clear anti-correlation between $\gamma^{\prime}$ and $f_{\mathrm{DM}}$ is seen for the IllustrisTNG ETG sample. The best linear fit to the $z=0$ IllustrisTNG ETG dataset gives $\partial\gamma^{\prime}/\partial f_{\mathrm{DM}} = -1.33 \pm 0.06$ and a Pearson correlation coefficient $r_{\mathrm{p}} = -0.70$, in good agreement with stellar-kinematic modeling dataset (upper panel). The simulation data of the Magneticum Pathfinder~\citep{2017MNRAS.464.3742R} gives $\partial\gamma^{\prime}/\partial f_{\mathrm{DM}}\approx -1$, which is close to our IllustrisTNG ETG result as shown (bottom panel). The best linear fit to the $z=0.2$ IllustrisTNG ETGs gives $\partial\gamma^{\prime}/\partial f_{\mathrm{DM}} = -1.32 \pm 0.05$ and a Pearson correlation coefficient $r_{\mathrm{p}} = -0.70$, in rough agreement with the strong lensing sample (middle panel) given that the lensing sample possesses systematics of the IMF inference. A Salpeter IMF (as favored by strong lensing observations) would result in lower central dark matter fractions for observed galaxies and mitigate the apparent mismatch of $f_{\mathrm{DM}}$ in the middle panel.

We point out that central dark matter fractions produced by IllustrisTNG are systematically larger than $\mathrm{ATLAS^{3D}}$ and Magneticum values in the upper and bottom panels of Fig.~\ref{fig:gam_fdm}, respectively. However, comparing to the strong lensing dataset at higher redshift, the larger dark matter fraction for IllustrisTNG ETGs seems to better match the observed $f_{\mathrm{DM}}$ values from strong lensing, despite being slightly shallower in $\gamma^{\prime}$ for similar $f_{\mathrm{DM}}$ values. These systematics in the central dark matter fraction have been quantified for the IllustrisTNG simulation in~\citet{2018MNRAS.481.1950L}: depending on the dataset of comparison, IllustrisTNG ETGs at $z=0$ may have an excess of dark matter in the innermost regions of galaxies ($\lesssim R_{\mathrm{eff}}$), but consistent with available measurements at larger apertures ($\lesssim 5\,R_{\mathrm{eff}}$), which is due to the extended sizes of the stellar component in the IllustrisTNG galaxies raising their central dark matter fraction (see Fig. 12 and discussions in \citealt{2018MNRAS.481.1950L}). The inclusion of black hole kinetic winds (low-accretion rate AGN feedback mode) and black hole thermal feedback (high accretion rate AGN feedback mode) increases the central dark matter fraction $f_{\mathrm{DM}}(r<R_{1/2})$ significantly compared with galactic winds (stellar feedback) or other model variations~(see Fig. 9 of \citealt{2018MNRAS.481.1950L}) in our IllustrisTNG ETG sample's halo mass range ($\mathrm{log}\,M_{200}\gtrsim10^{12}\,\mathrm{M_{\astrosun}}$, see Fig.~\ref{fig:gam_m200_fp}). So the effects of stronger (weaker) AGN feedback in more (less) massive IllustrisTNG ETGs could lead to larger (smaller) sizes and central dark matter fractions that result in shallower (steeper) $\gamma^{\prime}$. We will discuss how the individual effects of AGN and stellar feedback affect the $\gamma^{\prime}-f_{\mathrm{DM}}$ correlation and the other four correlations mentioned above in Section~\ref{sec:4}. The time evolution effects of AGN feedback energy on $\gamma^{\prime}$ will be shown in more detail along with the effects from galaxy mergers in an upcoming paper (Wang et al. in prep.).
 
\subsubsection{The correlation with the in-situ-formed stellar mass ratio}
\label{sec:3.2.7}

We use the in-situ-formed stellar mass ratio $f_{\mathrm{in-situ}}$ as determined by the \textsc{sublink} merger tree~\citep{2015MNRAS.449...49R,2016MNRAS.458.2371R} of each galaxy in our IllustrisTNG ETG sample, and study the correlation between $\gamma^{\prime}$ and $f_{\mathrm{in-situ}}$. $f_{\mathrm{in-situ}}$ is defined as the stellar mass of stars formed within the main progenitor branch of the galaxy versus the total stellar mass of the galaxy at $z=0$.

The total power-law density slope has been found to be positively correlated with $f_{\mathrm{in-situ}}$ in zoom-in and cosmological simulations~\citep{2013ApJ...766...71R,2017MNRAS.464.3742R,2018MNRAS.476.4543B}. \citet{2013ApJ...766...71R} showed that the total density slope $\gamma^{\prime}$ positively correlates with $f_{\mathrm{in-situ}}$ for CosmoZoom Simulation ETGs.  \citet{2018MNRAS.476.4543B} found $\partial\gamma^{\prime}/\partial f_{\mathrm{in-situ}} = 0.44$ for Magneticum ETGs with $M_{\ast} > 10^{10.7}\mathrm{M}_{\astrosun}$. For Oser ETGs, \citet{2017MNRAS.464.3742R} also found a negative correlation between $f_{\mathrm{in-situ}}$ and the central dark matter fraction which indicates a positive correlation between $\gamma^{\prime}$ and $f_{\mathrm{in-situ}}$. As demonstrated by \citealt{2013MNRAS.433.3297D}, AGN feedback quenches in-situ star formation and enhances the proportion of accreted stellar populations of the central ETG decreasing its $f_{\mathrm{in-situ}}$, while AGN feedback also flattens its total density profile leading to the positive $\gamma^{\prime}-f_{\mathrm{in-situ}}$ correlation. 

\begin{figure}
\includegraphics[width=\columnwidth]{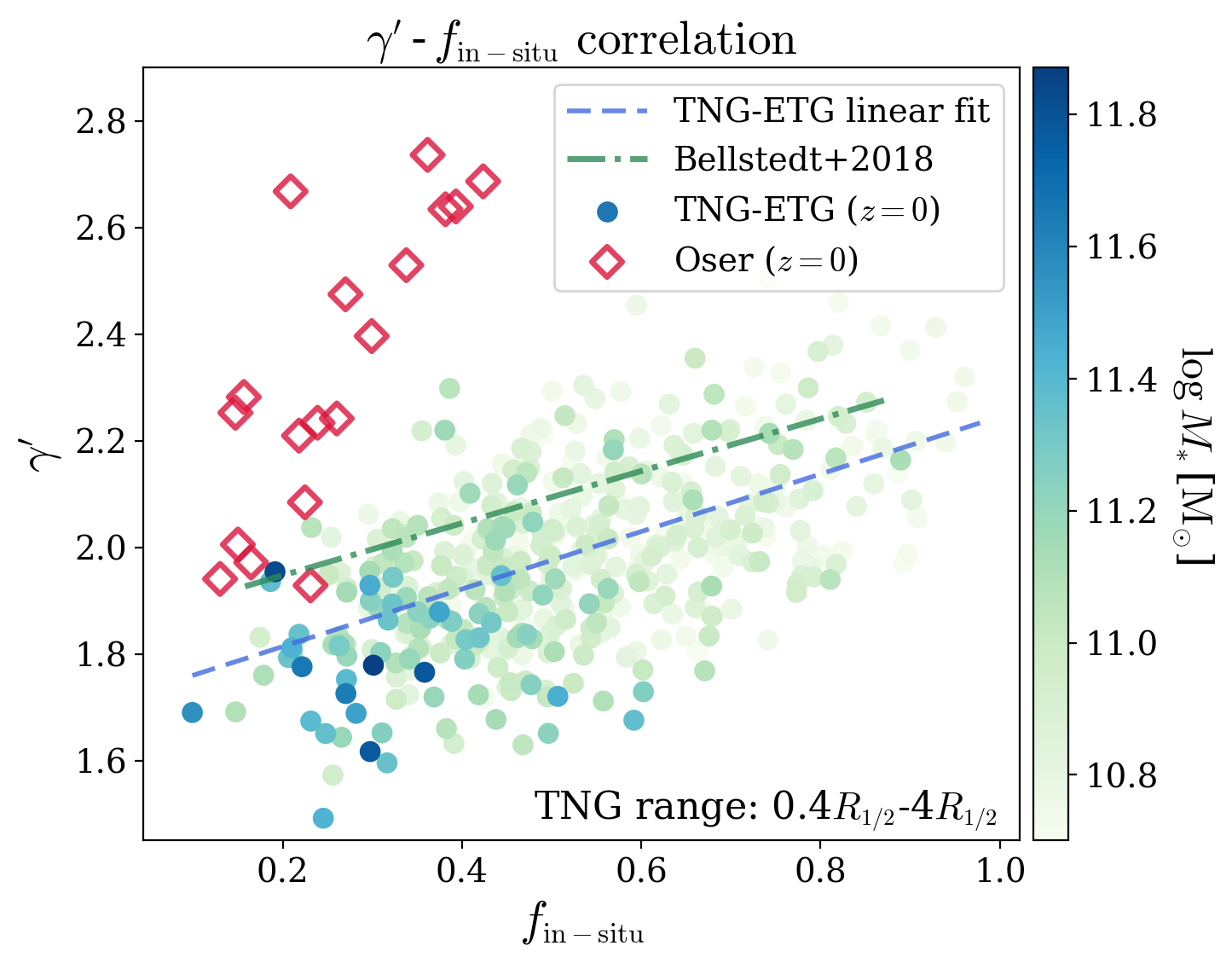}
\caption{The correlation of the total power-law density slope $\gamma^{\prime}$ and the in-situ-formed stellar mass ratio $f_{\mathrm{in-situ}}$ for the IllustrisTNG ETGs at $z = 0$. The IllustrisTNG ETGs are shown by the scattered dots, and the color index indicates the total stellar mass $M_{\ast}$. The dashed blue line is the best linear fit to the IllustrisTNG ETG data points, with $\partial\gamma^{\prime}/\partial f_{\mathrm{in-situ}} = 0.54 \pm 0.03$ and $r_{\mathrm{p}} = 0.58$. The best linear fit of the $\gamma^{\prime}-f_{\mathrm{in-situ}}$ correlation for Magneticum ETGs~\citep{2018MNRAS.476.4543B} with stellar mass $\mathrm{log}\,(M_{\ast}/\mathrm{M_{\astrosun}})>10.7$ is shown by the green dotted-dashed line. Data of Oser ETGs from \citet{2017MNRAS.464.3742R} are denoted by the red scattered diamonds.}
\label{fig:gam_f_insitu}
\end{figure}

The correlation between the total power-law density slope $\gamma^{\prime}$ and the in-situ-formed stellar mass ratio $f_{\mathrm{in-situ}}$ for the IllustrisTNG ETGs at $z = 0$ is shown in Fig.~\ref{fig:gam_f_insitu}. The dashed blue line is the best linear fit to the IllustrisTNG ETG data points, with $\partial\gamma^{\prime}/\partial f_{\mathrm{in-situ}} = 0.54 \pm 0.03$ and a Pearson correlation coefficient $r_{\mathrm{p}} = 0.58$. The best linear fit of the $\gamma^{\prime}-f_{\mathrm{in-situ}}$ correlation for Magneticum ETGs~\citep{2018MNRAS.476.4543B} with stellar mass $\mathrm{log}\,(M_{\ast}/\mathrm{M_{\astrosun}})>10.7$ is shown by the green dotted-dashed line. Data of Oser ETGs from \citet{2017MNRAS.464.3742R} are denoted by the red scattered diamonds. The positive correlation between $f_{\mathrm{in-situ}}$ and $\gamma^{\prime}$ is well produced by our IllustrisTNG ETG sample, consistent with  Magneticum. The Oser ETGs show a consistent trend for the $\gamma^{\prime}-f_{\mathrm{in-situ}}$ correlation, but have too steep $\gamma^{\prime}$ for given $f_{\mathrm{in-situ}}$ values due to the lack of AGN feedback. In all, shallower (steeper) profiles correlating with lower (higher) in-situ-formed stellar mass ratios preferably in higher (lower) mass galaxies indicate the dominant role played by gas-poor dry galaxy mergers in the formation of ETGs below $z=2$. Dry mergers occur more often for higher mass galaxies and continuously build up the density of their outskirts which lead to lower in-situ-formed stellar mass ratios and shallower total density slopes~\citep{2007ApJ...658..710N,2009ApJS..182..216K,2009ApJ...699L.178N,2012ApJ...754..115J,2013MNRAS.433.3297D,2013ApJ...766...71R,2015MNRAS.449..528H}. 

As shown in the analysis above, the total power-law density slope of the IllustrisTNG ETGs are in broad agreement with observations considering all the correlations with galaxy parameters presented above. A summary of all the correlations with galaxy parameters for the IllustrisTNG ETGs is given in Table~\ref{tab:FITS}. 

\begin{table*}
		\begin{center}
		\begin{tabular}{lccccc}
			\hline
			X & $\partial\gamma^{\prime}/\partial\mathrm{X}$ $(z=0)$ & $r_{\mathrm{p}}$ $(z=0)$ & $\partial\gamma^{\prime}/\partial\mathrm{X}$ $(z=0.2)$ & $r_{\mathrm{p}}$ $(z=0.2)$\\
			\hline
			$\mathrm{log}\,M_{\ast}$ & $-0.41 \pm 0.03$ & $-0.58$ & $-0.40\pm 0.02$ & $-0.57$\\
			$\mathrm{log}\,R_{\mathrm{eff}}$ & $-0.64 \pm 0.02$ & $-0.80$ & $-0.65 \pm 0.02$ & $-0.81$ \\
			$\mathrm{log}\,\Sigma_{\ast}$ & $0.45 \pm 0.02$ & $0.73$ & $0.45 \pm 0.02$ & $0.75$ \\
			$\sigma_{\mathrm{e/2}}$ & $-0.0019 \pm 0.0002$ & $-0.37$ & $-0.0018 \pm 0.0002$ & $-0.33$ \\
            $f_{\mathrm{DM}}$ & $-1.33 \pm 0.06$ & $-0.70$ & $-1.32 \pm 0.05$ & $-0.70$ \\
            $f_{\mathrm{in-situ}}$ & $0.54 \pm 0.03$ & $0.58$ & $-$ & $-$ \\
			\hline
		\end{tabular}
        \end{center}
		\caption{The best linear fit of the correlations with galaxy parameters for our selected IllustrisTNG ETGs. X stands for the different galaxy parameters, $\partial\gamma^{\prime}/\partial\mathrm{X}$ is the slope of the best linear fit correlation, and $r_{\mathrm{p}}$ is the Pearson correlation coefficient of the corresponding best linear fit. A `$-$' is assigned to any field that is not applicable.}
		\label{tab:FITS}
\end{table*}

\subsection{Galaxy redshift dependence}
\label{sec:3.3}

\begin{figure*}
\includegraphics[width=1.9\columnwidth]{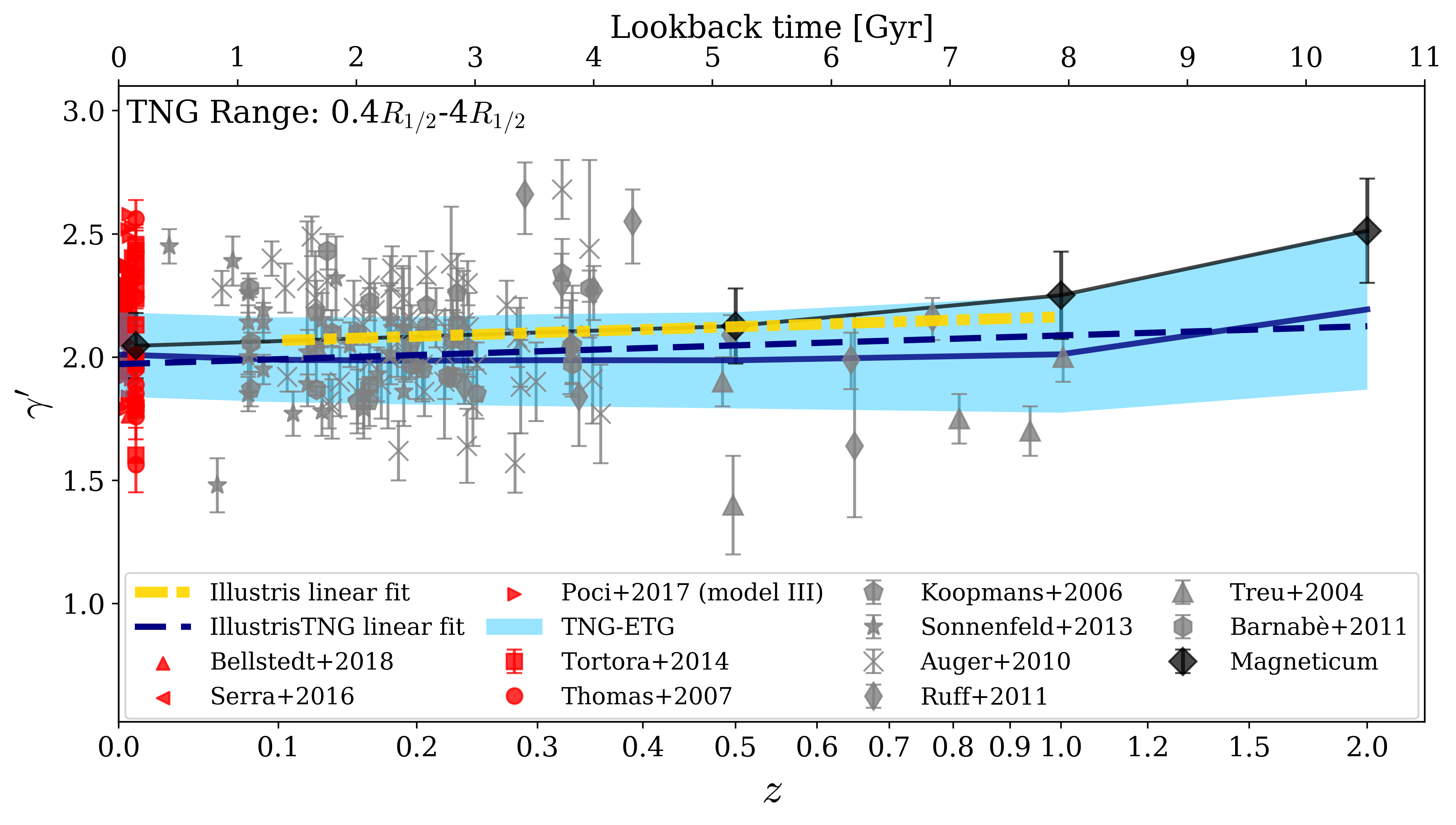}
\caption{The evolution of the total power-law density slope with redshift. The IllustrisTNG ETG redshift evolution is shown in blue, with the solid line denoting the mean, the shaded region denoting the standard deviation, and the dashed line denoting the best linear fit. The best linear fit of the IllustrisTNG ETGs gives $\partial\gamma^{\prime} / \partial z = 0.0116 \pm 0.0097$ and a Pearson correlation coefficient r = 0.0189. As for comparison, the dynamic modeling data are shown in red, and the strong lensing data are shown in grey. The Magneticum ETG redshift evolution from \citet{2017MNRAS.464.3742R} is shown in green, with the solid line denoting the mean and the shaded region denoting the standard deviation of the distribution. The best linear fit of the Illustris ETG power-law slope redshift evolution measured over the radial range [$0.5\,R_{\mathrm{eff}}$, $1.0\,R_{\mathrm{eff}}$] is shown by the magenta dashed line~\citep{2017MNRAS.469.1824X}. The IllustrisTNG ETGs show little to no evolution of the total density slope below $z=1$, with a mild increasing trend of the slope with increasing redshift.}
\label{fig:z_obs}
\end{figure*}

The redshift evolution of the total power-law density slope is shown in Fig.~\ref{fig:z_obs}. The Illustris ETG samples at $z=$~[0.1, 0.2, 0.3, 0.5, 1.0, 2.0] are selected using the same method as we used to select the ETGs at $z=0$ (see section 2.2 for details). The strong lensing dataset is shown by the grey scattered points with error bars along with the stellar-kinematic dataset at $z \approx 0$ shown in red in the same figure.

As it can be seen, the IllustrisTNG ETGs total density slope shows little to no evolution below $z = 1$, and displays a slight increase in the slope above $z=1$. The best linear fit of the IllustrisTNG ETGs gives $\partial\gamma^{\prime} / \partial z = 0.0768 \pm 0.0065$ and a Pearson correlation coefficient $r_{\mathrm{p}}$ = 0.1795.

For the observational datasets, a general trend of the total density slope becoming steeper with time is demonstrated. \citet{2011ApJ...727...96R} reported $\partial\gamma^{\prime}/\partial z_{d} = -0.25_{-0.12}^{+0.10}$ for 11 SL2S lens ETGs, suggesting the that dissipative processes steepen the density profile of ETGs since $z = 1$. A similar trend was found by \citet{2013ApJ...777...98S} for 36 confirmed strong lenses and 17 SL2S strong lens candidates, with $\partial\gamma^{\prime}/\partial z = -0.31 \pm 0.10$, and a full redshift evolution of the total density slope $d\gamma^{\prime}/dz = -0.10 \pm 0.12$ taking into account the redshift evolution of galaxy stellar mass and effective radius. Subsequent theoretical study of the evolutionary trend advocates the necessity for wet mergers involving cold gas to account for the steepening evolution of the total density profile at $z \leqslant 1$ ~\citep{2014ApJ...786...89S}. 

However, the latest cosmological simulations show tension with the currently observed redshift evolution trends. The redshift evolution trend of the Magneticum ETGs from \citet{2017MNRAS.464.3742R} with $\partial\gamma^{\prime}/\partial z = 0.21$ in the redshift range $z = 0 - 2$ is displayed in Fig.~\ref{fig:z_obs}. We neglect the Oser and Wind datasets for the $\gamma^{\prime}$ redshift evolution since they are zoom-in simulations with very different AGN feedback and galactic wind models, which produced unrealistically steep $\gamma^{\prime}$ values compared to observations as shown in the previous sections. The shallowing evolutionary trend of the slope towards low redshift is significant, and large deviation from the strong lensing data is visible for $z \geqslant 0.5$. Similar trends were also discovered by \citet{2012ApJ...754..115J,2013ApJ...766...71R,2017MNRAS.469.1824X} in other cosmological simulations. Such a trend is consistent with the scenario of gas-poor dry mergers dominating the mass and size growth of ETGs at $z \leqslant 2$, leading to decreases in $f_{\mathrm{in-situ}}$ and thus shallower total density slopes with time. Interestingly, \citet{2017MNRAS.469.1824X} found a steepening trend of $\gamma_{0}^{\mathrm{LD}}$ towards lower redshift, where $\gamma_{0}^{\mathrm{LD}}$ is the slope obtained by combining strong lensing and stellar dynamics and $\partial \gamma_{0}^{\mathrm{LD}}/\partial z = -0.03 \pm 0.01$ with $r_{\mathrm{p}} = -0.03$, while assuming isotropic stellar orbits for the Illustris ETGs. In contrast, a shallowing trend of the intrinsic power-law slope $\partial\gamma^{\mathrm{PL}}/\partial z = 0.11 \pm 0.01$ with $r_{\mathrm{p}} = 0.11$ was seen for the same Illustris ETG sample. Similarly, the steepening evolution trend of $\gamma^{\prime}$ with increasing redshift in \citet{2017MNRAS.464.3742R} vanishes if the slopes are deducted using mock-strong lensing pipelines, which changes into being constant with time. Since strong lensing slopes involve various model assumptions including isotropic stellar orbits~\citep{2013ApJ...777...98S,2017MNRAS.464.3742R} and power-law models~\citep{2016MNRAS.456..739X} which inevitably suffer from systematic biases, the slope redshift evolution discrepancy might be rooted in these strong model assumptions, although it is not clear if the discrepancy also involves simulation limitations. Furthermore, the strong lensing selection bias of ETGs with steeper inner slope projection effects could also add to the discrepancy between simulated and observed $\gamma^{\prime}$ redshift evolution trends~\citep{2017MNRAS.464.3742R,2017MNRAS.469.1824X}. 

In general, the redshift evolution of $\gamma^{\prime}$ of the IllustrisTNG ETGs is in line with other numerical simulations in comparison, demonstrating a decrease in the total power-law density slope with time since $z=2$. This trend still exhibits some tension with the strong lensing observation dataset, which suggests a seemingly increasing slope with time. In our upcoming paper (Wang et al. in preparation), we will aim at quantifying the effects of galaxy mergers, star formation activities, and feedback processes on the formation and evolution of the isothermal density profile.

\section{Effects of feedback model variations on $\gamma^{\prime}$}
\label{sec:4}

\subsection{Methods}
\label{sec:4.1}

In this subsection, we present the galaxy correlations under various simulation model variations to demonstrate the effects of different physical prescriptions on $\gamma^{\prime}$. These variation runs are conducted in a smaller box with a side length of $25\, h^{-1}\,\mathrm{Mpc}$ containing $2\times 512^{3}$ resolution elements for a mass resolution that is similar to TNG100. A detailed description of the various model variations can be found in the TNG methods paper~(\citealt{2018MNRAS.473.4077P}, and references therein). 

We cconsider four different model variations apart from the IllustrisTNG fiducial model (same as the TNG100 model as described in Section~\ref{sec:2.1}, labeled as `TNG' for short in tables and figures), each with only one modification to the IllustrisTNG default model: \textbf{i)} Stronger winds with higher energy, labeled as `StrongerWinds' for short; \textbf{ii)} No stellar wind feedback, metal cooling and AGN feedback on, labeled as `NoWinds' for short; \textbf{iii)} No black holes, AGN feedback turned off, metal cooling and stellar winds on, labeled as `NoBHs'; \textbf{iv)} No black hole low accretion rate kinetic wind feedback, thermal quasar mode AGN feedback acts at all accretion rates, labeled as `NoKineticBHwinds' for short.

For the model variation ETG samples, we first select massive central galaxies with total stellar mass in the range of [$10^{10.7}\mathrm{M_{\astrosun}}$, $10^{11.9}\mathrm{M_{\astrosun}}$] in the TNG model run. Then, the ETG sample is selected with the same criteria described in Section~\ref{sec:2.2} in these massive central galaxies. This results in a sample of 32 central ETGs in the TNG model run. Since the dark matter particles share the same IDs and the same initial conditions across different model variations, we match the ETG sample in the TNG run with galaxies in other model variation runs through the dark matter particles that they have in common. The total density profile power-law slope $\gamma^{\prime}$ is calculated in the radial range [$0.4R_{1/2}$, $4R_{1/2}$], where $R_{1/2}$ is the stellar half mass radius of the ETGs in the corresponding simulation. 

We have checked that the gas component is subdominant compared the stellar and dark matter components across the samples in all models that we have selected, although we do not explicitly show it here for brevity. In addition, a single power-law fit provides a reasonable fit to the total density profiles in all the different models, with the strongest deviation from a single power-law occurring in the `NoBHs' run where a central stellar cusp is present due to starburst in the absence of AGN feedback. In such a case, the outskirts of the galaxies are still well-described by a single power-law total density profile, and $\gamma^{\prime}$ still reflects the compactness of the matter distribution for stars and dark matter combined. 

\subsection{Results}
\label{sec:4.2}

\begin{figure*}
\includegraphics[width=1.9\columnwidth]{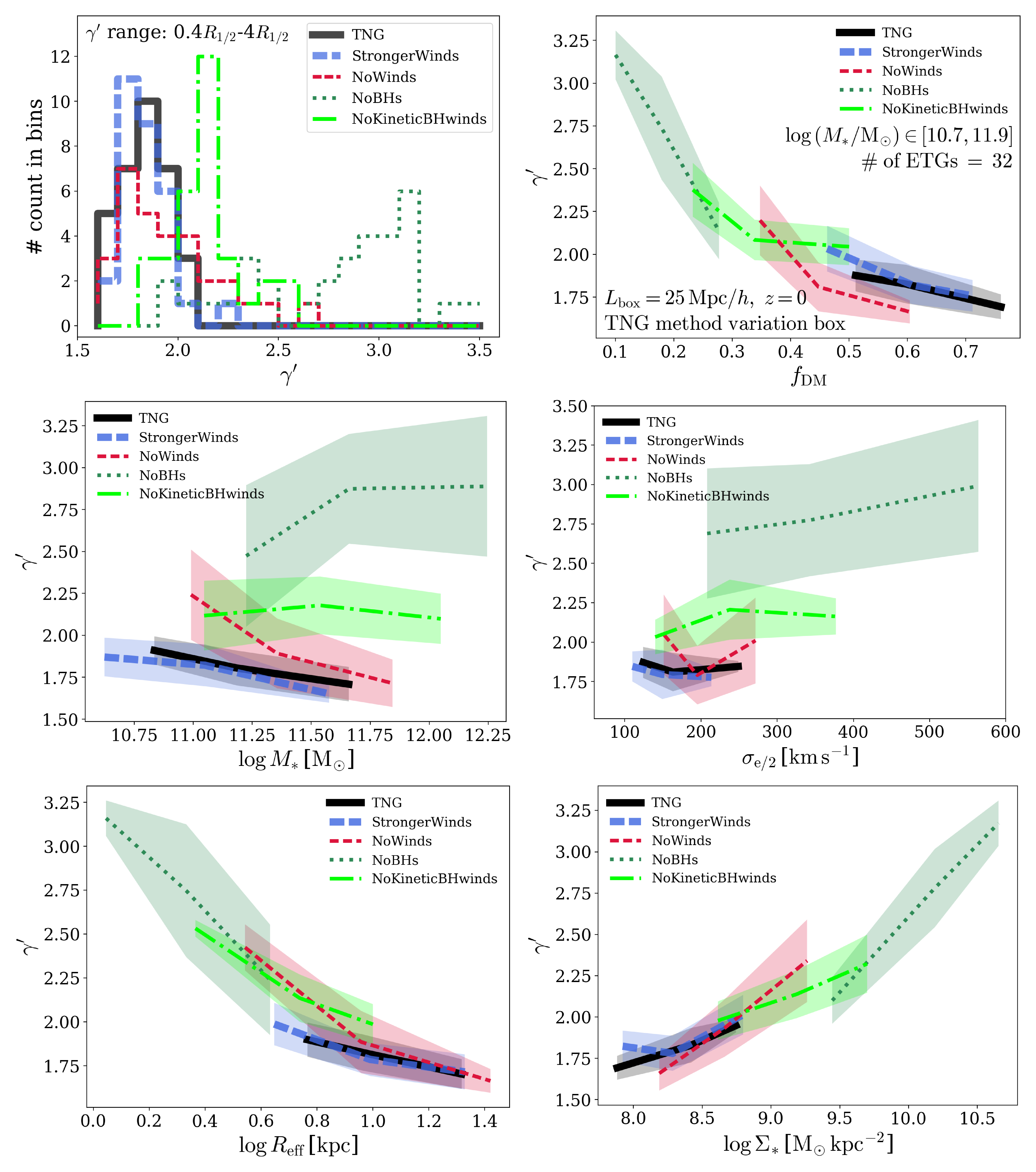}
\caption{TNG model variations. Top left: histograms of the $\gamma^{\prime}$ distributions for different model variations. Top right: mean $\gamma^{\prime}-f_{\mathrm{DM}}$ correlation. Middle left: mean $\gamma^{\prime}-M_{\ast}$ correlation. Middle right: mean $\gamma^{\prime}-\sigma_{\mathrm{e/2}}$ correlation. Bottom left: mean $\gamma^{\prime}-R_{\mathrm{eff}}$ correlation. Bottom right: mean $\gamma^{\prime}-\Sigma_{\ast}$ correlation.  All the mean correlations are binned in three equal particle bins with respect to the five different galaxy properties. The shaded regions indicate the $1-\sigma$ scatter of the correlations.}
\label{fig:vary}
\end{figure*}

\begin{table}
		\begin{center}
		\begin{tabular}{lcc}
			\hline
			Model & $\langle\gamma^{\prime}\rangle$ & $\sigma_{\gamma^{\prime}}$ \\
			\hline
			TNG & $1.835 \pm 0.020$ & $0.114$ \\
			StrongerWinds & $1.816 \pm 0.023$ & $0.131$ \\
			NoWinds & $1.924 \pm 0.046$ & $0.259$ \\
			NoBHs & $2.763 \pm 0.073$ & $0.411$ \\
			NoKineticBHwinds & $2.153 \pm 0.032$ & $0.182$ \\
            \hline
		\end{tabular}
        \end{center}
		\caption{The mean and standard deviation of the total density profile power-law slope $\gamma^{\prime}$ in the five model variations. All $\gamma^{\prime}$ are calculated in the radial range [$0.4R_{1/2}$, $4R_{1/2}$], where $R_{1/2}$ is the stellar half mass radius of the corresponding galaxies in each model variation.}
		\label{tab:vary}
\end{table}

In the top left panel of Fig.~\ref{fig:vary} we show the histograms for the $\gamma^{\prime}$ distribution in the five model variations. We summarize the mean and the standard deviation of the distributions in Table~\ref{tab:vary}. The first piece of information to notice is that with stronger winds (`StrongerWinds'), $\gamma^{\prime}$ is marginally shallower than in the TNG model, while reducing feedback, either stellar winds (`NoWinds') or black hole kinetic winds (`NoKineticBHwinds') increase $\gamma^{\prime}$. The complete removal of blacks holes and subsequent AGN feedback in the `NoBHs' case has the most significant impact and creates the steepest $\gamma^{\prime}$ with a large scatter. Overall, the effects of BH related feedback has a stronger impact on $\gamma^{\prime}$ in comparison with the stellar feedback. We point out that for the TNG model, the $\gamma^{\prime}$ values are slightly shallower than the TNG100 values as we have shown in the previous section. This is due to the slightly coarser simulation resolution in the model variation box smearing out the density profile, and this effect is stronger for lower-mass galaxies which in general have fewer resolution elements and steeper $\gamma^{\prime}$ intrinsically.  

In the other five panels of Fig.~\ref{fig:vary}, we show the mean correlations between $\gamma^{\prime}$ and the total stellar mass, the effect radius, the stellar surface density, the central velocity dispersion, and the central dark matter fraction across different model variations. All the mean correlations are binned in three equal particle bins with respect to the five different galaxy properties. The shaded regions indicate the $1-\sigma$ standard deviation of the five correlations. 

In the middle left panel of Fig.~\ref{fig:vary}, the $\gamma^{\prime}-M_{\ast}$ correlation is shown for all the model variations. With stronger stellar wind feedback, both the total stellar mass and $\gamma^{\prime}$ are slightly reduced. Opposite behavior with larger stellar mass and steeper $\gamma^{\prime}$ happens for the `NoWinds' model, and the deviation from the TNG model is more significant at the lower mass end. However, the anti-correlation of $\gamma^{\prime}-M_{\ast}$ is preserved with the presence of the TNG black hole models in these two stellar wind model variations. In contrast, the `NoBHs' case shows a much higher total stellar mass due to missing AGN feedback, and a reversed $\gamma^{\prime}-M_{\ast}$ trend which indicates more active star formation that leads to higher stellar fractions in the galactic central regions that eventually lead to much steeper $\gamma^{\prime}$ towards the higher-mass end. In the `NoKineticBHwinds' case, the total stellar mass range is slightly larger than the `NoWinds' case and produces a flat $\gamma^{\prime}-M_{\ast}$ correlation. This accounts for the differences of the $\gamma^{\prime}-M_{\ast}$ correlation compared to Magneticum shown in Fig.~\ref{fig:gam_mstar} of Section~\ref{sec:3.2.2}, in which IllustrisTNG includes black hole kinetic winds and has a steeper $\gamma^{\prime}-M_{\ast}$ anti-correlation, while the Magneticum model~\citep{2014MNRAS.442.2304H} does not include the black hole kinetic winds and produces a flatter $\gamma^{\prime}-M_{\ast}$ trend. 

In the middle right panel of Fig.~\ref{fig:vary}, the different model variations for the $\gamma^{\prime}-\sigma_{\mathrm{e/2}}$ correlation have similar trends as the $\gamma^{\prime}-M_{\ast}$ correlation. Stronger stellar winds produce slightly lower central velocity dispersions, while removing winds, black hole kinetic winds, or black holes completely all increase the central velocity dispersion and $\gamma^{\prime}$. The `NoWinds' model produces a non-monotonic $\gamma^{\prime}-\sigma_{\mathrm{e/2}}$ correlation with some scatter that is still marginally consistent with a weak negative correlation. Although the `NoKineticBHwinds' seems to produce a more consistent $\gamma^{\prime}-\sigma_{\mathrm{e/2}}$ trend compared with the dynamical modeling dataset shown in Fig.~\ref{fig:gam_sv} in Section~\ref{sec:3.2.5}, it still produces higher $\gamma^{\prime}$ than the observed density slopes and fails to match the strong lensing dataset. This indicates that although the inclusion of kinetic BH winds is crucial to reducing $\gamma^{\prime}$ values and matching with observations, a more refined treatment of the model implementation is required to further recover the realistic $\gamma^{\prime}-\sigma_{\mathrm{e/2}}$ correlation trend. 

We show the $\gamma^{\prime}-R_{\mathrm{eff}}$ and $\gamma^{\prime}-\Sigma_{\ast}$ correlations in the bottom panels of Fig.~\ref{fig:vary}. The `StrongerWinds' model has a $R_{\mathrm{eff}}$ range that is similar to the TNG model, with slightly more compact sizes and an almost identical $\gamma^{\prime}-R_{\mathrm{eff}}$ trend. The `NoWinds' model also covers the $R_{\mathrm{eff}}$ range of the TNG model, extending to smaller sizes more compared to the `StrongerWinds' model and produces steeper $\gamma^{\prime}$ towards the more compact end. Since the `StrongerWinds' model has smaller $M_{\ast}$ and slightly lower $R_{\mathrm{eff}}$ compared to the TNG model, the net $\gamma^{\prime}-\Sigma_{\ast}$ trend is similar to the TNG model. The `NoWinds' model has higher $M_{\ast}$ and a $R_{\mathrm{eff}}$ range that extends to smaller sizes, which lead to higher $\Sigma_{\ast}$ values and a steeper $\gamma^{\prime}-\Sigma_{\ast}$ correlation. In contrast, the two black hole model variations `NoBHs' and `NoKineticBHwinds' demonstrate a more significant systematic reduction in the $R_{\mathrm{eff}}$ values, combined with an enhanced stellar mass in these two models, leading to a systematic increase in $\Sigma_{\ast}$ with the `NoBHs' model having stronger impact than the `NoKineticBHwinds' model, while preserving the negative $\gamma^{\prime}-R_{\mathrm{eff}}$ trend and positive $\gamma^{\prime}-\Sigma_{\ast}$ trend. Overall, both stellar feedback and AGN feedback can alter $R_{\mathrm{eff}}$ and $\Sigma_{\ast}$ of ETGs, however the correlation trends of these two quantities are preserved. AGN feedback is more effective in reducing effective sizes systematically compared with stellar feedback, and stronger stellar feedback does not enlarge effective sizes. This indicates that the mild discrepancies at the larger size end of the $\gamma^{\prime}-R_{\mathrm{eff}}$ correlation (Fig.~\ref{fig:gam_reff}) and the smaller stellar surface density end of the $\gamma^{\prime}-\Sigma_{\ast}$  correlation (Fig.~\ref{fig:gam_sstar}) could indeed be attributed to limitations in the AGN feedback model. 

The dark matter fraction in different runs (upper right panel in Fig.~\ref{fig:vary}) shows that different feedback prescriptions result in different baryon-dark matter fractions in the central regions of the ETGs. All model variations preserve the negative $\gamma^{\prime}-f_{\mathrm{DM}}$ correlation. The change of $\gamma^{\prime}$ in different models is also a reflection of the change in the relative proportions of baryons to dark matter through the variations of the $\gamma^{\prime}-f_{\mathrm{DM}}$ correlation. The two black hole model variations both show lower $f_{\mathrm{DM}}$ and higher $\gamma^{\prime}$ compared with the TNG model, especially for the `NoBHs' model which has the lowest $f_{\mathrm{DM}}$ and the steepest $\gamma^{\prime}-f_{\mathrm{DM}}$ trend out of all the model variations. In the `NoWinds' model, the dark matter fractions are lower than the TNG model but higher than the `NoKineticBHwinds' model, consistent with the $\gamma^{\prime}-M_{\ast}$ and the $\gamma^{\prime}-\Sigma_{\ast}$ trends in the middle and bottom panels of Fig.~\ref{fig:vary}. However, even in the `StrongerWinds' model, the mean $\gamma^{\prime}-f_{\mathrm{DM}}$ correlation is shifted towards lower $f_{\mathrm{DM}}$ and higher $\gamma^{\prime}$. Although this seems counter intuitive, this is a net effect due to the simultaneous decrease of the total stellar mass and the effective radius in the `StrongerWinds' model that led to a slightly smaller $f_{\mathrm{DM}}$ which is calculated in a smaller aperture (smaller $R_{\mathrm{eff}}$) compared to the TNG model. Overall, AGN feedback still plays a more important role than stellar feedback in altering the $\gamma^{\prime}-f_{\mathrm{DM}}$ correlation, while the inclusion of black hole kinetic winds in the AGN feedback model is crucial to reducing $\gamma^{\prime}$ and increasing $f_{\mathrm{DM}}$ coherently. This also explains why the comparison of IllustrisTNG ETGs with the Magneticum ETGs in the same stellar mass range and similar $\gamma^{\prime}$ value range, shows that the IllustrisTNG ETGs extend further out to larger $f_{\mathrm{DM}}$ values with the inclusion of black hole kinetic winds in the lower panel of Fig.~\ref{fig:gam_fdm}, which improves the agreement with the strong lensing dataset. The $f_{\mathrm{DM}}$ values of IllustrisTNG ETGs with similar $\gamma^{\prime}$ values are still systematically higher than the dynamical modeling dataset, so further refinement of the black hole kinetic wind model is required to better match observational results from diverse modeling methods and redshift ranges.

In summary, AGN feedback dominates over stellar feedback in altering $\gamma^{\prime}$. A new finding is that the low accretion rate black hole kinetic wind feedback mechanism in the IllustrisTNG AGN feedback model~\citep{2017MNRAS.465.3291W} is a crucial component that effectively reduces $\gamma^{\prime}$ and forms realistic correlations with galaxy properties that match observations.  

\subsection{Discussion}
\label{sec:4.3}

In this subsection, we will compare the TNG model variation results with previous studies, and discuss the physics of how AGN and stellar feedback affect the total density profile. As a brief summary to previous studies regarding the impact of AGN feedback on $\gamma^{\prime}$, \citet{2010MNRAS.405.2161D} found that AGN feedback is necessary to produce stellar fractions that match observed values in groups and clusters. AGN feedback consequently significantly reduces baryon fractions in galaxies and the inner regions of clusters, leading to shallower-than-isothermal inner density profiles and reduced halo concentrations. \citet{2012MNRAS.420.2859M} found that AGN feedback is crucial to forming brightest cluster galaxies with realistic density profiles, stellar mass fractions, velocity dispersion profiles, and spin. They also propose a new mechanism for AGN feedback to form cored stellar profiles, allowing the creation of near-isothermal total density profiles in luminous elliptical galaxies with a combination of dynamical friction heating and gaseous ejections from AGN feedback. \citet{2012MNRAS.422.3081M} further confirms that AGN feedback reduces the steepness of the stellar density profile and the amount of cold gas available for star formation at low redshifts. They point out that apart from dynamical friction and gaseous ejections from AGNs, the slow expulsion of gas in the quiescent phase of AGN activity can also lead to adiabatic expansion that leads to further reduction of $\gamma^{\prime}$. Through direct simulation of AGN feedback, \citet{2013MNRAS.433.3297D} found that AGN feedback effectively quenches in-situ star formation, randomizes stellar orbits, enhances accreted stellar populations, grows galaxy effective sizes, and drives galaxy morphological transformation from late to early type. 

These effects of AGN feedback on $\gamma^{\prime}$ have been well-established through the above-mentioned numerical experiments, and are clearly consistent with the results from the model variations shown in Section~\ref{sec:4.2} where we compare the `NoBHs' model with the TNG model. However, not all of the above-mentioned models operate in different modes during high and low accretion rates, and the quantitative effects on $\gamma^{\prime}$ of these two AGN feedback modes have not been thoroughly explored. We have shown in the analysis of Section~\ref{sec:4.2} that the `NoKineticBHwinds' model shows a clearly different effect compared to the TNG model resulting in steeper $\gamma^{\prime}$. This explicitly indicates that a change in the physical model of the low accretion rate AGN feedback mode substantially alters the total density profile and its various correlations with galaxy properties. This also indicates that the AGN feedback model improvement made in \citet{2017MNRAS.465.3291W} where the low accretion rate feedback mode is changed from depositing thermal energy in radio bubbles~\citep{2013MNRAS.436.3031V,2014MNRAS.438.1985T} to the AGN ejecting momentum and kinetic energy into its surrounding ISM in random directions that better matches the observed galaxy scaling relations~\citep{2018MNRAS.473.4077P} is a crucial step towards simulating self-consistent near-isothermal $\gamma^{\prime}$ values and various galaxy properties at the same time. Nonetheless, we have to keep in mind that the anti-correlations of $\gamma^{\prime}-M_{\ast}$ and $\gamma^{\prime}-\sigma_{\mathrm{e/2}}$, as well as the systematics for IllusrisTNG galaxy sizes and central dark matter fractions as we have discussed in Section~\ref{sec:3.2} which are slightly off from the observed results, calls for further improvements of the black hole kinetic wind model calibration. A promising direction may lie in considering black hole spin and adopting a spin-dependent switch between the high and low accretion rate AGN feedback modes as suggested by \citet{2019arXiv190204651B}. 

Another important result from the TNG model variations analysis in the previous subsection is that while stellar feedback is subdominant, it works in the same direction in changing $\gamma^{\prime}$ as AGN feedback. The dominance of AGN feedback over stellar feedback with increasing stellar mass is a manifestation of the stellar mass-black hole mass correlation~\citep{2013ARA&A..51..511K}, which suggests the coevolution of ellipticals and the bulges of lenticulars with their supermassive blackholes leading to more powerful AGN feedback in higher mass ETGs. However, stronger winds causing shallower $\gamma^{\prime}$ seems to be {\it at odds} with some claims in the literature. \citet{2013MNRAS.436.2929H} showed that stellar wind feedback leads to higher in-situ-formed stellar mass ratio in $z=0$ galaxies and thus correlates with steeper $\gamma^{\prime}$~\citep{2013ApJ...766...71R}. They suggest that the inclusion of stellar feedback makes $\gamma^{\prime}$ steeper, compensating the effects of AGN feedback. This is the same argument made by \citet{2017MNRAS.464.3742R} while comparing the Oser and Wind ETGs galaxy-total density profile correlations. Notwithstanding, we believe that their argument is limited by the assumption that the effects of stellar wind and AGN feedback on $\gamma^{\prime}$ are independent. In both the numerical experiments of \citet{2013MNRAS.436.2929H} and \citet{2017MNRAS.464.3742R}, stellar wind variations were conducted with no AGN feedback. This leads to an increased amount of hot gas in galaxies and more dissipation during galaxy mergers driving enhanced post-merger in-situ star formation and steeper $\gamma^{\prime}$. However, our stellar wind-related model variations were conducted with the inclusion of AGN feedback, and arrive at opposite results with \citet{2013MNRAS.436.2929H} and \citet{2017MNRAS.464.3742R}. This suggests that the effect of stellar winds on $\gamma^{\prime}$ is dependent on whether or not AGN feedback is present. In the IllustrisTNG AGN feedback model, the high accretion rate thermal mode keeps a hot ISM, and the low accretion rate black hole kinetic winds efficiently expels gas to larger radii. In this case, stronger stellar winds increase the amount of hot gas in the galaxy which serves as the `target medium' for AGN feedback, thus acting in concordance with AGN feedback to further enhance ex-situ-formed stellar populations and reduce $\gamma^{\prime}$. Hence, another new finding of our TNG model variation study together with the comparison to previous works is that the effect of stellar feedback is coupled to AGN feedback in a complex way, and specifically, stellar feedback induces shallower (steeper) $\gamma^{\prime}$ with (without) AGN feedback.

Futhermore, the presence of baryons and their related feedback processes not only affect the fraction and distribution of stars and gas within the galaxy, but may also back-react on the density profiles of the dark matter halo hosting them, further altering the total density profile. \citet{2015MNRAS.451.1247S} found although a standard NFW profile provides a good universal fit for EAGLE halos, the presence of baryons creates cuspier dark matter density profiles. \citet{2015MNRAS.453.2447D} found that forced star formation quenching mechanisms in simulations can lead to contracted dark matter halos. This seems to be at odds with simulations including AGN feedback that support an expanded or standard NFW halo~\citep{2010MNRAS.405.2161D,2015MNRAS.451.1247S}. \citet{2016MNRAS.461.2658D} further suggested that stellar feedback can drive cycles of gas inflows and outflows that results in mass-dependent halo responses (contraction or expansion). \citet{2018MNRAS.481.1950L} find that for $\sim 10^{12}\,\mathrm{M_{\astrosun}}$ IllustrisTNG halos, adiabatic contraction~\citep{1986ApJ...301...27B,2004ApJ...616...16G} is present from the inner most regions of the galaxies out to $20\,\mathrm{kpc}$, whereas halo expansion at small radii for similar Illustris halos occur. The Illustris halos with higher star formation efficiency are expected to have higher contraction levels than the IllustrisTNG halos according to \citet{2016MNRAS.461.2658D}, so the agreement with \citet{2016MNRAS.461.2658D} breaks at smaller radii. All of these can be summarized by the finding that higher baryonic fraction often leads to higher halo contraction, while the subsequent feedback processes (depending how they are implemented) may expand or contract the halo further in a mass-dependent fashion. We will elaborate on the effects of baryons on the dark matter profile for IllustrisTNG ETGs further in Section~\ref{sec:5}.

\section{The effects of baryons on dark matter}
\label{sec:5}

Since neither the individual density profile of dark matter nor baryonic matter is isothermal, it is of great importance to understand how baryonic effects shape the dark matter halo and how their interplay `conspired' to form the observed near-isothermal density profiles. In this section we present comparisons of dark matter profiles made between the TNG100-full physics (FP) and the TNG100-dark matter only (DMO) runs in order to investigate the impact of baryons on dark matter. The DMO run has the same total mass contained in the simulation box, as well as the number of dark matter tracer particles as the FP run. The mass of all baryonic particles (stellar, gas and BH particles) in the FP run are absorbed into the particle mass of collisionless dark matter particles in the DMO run. All baryonic processes (AGN feedback, galactic winds, star formation etc.) are also ignored in the DMO run. Dark matter halos in the DMO run are identified with their FP halo counterparts using the \textsc{sublink} algorithm. Although not all halos in the FP run have DMO counterparts, our choice of `central' ETGs in the FP run of TNG100 maximally mitigates this problem. However, we point out that out of the 559 DMO counterparts, only 545 are `central' halos, and 14 are `satellite' halos. 

\subsection{The power-law density slope of the dark matter component $\gamma_{\mathrm{DM}}^{\prime}$}
\label{sec:5.1}

\begin{figure*}
\includegraphics[width=1.9\columnwidth]{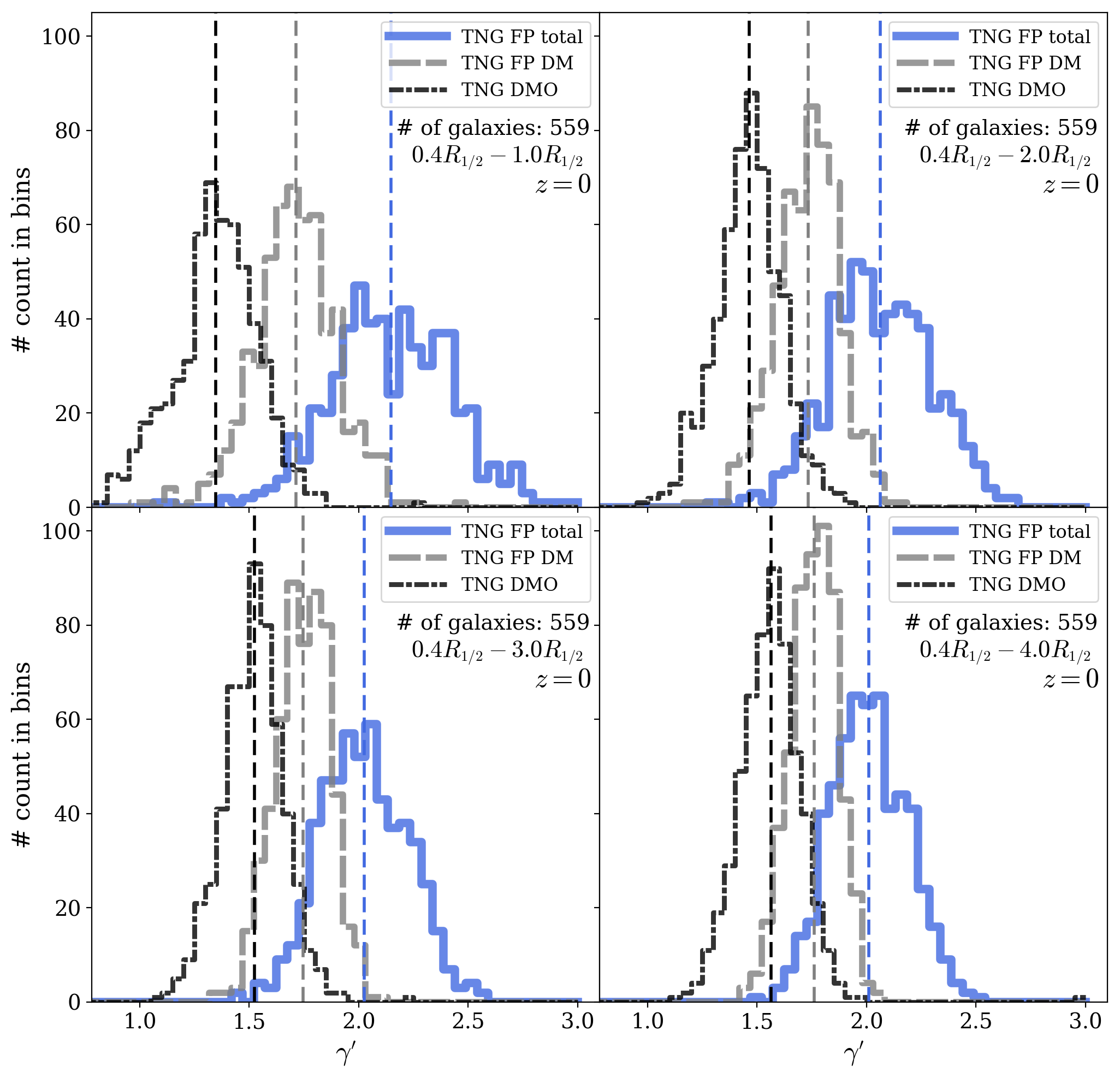}
\caption{Comparison between the power-law density slopes in the IllustrisTNG DMO (dark matter only) and the IllustrisTNG FP (full physics) ETGs at $z = 0$. The slope distribution of the 559 ETGs in four slope radial ranges are shown in the four subplots. In each subplot, the DMO slope distribution is shown by the black histogram, whereas the total and dark matter slopes of their FP counterparts are shown by the blue and the grey histograms, respectively. The dashed lines represent the mean of the slope distributions and with the same color legend as the solid histograms. The slopes of the DMO halos ($\langle\gamma_{\mathrm{DM,\,DMO}}^{\prime}\rangle \approx 1.5$, in agreement with \citealt{2013ApJ...766...71R}) are shallower than the dark matter component in their FP counterparts ($\langle\gamma_{\mathrm{DM,\,DMO}}^{\prime}\rangle \approx 1.7$) within all four radial ranges investigated here.}
\label{fig:denprof_DMO}
\end{figure*}

The fiducial `stellar half mass radius' and the four radial ranges over which we measure the power-law density slope for the DMO run halo is chosen to be identical to its corresponding FP halo (see details in Section~\ref{sec:3.1}). The slope distribution of the four radial ranges are shown in the four subplots of Fig.~\ref{fig:denprof_DMO}. In each subplot, the DMO slope distribution is shown by the black histogram, whereas the total and dark matter slopes of their FP counterparts are shown by the blue and the grey histograms, respectively. The dashed lines represent the mean of the slope distributions, with the same color legend as the solid histograms. The mean and standard deviation of the slope distributions are summarized in Table~\ref{tab:denprof_DMO}. 

As it can be seen from the figure, the slopes of the DMO halos are generally shallower than the dark matter slope of their counterparts in the FP run within all four radial ranges investigated here. The DMO slopes also possess larger scatter than the FP slopes (see Table~\ref{tab:denprof_DMO}). The DMO slopes~($\langle\gamma_{\mathrm{DM,\,DMO}}^{\prime}\rangle \approx 1.5$) are in good agreement with \citet{2013ApJ...766...71R}, in which the dark matter power-law density slope is measured over the radial range [$0.3\,R_{\mathrm{1/2}}$,$4\,R_{\mathrm{1/2}}$], very similar to our IllustrisTNG ETG radial range. This suggests that the presence of baryons and the baryonic processes steepen both the total and the dark matter power-law density slopes simultaneously. 

\begin{table}
		\begin{center}
		\begin{tabular}{lccc}
			\hline
			Run & Radial range & $\langle\gamma_{\mathrm{DM}}^{\prime}\rangle$ & $\sigma_{\gamma_{\mathrm{DM}}^{\prime}}$ \\
			\hline
			DMO & $0.4\,R_{\mathrm{1/2}}-1\,R_{\mathrm{1/2}}$ & $1.346 \pm 0.008$ & $0.193$ \\
			DMO & $0.4\,R_{\mathrm{1/2}}-2\,R_{\mathrm{1/2}}$ & $1.465 \pm 0.008$ & $0.190$ \\
			DMO & $0.4\,R_{\mathrm{1/2}}-3\,R_{\mathrm{1/2}}$ & $1.523 \pm 0.008$ & $0.198$ \\
			DMO & $0.4\,R_{\mathrm{1/2}}-4\,R_{\mathrm{1/2}}$ & $1.564 \pm 0.009$ & $0.202$ \\
			\hline
            FP & $0.4\,R_{\mathrm{1/2}}-1\,R_{\mathrm{1/2}}$ & $1.713 \pm 0.008$ & $0.188$ \\
			FP & $0.4\,R_{\mathrm{1/2}}-2\,R_{\mathrm{1/2}}$ & $1.732 \pm 0.006$ & $0.145$ \\
			FP & $0.4\,R_{\mathrm{1/2}}-3\,R_{\mathrm{1/2}}$ & $1.745 \pm 0.005$ & $0.123$ \\
			FP & $0.4\,R_{\mathrm{1/2}}-4\,R_{\mathrm{1/2}}$ & $1.760 \pm 0.005$ & $0.108$ \\
            \hline
		\end{tabular}
        \end{center}
		\caption{The mean and standard deviation of the dark matter power-law density slope $\gamma_{\mathrm{DM}}^{\prime}$ of the four radial ranges on which we measure the slope for the IllustrisTNG DMO halos and their FP counter parts. The inner radius is set to 0.4$\,R_{\mathrm{1/2}}$ and we select the different outer radii as $R_{\mathrm{1/2}}$, $2\,R_{\mathrm{1/2}}$, $3\,R_{\mathrm{1/2}}$, and $4\,R_{\mathrm{1/2}}$ following Section \ref{sec:3.1}. The `Run' column corresponds to the dark matter only (DMO) run and the full physics (FP) run, respectively. }
		\label{tab:denprof_DMO}
\end{table}

\subsection{The inner slope of the gNFW profile}
\label{sec:5.2}

\begin{figure}
\includegraphics[width=\columnwidth]{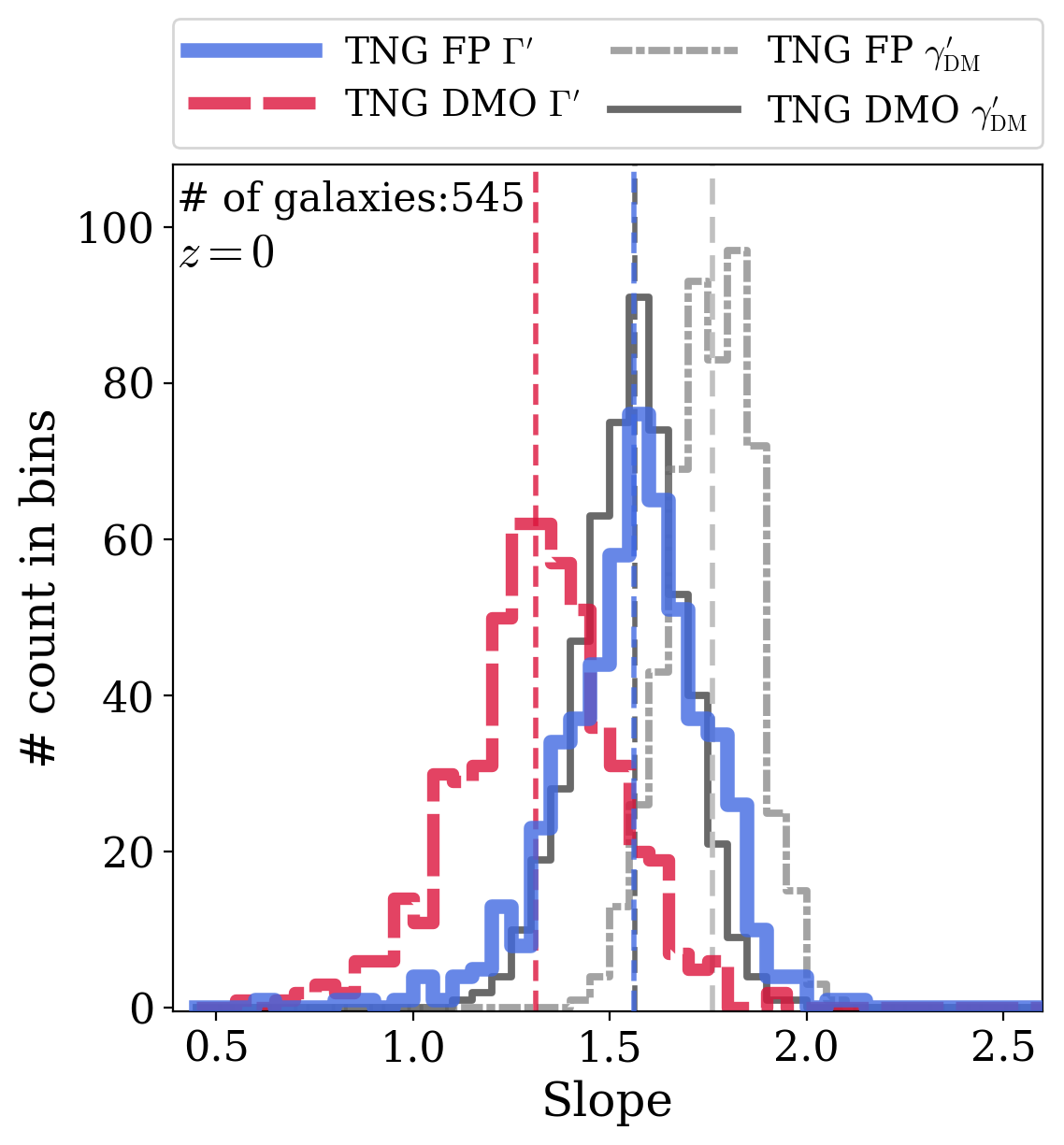}
\caption{Comparison between the gNFW inner slope $\Gamma^{\prime}$ and the dark matter power-law density slope $\gamma_{\mathrm{DM}}^{\prime}$ in the DMO and the FP runs at $z = 0$. Shown here are only the 545 `central' halos in the DMO run along with their FP counterparts. $\gamma_{\mathrm{DM}}^{\prime}$ is measured over the radial range [$0.4\,R_{1/2}$, $4\,R_{1/2}$]. The inner slopes of the gNFW profile in the FP and the DMO runs are shown by the blue and the red histograms, while the dark matter power-law density slopes in the FP and the DMO runs are given by the light and dark grey histograms. The dashed lines represent the mean of the distributions, with the same color legend as the solid histograms.}
\label{fig:beta_gam}
\end{figure}

Since dark matter halos are well-modeled by the NFW profile~\citep{1997ApJ...490..493N} instead of the power-law model, and in order to make fair comparisons with observations, we also fit a generalized NFW (gNFW) profile to the dark matter component in both the FP and the DMO runs with a variable inner slope $\Gamma^{\prime}$ ~\citep{1996MNRAS.278..488Z}:
\begin{equation}
\rho(r) = \rho_{0} \left(\frac{r}{r_{\mathrm{s}}}\right)^{-\Gamma^{\prime}} \left(1+\frac{r}{r_{\mathrm{s}}}\right)^{-3+\Gamma^{\prime}}\,,
\end{equation}
where $\rho_{0}$ is the characteristic density and $r_{\mathrm{s}}$ is the scale radius. We fit (with equal radial weighting) the gNFW profile only to the 545 `central' halos in the DMO run and their corresponding FP ETGs within [$0.01\,R_{200}$, $R_{200}$] of each halo (ETG). The distribution of the inner slope $\Gamma^{\prime}$ compared with the power-law density slope of the dark matter component $\gamma_{\mathrm{DM}}^{\prime}$ over the radial range [$0.4\,R_{1/2}$, $4\,R_{1/2}$] is shown in Fig.~\ref{fig:beta_gam}. The mean and the standard deviation of the inner slope are summarized in Table~\ref{tab:beta}.

\begin{table}
		\begin{center}
		\begin{tabular}{lcc}
			\hline
			Run & $\langle\Gamma^{\prime}\rangle$ & $\sigma_{\Gamma^{\prime}}$ \\
			\hline
            DMO & $1.312 \pm 0.009$ & $0.202$ \\
            FP & $1.561 \pm 0.008$ & $0.185$ \\
            \hline
		\end{tabular}
        \end{center}
		\caption{The mean and the standard deviation of the inner slope $\Gamma^{\prime}$ of the best fit gNFW profile to the DMO halos and the dark matter component in their FP counterparts. The `Run' column indicates the type of the simulation: the dark matter only (DMO) run and full physics (FP) run. The mean $\langle\Gamma^{\prime}\rangle$ is shown along with its $1\,\sigma$ error and does $not$ take into account any weighting of the global galactic properties, while the scatter $\sigma_{\Gamma^{\prime}}$ shows the standard deviation of the distribution. }
		\label{tab:beta}
\end{table}

It can be seen from Fig.~\ref{fig:beta_gam} that $\Gamma^{\prime}$ in both the FP and the DMO runs is shallower than the dark matter power-law slope $\gamma_{\mathrm{DM}}^{\prime}$ measured within the investigated radial ranges in the corresponding run. The more important aspect of the slope distribution is that the inner slopes $\Gamma^{\prime}$ of the TNG FP ETGs are much steeper than their DMO counterparts, whose $\Gamma^{\prime}$ are closer to the standard NFW inner slope $\Gamma^{\prime} = 1$. The steepening reflects dark matter halo contraction due to the presence of baryons and dissipation processes.

\subsection{The correlation of the inner slope $\Gamma^{\prime}$ with the halo mass $M_{200}$}
\label{sec:5.3}

\begin{figure}
\includegraphics[width=\columnwidth]{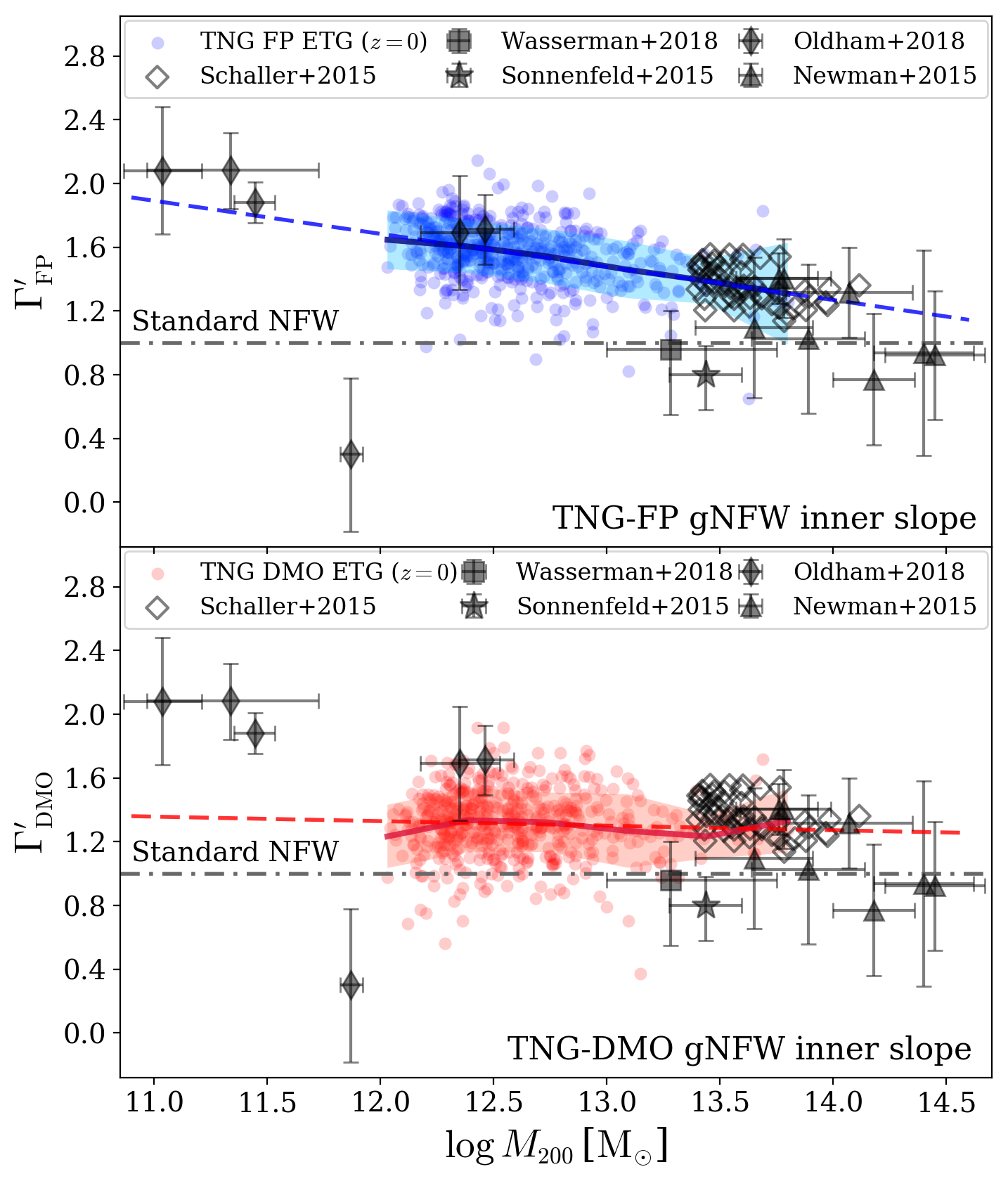}
\caption{The correlation of the gNFW inner slope $\Gamma^{\prime}$ with the halo mass $M_{200}$ in the FP ETGs versus their halo mass at $z = 0$. The IllustrisTNG FP ETGs are shown by the blue scattered dots in the upper panel, and their corresponding DMO halos are shown by the red scattered dots in the lower panel. The solid curves in each panel give the mean of the inner slopes, and the shaded regions show the standard deviations of the inner slope distribution. The dashed lines are the best fits to the $\Gamma^{\prime}-M_{200}$ correlation, with $
\partial\Gamma_{\mathrm{FP}}^{\prime}/\partial\mathrm{log} M_{200} = -0.21 \pm 0.02$ and a Pearson correlation coefficient $r_{\mathrm{p}}=-0.37$ for the FP dark matter components, and $\partial\Gamma_{\mathrm{DMO}}^{\prime}/\partial\mathrm{log} M_{200} = -0.03 \pm 0.03$ with a Pearson correlation coefficient $r_{\mathrm{p}}=-0.05$ for the DMO dark matter halos. The comparison datasets are shown in black and are identical in the two panels. The horizontal dot dashed line in the two panels indicate the standard NFW inner slope $\Gamma^{\prime}=1$. Note that the halo mass used for the DMO halos is the $M_{200}$ of their FP counterparts, for consistency.}
\label{fig:gam_m200_fp}
\end{figure}

The correlation of the gNFW inner slope $\Gamma^{\prime}$ and the halo mass $M_{200}$  is shown in Fig.~\ref{fig:gam_m200_fp}. The IllustrisTNG FP ETGs are shown by the blue scattered dots in the upper panel, and their corresponding DMO halos are shown by the red scattered dots in the lower panel. The solid curve in each panel shows the mean of the inner slopes, and the shaded region shows the standard deviation of the inner slope distribution.

The dashed line in each subplot is the best linear fit to the $\Gamma^{\prime}-M_{200}$ correlation, with $\partial\Gamma_{\mathrm{FP}}^{\prime}/\partial\mathrm{log} M_{200} = -0.21 \pm 0.02$ and a Pearson correlation coefficient $r_{\mathrm{p}}=-0.37$ for the FP dark matter components, and $\partial\Gamma_{\mathrm{DMO}}^{\prime}/\partial\mathrm{log} M_{200} = -0.03 \pm 0.03$ with a Pearson correlation coefficient $r_{\mathrm{p}}=-0.05$ for the DMO dark matter halos. The halo masses of the DMO halos are approximated by the $M_{200}$ of their FP counterparts for consistency ($\mathrm{log}\,M_{\mathrm{200, FP}}/\mathrm{log}\,M_{\mathrm{200, DMO}}$ has a mean of 0.995 and a scatter of 0.003 for our sample). While the inner slopes $\Gamma_{\mathrm{DMO}}^{\prime}$ of the DMO halos is almost constant with halo mass, $\Gamma_{\mathrm{FP}}^{\prime}$ of the FP halos steepens as the halo mass $M_{200}$ decreases, indicating that the presence of baryons and baryonic processes is essential to steepen the inner slope of the dark matter halo, especially in lower-mass galaxies, and for forming the observed negative trend of the $\Gamma^{\prime}-M_{200}$ correlation. This is also consistent with the fact that lower-mass ETGs also possess a higher (lower) central baryonic (dark matter) fraction (also see Fig.~11 in \citet{2017MNRAS.469.1824X} and Fig.~9 in \citealt{2018MNRAS.481.1950L}).  

We compare the IllustrisTNG $\Gamma^{\prime}-M_{200}$ correlation with the observed and simulated results. Over-plotted in Fig.~\ref{fig:gam_m200_fp} are measurements for observed and simulated galaxies modeled with gNFW profiles that allow a variable dark matter profile inner slope. \citet{2015ApJ...814...26N} modeled 10 group scale lenses and inferred the dark matter power-law density slope $\gamma_{\mathrm{DM}}^{\prime}$ within the effective radius  combining dynamical constraints (we only show 8 of the 10 lenses with $M_{200}$ provided). \citet{2015ApJ...800...94S} selected 81 strong lenses from the SL2S and SLACS surveys and modeled their inner slope $\Gamma^{\prime}$ of the dark matter profile using joint lensing and stellar dynamics method. The average dark matter inner slope of the 81 lenses combined has a mean of $\langle\gamma^{\prime}_{\mathrm{DM}}\rangle = 0.80^{+0.18}_{-0.22}$ which is shown in the figure. It is consistent with the standard NFW model but has a large uncertainty. \citet{2018MNRAS.476..133O} modeled strong lenses combined with stellar kinematics to determine the dark matter inner slope. \citet{2018ApJ...863..130W} modeled a single ETG NGC1407 using Jeans modeling with varying radial IMF and kinematic tracers to constrain the dark matter inner slope. As for the simulated values, \citet{2015MNRAS.452..343S} selected halos from the EAGLE simulation~\citep{2015MNRAS.446..521S} and modeled the inner slope $\Gamma^{\prime}$ of their dark matter halos. We only include their small-mass halos which are similar in mass to our selected IllustrisTNG ETGs. We refer the reader to Appendix \ref{sec:AB} for more details on the comparison dataset included in this section.

The IllustrisTNG FP data (upper panel) is in agreement with the EAGLE Simulations results for a steeper-than-NFW inner slope and a mild negative correlation of the inner slope with the halo mass. However, although the datasets from observations also show a mild negative correlation of $\Gamma^{\prime}$ versus $M_{200}$, the observations generally favor a standard NFW inner slope towards the high-mass end which is shallower than the inner slopes derived in IllustrisTNG and EAGLE. Nevertheless, the large uncertainties in the observed inner slopes does not rule out the agreement with simulated values for halos with $\mathrm{log}\,M_{200} \gtrsim 13.5\mathrm{M_{\astrosun}}$. Extrapolation of the IllustrisTNG FP $\Gamma^{\prime}-M_{200}$ correlation to the low-mass end also suggest an agreement with observation~(\citealt{2018MNRAS.476..133O}, data from \citealt{2018ApJ...863..130W}). 

\begin{figure}
\includegraphics[width=\columnwidth]{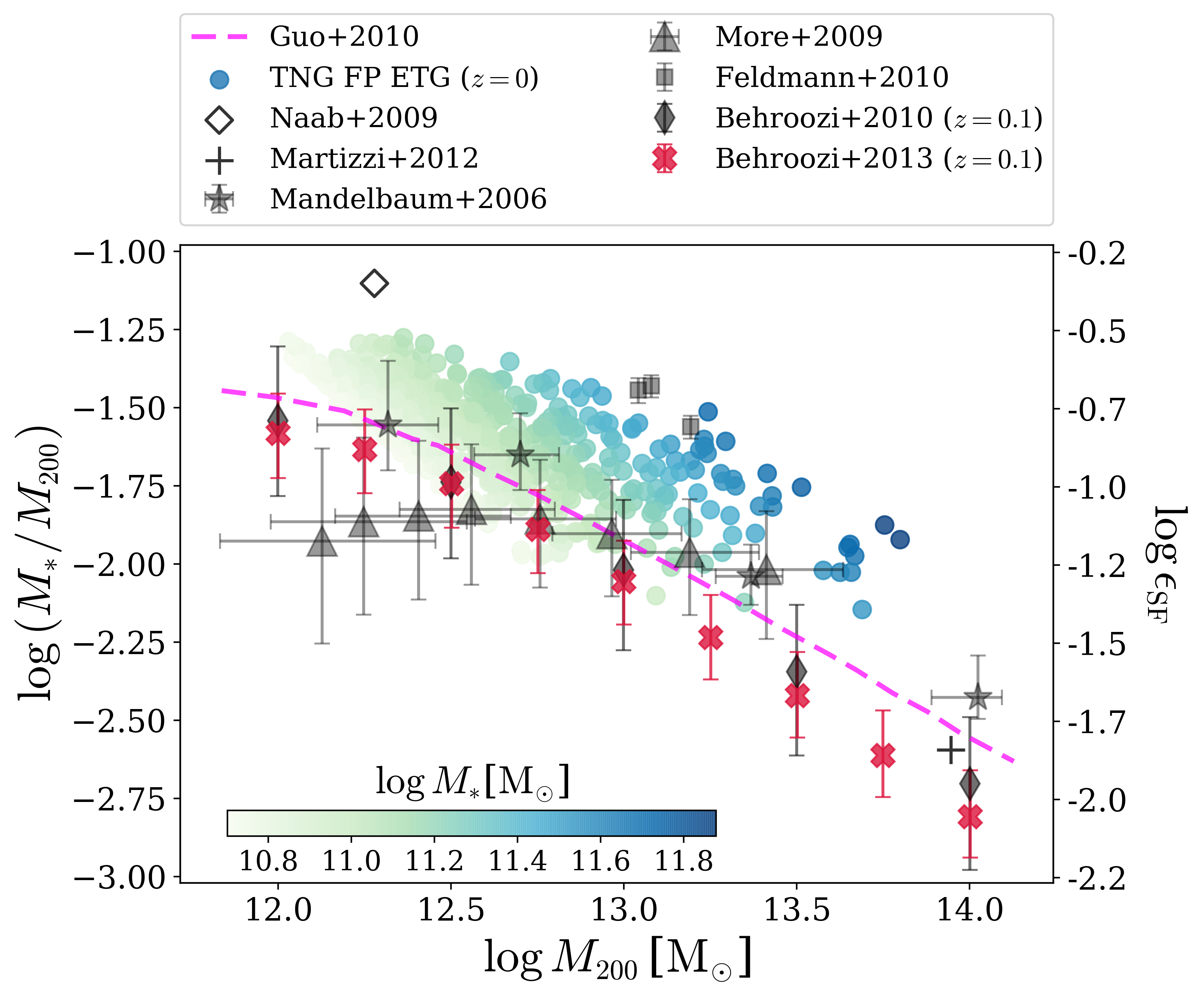}
\caption{The halo baryon fraction $M_{\ast}/M_{200}$ and the integrated star formation efficiency (a measure for baryon fraction), $\epsilon_{\mathrm{SF}} = (M_{\ast}/M_{200})/(\Omega_{\mathrm{b}} / \Omega_{\mathrm{m}})$, of IllustrisTNG FP ETGs is shown by the gradient-colored dots in the figure as a function of halo mass $M_{200}$. The scattered dots are color-coded by the ETGs' stellar mass $M_{\ast}$ indicated by the colorbar. There is higher baryon fraction in lower halo (stellar) mass systems and vice versa, suggesting more significant baryon impacts in such systems. Data from observations, cosmological hydrodynamic simulations, and abundance matching models are shown for comparison (taken from the references listed in the legend).}
\label{fig:esf}
\end{figure}

As a cosmological simulation, the IllustrisTNG FP data covers a large halo mass range from galaxy-size halos ($\lesssim 10^{12}\mathrm{M}_{\astrosun}$) to group-size halos ($\gtrsim 10^{14}\mathrm{M}_{\astrosun}$) giving a tight $\Gamma^{\prime}-M_{200}$ correlation. We show the baryon fraction $M_{\ast}/M_{200}$ and the integrated star formation efficiency (a measure for baryon fraction), $\epsilon_{\mathrm{SF}} = (M_{\ast}/M_{200})/(\Omega_{\mathrm{b}} / \Omega_{\mathrm{m}})$, as a function of halo mass $M_{200}$ in Fig.~\ref{fig:esf} for IllustrisTNG FP ETGs to demonstrate the different significance of the baryonic component in this large dynamic range of halo mass. The scattered dots are color-coded by the ETGs' stellar mass $M_{\ast}$ indicated by the colorbar. As shown in the figure, the integrated star formation efficiency is higher (lower) in low-mass (high-mass) halos, suggesting a more (less) significant baryon impact on the dark matter profile inner slope in these systems, which is consistent with the $\Gamma^{\prime}-M_{200}$ anti-correlation for the IllustrisTNG FP halos displayed in the upper panel of Fig.~\ref{fig:gam_m200_fp}. In Figure \ref{fig:esf} we also include data from weak lensing observation~\citep{2006MNRAS.368..715M}, stacked satellite kinematics observation~\citep{2009MNRAS.392..801M}, cosmological hydrodynamic simulations~\citep{2009ApJ...699L.178N,2010ApJ...709..218F,2010MNRAS.404.1111G,2012MNRAS.420.2859M}, and abundance matching models~\citep{2010ApJ...717..379B,2013ApJ...770...57B} for a comparison. IllustrisTNG ETGs have consistent total baryon fractions with results (regardless of galaxy morphology) derived from a wide range of methods, especially with the ETG subsample from~\citet{2010ApJ...709..218F} shown in the figure.

\begin{figure}
\includegraphics[width=\columnwidth]{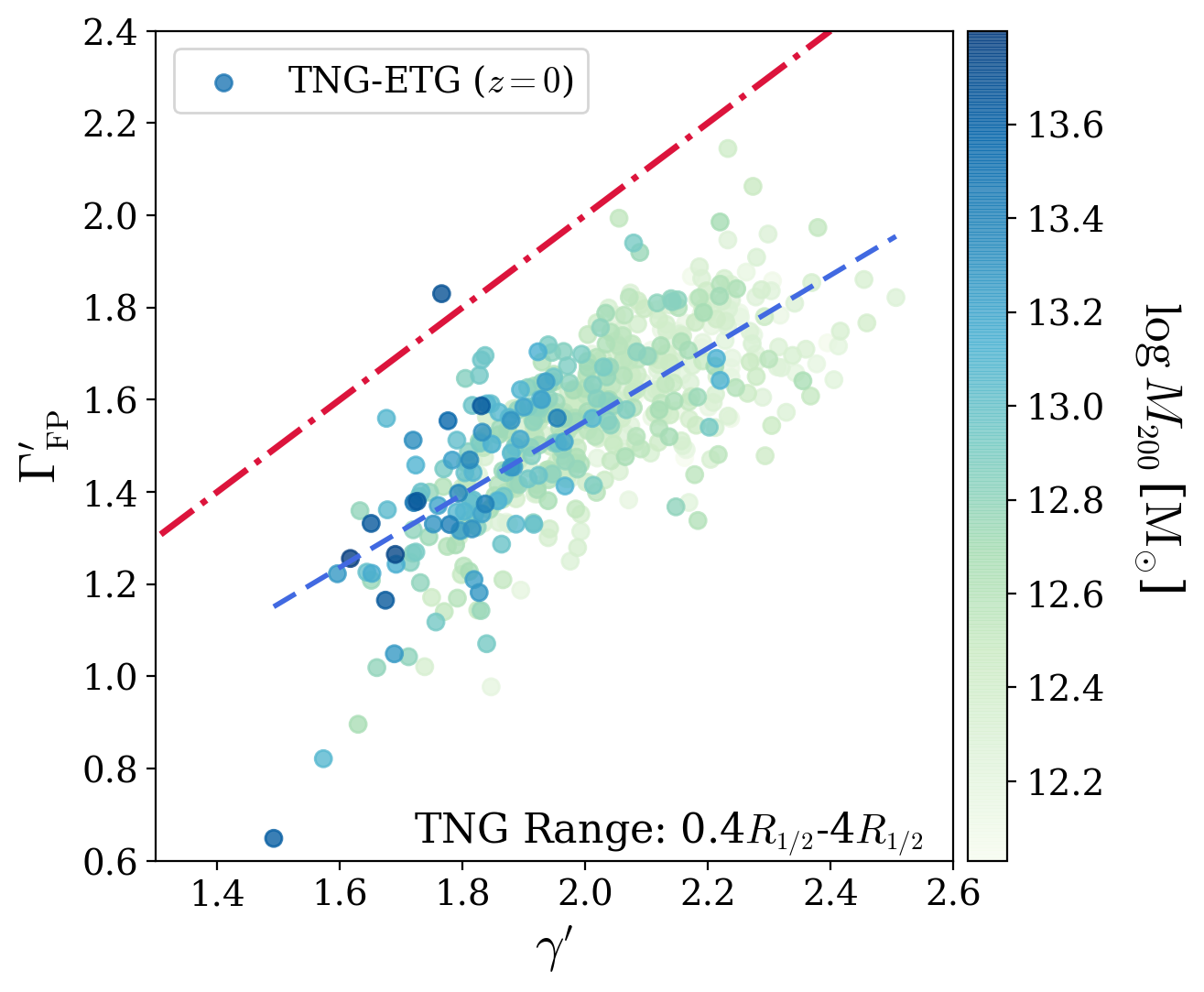}
\caption{The correlation of the gNFW inner slope $\Gamma_{\mathrm{FP}}^{\prime}$ with the total power-law density slope $\gamma^{\prime}$. The scattered dots denoting the IllustrisTNG ETGs are colored by the halo mass $M_{200}$ of each galaxy. The red dotted dashed line is the line where $\Gamma_{\mathrm{FP}}^{\prime} = \gamma^{\prime}$. The blue dashed line is the best linear fit to the correlation, with $\partial \Gamma_{\mathrm{FP}}^{\prime} / \partial\gamma^{\prime} = 0.79 \pm 0.03$ and a Pearson correlation coefficient $r_{\mathrm{p}} = 0.73$. The total density slope $\gamma^{\prime}$ is also anti-correlated with the halo mass $M_{200}$ as seen from the figure.}
\label{fig:Gg}
\end{figure}

We note that the total power-law density slope $\gamma^{\prime}$ of the IllustrisTNG ETGs is also positively correlated with the FP dark matter inner slope $\Gamma_{\mathrm{FP}}^{\prime}$ at $z = 0$ as shown in Fig.~\ref{fig:Gg}. The scattered dots denoting the IllustrisTNG ETGs are colored by the halo mass $M_{200}$ of each galaxy. The best linear fit of the $\Gamma_{\mathrm{FP}}^{\prime}-\gamma^{\prime}$ correlation gives $\partial \Gamma_{\mathrm{FP}}^{\prime} / \partial\gamma^{\prime} = 0.79 \pm 0.03$ and a Pearson correlation coefficient $r_{\mathrm{p}} = 0.73$. The total density slope $\gamma^{\prime}$ is also anti-correlated with the halo mass $M_{200}$ as seen from the figure. The halo masses of the simulated galaxies are significantly smaller than those derived for observed galaxies, which points to an overestimation of baryonic sizes of galaxies, consistent with the excess of central dark matter fractions as found in \citet{2018MNRAS.481.1950L}.

We also note that some previous studies which built the dynamic model through the Jeans equations set priors on the dark matter inner slope $\Gamma^{\prime}$ from 0 to $1.6$ (e.g. \citealt{2017ApJ...838...77L}), or even smaller (0 to $1.2$, \citealt{2013MNRAS.432.1862C}). Our analysis above suggests that this prior is better broadened to [0, 2].

\begin{table}
		\begin{center}
		\begin{tabular}{lcc}
			\hline
			Run & $\partial\Gamma^{\prime}/\partial\mathrm{log}\,M_{200}$ & $r_{\mathrm{p}}$ \\
			\hline
			FP & $-0.21 \pm 0.02$ & $-0.37$  \\
			DMO & $-0.03 \pm 0.03$ & $-0.05$ \\
			\hline
		\end{tabular}
        \end{center}
		\caption{The best linear fit of the $\Gamma^{\prime}-M_{200}$ correlations presented in Section \ref{sec:5.3}. `Run' stands for the different galaxy parameters, $\partial\Gamma^{\prime}/\partial\mathrm{log}\,M_{200}$ is the slope of the best linear fit to the correlation, and $r_{\mathrm{p}}$ is the Pearson correlation coefficient of the corresponding best linear fit. All values are given for $z=0$.}
		\label{tab:FITS_dark}
\end{table}

A summary of the $\Gamma^{\prime}-M_{200}$ correlations presented in this section for the IllustrisTNG ETGs is given in Table~\ref{tab:FITS_dark}.

\subsection{The correlation of the inner slope $\Gamma^{\prime}$ with the halo concentration $c_{200}$}
\label{sec:5.4}

\begin{figure}
\includegraphics[width=\columnwidth]{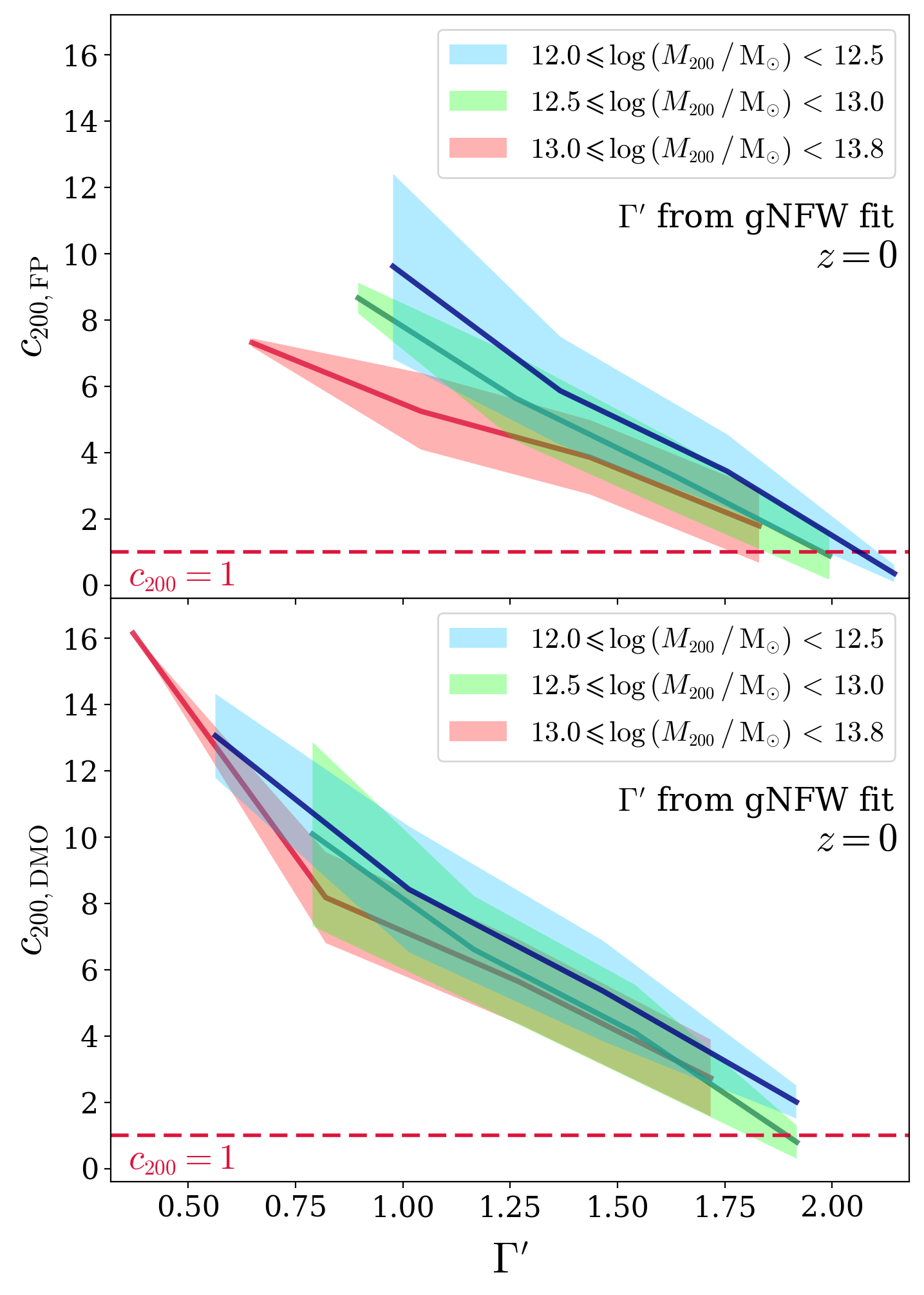}
\caption{The correlation of the inner slope $\Gamma^{\prime}$ with the halo concentration $c_{200}$ in the FP run (upper panel) and DMO run (lower panel) at $z = 0$. The halos are divided into three mass bins, labeled in blue, green and red as indicated in the legend box, containing 260, 220 and 65 ETGs, respectively. The solid lines represent the mean while the shaded regions represent the standard deviation of the distribution. It is seen from the figure that $c_{200}$ decreases with increasing inner slope $\Gamma^{\prime}$ in both the FP and the DMO run. The red dashed line in each subplot indicates where $c_{200}=1$ and exposes extreme outliers with $c_{200}<1$ at the large $\Gamma^{\prime}$ end in both the FP and DMO run. These halos have concentrations with $c_{200} > 1$ if we perform a standard NFW fit. Note that the halo mass used for the DMO halos is the $M_{200}$ of their FP counterparts, for consistency.}
\label{fig:gam_m200_DMO}
\end{figure}

We show the correlation of the inner slope $\Gamma^{\prime}$ with the halo concentration $c_{200}$ at $z = 0$ in Fig.~\ref{fig:gam_m200_DMO}. The upper and bottom panels represent the FP and the DMO cases, respectively. The halos are divided into three mass (FP halo mass $M_{200}$) bins, namely $12.0\,\leqslant\, \mathrm{log}\,(M_{200}\,/\,\mathrm{M_{\odot}})\,<\, 12.5$, $12.5\,\leqslant\, \mathrm{log}\,(M_{200}\,/\,\mathrm{M_{\odot}})\,<\, 13.0$, $13.0\,\leqslant\, \mathrm{log}\,(M_{200}\,/\,\mathrm{M_{\odot}})\,<\, 13.8$, containing 260, 220 and 65 ETGs, respectively. We use the halo mass of the FP counterparts for the DMO halos for consistency. 

It is seen from the figure that $c_{200}$ decreases with increasing inner slope $\Gamma^{\prime}$ in both the FP and the DMO run, regardless of halo mass range. Also, there are a few extreme outliers with $c_{200} < 1$ in both the FP and the DMO run. These halos have concentrations with $c_{200} > 1$ if we perform a standard NFW fit ($c_{\mathrm{200,\,FP}}>3.48$ and $c_{\mathrm{200,\,DMO}}>4.40$). Hence, this issue is mainly caused by our choice of the gNFW profile model. Furthermore, the concentration parameter $c_{200}$ decreases with increasing halo mass range. This is consistent with the $c_{200}-M_{200}$ correlation compared to observations and other simulations~\citep{1997ApJ...490..493N,2000ApJ...535...30J,2001MNRAS.321..559B,2001ApJ...554..114E,2003MNRAS.339...12Z,2007MNRAS.378...55M,2007MNRAS.381.1450N,2008MNRAS.390L..64D,2008MNRAS.387..536G,2014MNRAS.441.3359D,2015MNRAS.451.1247S}, and it will be elucidated in more detail in an upcoming IllustrisTNG paper (Pillepich et al. in preparation).

The fact that the dark matter inner slope decreases with increasing $c_{200}$ is more significant for the lower-mass systems. For a standard NFW profile, a higher $c_{200}$ indicates a smaller scale radius and higher concentration. However, larger $\Gamma^{\prime}$ also indicates a cuspy dark matter core and probable halo contraction. This suggests that when dark matter halos are highly concentrated, the standard NFW $c_{200}$ might not provide an objective measure of the halo concentration, and one must combine the values of $c_{200}$ and $\Gamma^{\prime}$ of a gNFW profile to determine the concentration of dark matter halos. This is also in line with the suggestion of modeling steeper dark matter profiles in observations proposed in \citet{2018MNRAS.481.1950L}.

\begin{figure*}
\includegraphics[width=1.9\columnwidth]{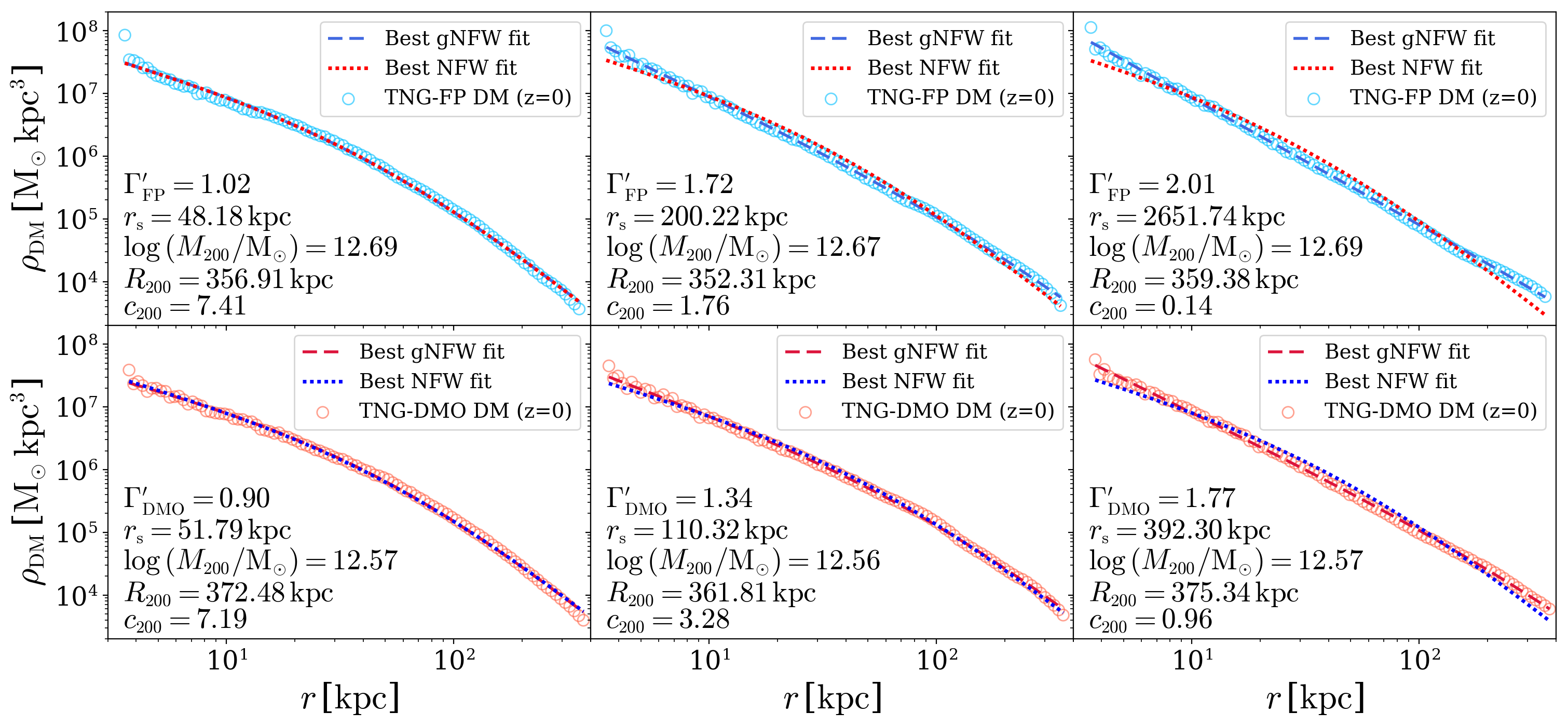}
\caption{Three dark matter halo profiles from the FP run (upper panels) with similar $R_{200}$ ($M_{200}$) which have dramatically different $c_{200}$ as an illustration of the $c_{200}-\Gamma^{\prime}$ anti-correlation. Their DMO counterparts are shown in the lower panels. The scattered dots represent the measured 3D dark matter radial density, the dashed curves show the best gNFW fit to the dark matter radial density profile, and the dotted curve show the best standard NFW fit. The upper left panel shows a halo with dark matter inner slope close to the standard NFW inner slope ($\Gamma^{\prime} \approx 1$), the upper middle panel shows a halo with inner slope steeper than the NFW inner slope ($\Gamma^{\prime} \approx 1.7$), and the upper right panel displays a case with an extremely steep inner slope ($\Gamma^{\prime} \approx 2$). Their DMO counterparts in the lower panels have shallower gNFW inner slopes. It is obvious that increasing inner slope leads to larger best-fit scale radius $r_{\mathrm{s}}$ and hence lower $c_{200}$, indicating stronger halo contraction. The deviation from a standard NFW dark matter profile also increases from the left to the right in the figure.}
\label{fig:CM}
\end{figure*}

As an illustration of this issue, we show three halos from the FP run (upper panels) that have similar $R_{200}$ ($M_{200}$) but dramatically different $c_{200}$ along with their DMO counterparts (lower panels) in Fig.~\ref{fig:CM}. In the upper and lower panels, the scattered dots represent the measured 3D dark matter radial density, and dashed curves are the best gNFW fit to the dark matter radial density profile, and the dotted curves are the best standard NFW fit to the dark matter radial density profile. The upper left panel shows a halo with dark matter inner slope close to the standard NFW inner slope ($\Gamma^{\prime} \approx 1$), the upper middle panel shows a halo with inner slope steeper than the NFW inner slope ($\Gamma^{\prime} \approx 1.7$), and the upper right panel shows an extremely steep inner slope case ($\Gamma^{\prime} \approx 2$). Their DMO counterparts in the lower panels have shallower gNFW inner slopes. The deviation from a standard NFW dark matter profile increases from the left to the right in the figure. Since the gNFW profile enforces a constant outer slope of $3$, larger inner slopes correspond to larger scale radii $r_{\mathrm{s}}$ and hence lower concentration parameters $c_{200}$, which is present in both the FP and the DMO cases. It is not clear, however, why halos of similar masses can have markedly different concentrations (i.e. the large range of $c_{200}$ in Fig.~\ref{fig:gam_m200_DMO}). Environment and merger histories might play important roles, although we leave this issue to future work to address. 

\subsection{The correlation of the mass-weighted slope $\gamma_{\mathrm{mw}}^{\prime}$ with central dark matter fraction $f_{\mathrm{DM}}$}
\label{sec:5.5}

The steeper dark matter inner slopes of IllustrisTNG-FP ETGs compared with their DMO counterparts~(Fig.~\ref{fig:beta_gam} and Fig.~\ref{fig:gam_m200_fp}) and the anti-correlation of $c_{200}-\Gamma^{\prime}$~(Fig.~\ref{fig:gam_m200_DMO} and Fig.~\ref{fig:CM}) are essentially in agreement with the predictions of dark matter halo contraction. 

Adiabatic halo contraction originally proposed that the dissipative infall of gas contracts dark matter and creates dense cores in the center of halos~\citep{1986ApJ...301...27B}. However, subsequent studies of gas cooling in hydrodynamical simulations favored less contraction than the prediction of the adiabatic contraction scheme, but still found halos to be contracted and the dark matter profiles in the inner region to be steeper than the standard NFW profile~\citep{2004ApJ...616...16G,2010MNRAS.407..435A}. These findings are also self-consistent in predicting the transformation of prolate halos to oblate ones through dissipation which matches the shape distribution of the observed early-type galaxies~\citep{1994ApJ...431..617D,2010MNRAS.407..435A}. 

Nevertheless, subsequent observations exposed tension about the level of contraction of dark matter halos. \citet{2012ApJ...752..163S} measured the dark matter slope of SDSSJ0946+1006 giving $\gamma_{\mathrm{DM}}^{\prime} = 1.7 \pm 0.2$, suggesting strong contraction in concordance with simulations~\citep{2010MNRAS.405.2161D}. \citet{2012ApJ...747L..15G} measured average dark matter slopes for SLACS lenses of $\langle\gamma_{\mathrm{DM}}'\rangle = 1.7 \pm 0.5$, which was later corrected by \citet{2014MNRAS.438.3594D} to $1.40_{-0.26}^{+0.15}$ assuming a Salpeter IMF, favoring mild contraction. Similarly, \citet{2015ApJ...814...26N} obtained $\langle\gamma_{\mathrm{DM}}^{\prime}\rangle = 1.35 \pm 0.09$ for 10 group-scale lenses, with mild contraction in agreement with \citet{2004ApJ...616...16G}. Interestingly, \citet{2014MNRAS.439.2494O} found the best-fit dark matter slope for 85 SLACS strong lenses~\citep{2009ApJ...705.1099A} to be $\gamma_{\mathrm{DM}}^{\prime} = 1.60_{-0.18}^{+0.13}$, steeper than \citet{2015ApJ...814...26N} but favoring the standard NFW profile without contraction combined with the radial distribution of dark matter fraction. Hydrodynamical simulations~\citep{2012ApJ...744...63O,2013ApJ...766...71R} favor dark matter component slopes $\gamma_{\mathrm{DM}}^{\prime} \leqslant 1.5$ for ETGs, implying little to no contraction in the dark matter halo. Thus, the dark matter slope does not directly indicate the level of contraction, and one must also account for the central dark matter fraction to constrain the level of contraction.

To further quantify the level of contraction of the IllustrisTNG ETG dark matter halos, we compare the correlation of the mass weighted slope $\gamma^{\prime}_{\mathrm{mw}}$ at $R_{\mathrm{eff}}$  and the central dark matter fraction $f_{\mathrm{DM}}$ with the semi-empirical models presented in \citet{2017ApJ...840...34S}. The mass weighted slope at $\gamma^{\prime}_{\mathrm{mw}}\,(R_{\mathrm{eff}})$ is defined in Equation~\ref{equ:gmw} in Section~\ref{sec:3.2.5}.

\begin{figure}
\includegraphics[width=\columnwidth]{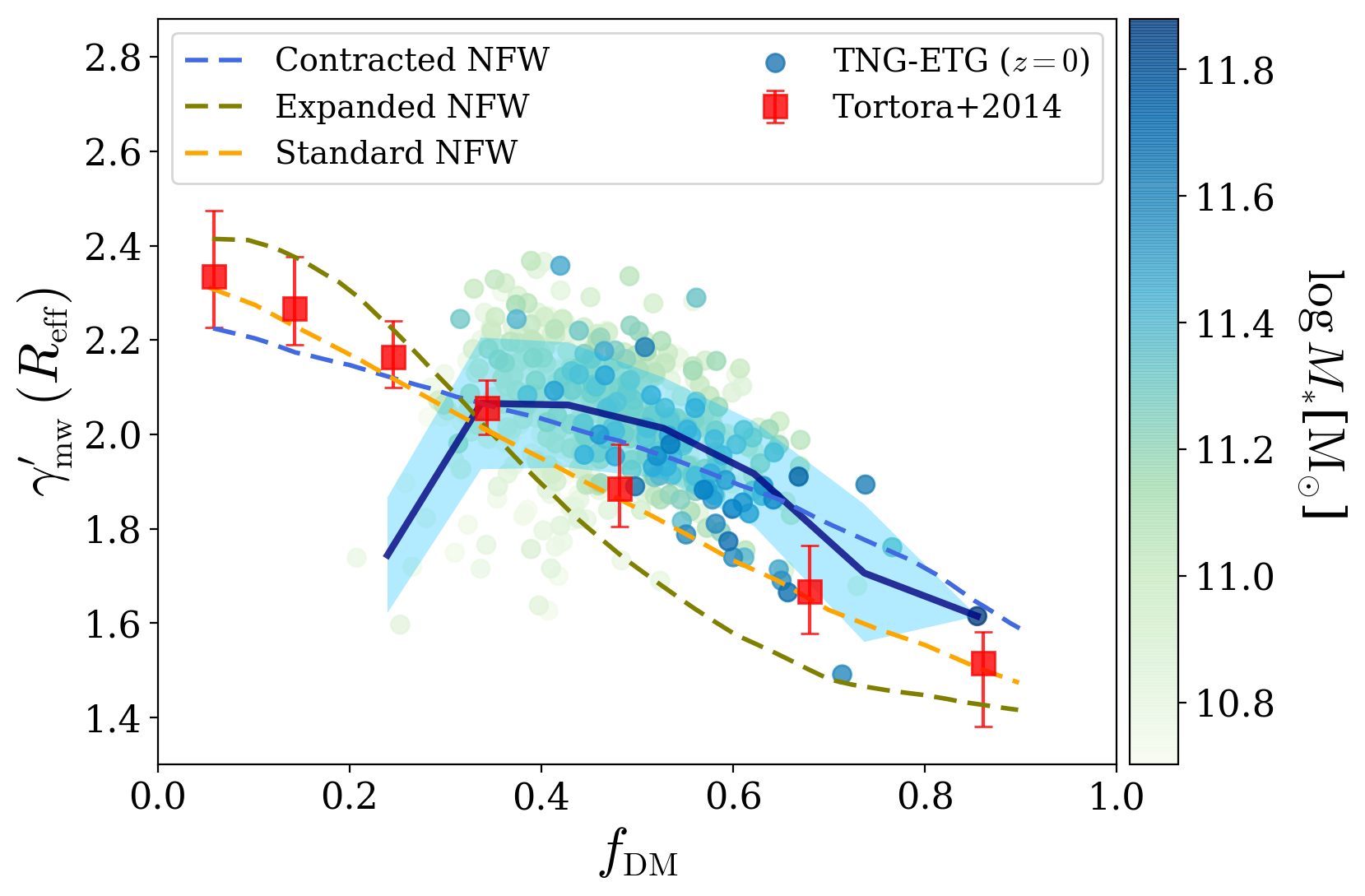}
\caption{The $\gamma_{\mathrm{mw}}^{\prime}-f_{\mathrm{DM}}$ correlation in comparison with semi-empirical  models~\citep{2017ApJ...840...34S} and observations~\citep{2014MNRAS.445..115T}. The mass-weighted slope is measured at the effective radius. The semi-empirical models with standard NFW profile, with halo contraction, and with halo expansion are shown by orange, blue and green dashed curves, respectively. The binned values of SPIDER and $\mathrm{ATLAS^{3D}}$~\citep{2014MNRAS.445..115T} data are shown by the red squares with error bars (same as Fig.~\ref{fig:gam_fdm}). The mass weighted slopes at the effective radius ($R_{\mathrm{eff}}$) versus the central dark matter fraction of the IllustrisTNG ETGs are shown by the colored scattered dots, with the color bar indicating the total stellar mass. The navy curve and the shaded blue region are the mean and the standard deviation of the IllustrisTNG ETG slope distribution.}
\label{fig:AC}
\end{figure}

Utilizing a `S$\mathrm{\acute{e}}$rsic-NFW' model and comparing their predictions with observational data~\citep{2013ApJ...765...25N,2015ApJ...800...94S}, \citet{2017ApJ...840...34S} have ruled out at $\gtrsim 2-3\,\sigma$ deviations from a S$\mathrm{\acute{e}}$rsic stellar profile and an uncontracted NFW dark matter profile. The $\gamma_{\mathrm{mw}}^{\prime}-f_{\mathrm{DM}}$ correlation predicted by their standard NFW model, contracted NFW model and expanded NFW model are shown along with our IllustrisTNG ETG sample and binned values of the $\gamma_{\mathrm{mw}}^{\prime}-f_{\mathrm{DM}}$ correlation from SPIDER and $\mathrm{ATLAS^{3D}}$~\citep{2014MNRAS.445..115T} in Fig.~\ref{fig:AC}. The total density slope values from the \citet{2017ApJ...840...34S} models and \citet{2014MNRAS.445..115T} are the mass-weighted density slopes~\citep{2014MNRAS.438.3594D} measured at the effective radius of each galaxy. The contraction and expansion levels of the semi-empirical models are less extreme than the adiabatic case~\citep{2017ApJ...840...34S}.

As shown in Fig.~\ref{fig:AC}, the semi-empirical models with standard NFW profile, with halo contraction, and with halo expansion are shown by the orange, blue and green dashed curves, respectively. The binned values of SPIDER and $\mathrm{ATLAS^{3D}}$~\citep{2014MNRAS.445..115T} data are shown by the red squares with error bars (same as in Fig.~\ref{fig:gam_fdm}), which have assumed an underlying NFW model for the dark matter profile. The mass weighted slope at $R_{\mathrm{eff}}$ versus the central dark matter fraction of the IllustrisTNG ETGs are shown by the colored scattered dots, with the color index indicating the total stellar mass.  The navy curve and the shaded blue region are the mean and the standard deviation of the IllustrisTNG ETG slope distribution. It is obvious that the mean of the IllustrisTNG ETG $\gamma_{\mathrm{mw}}^{\prime}-f_{\mathrm{DM}}$ correlation is in better agreement with the contracted NFW model, which corroborates the steeper dark matter slopes in the IllustrisTNG ETGs aforementioned. The low mass end of our sample tends to have a steeper mass-weighted slope $\gamma^{\prime}_{\mathrm{mw}}$ that agrees with the contracted NFW model, while the IllustrisTNG ETGs with larger stellar mass agrees better with the standard NFW model as well as the observation data of SPIDER and $\mathrm{ATLAS^{3D}}$~\citep{2014MNRAS.445..115T}. This is in line with the anti-correlation of $\Gamma^{\prime}-M_{200}$ in the upper panel of Fig.~\ref{fig:gam_m200_fp}, where the high-mass end has a shallower dark matter inner slope and agrees better with observations compared to the low-mass end.

Also noticable in Fig.~\ref{fig:AC} is a small number of halos showing expansion in comparison with the models at all $f_{\mathrm{DM}}$ scales, especially towards the lower mass end. By visually checking the $\gamma^{\prime}_{\mathrm{mw}}$ values as a function of radius in these galaxies, we found that they have flattened central stellar density profiles in common, which leads to a shallower $\gamma^{\prime}_{\mathrm{mw}}$ value at $R_{\mathrm{eff}}$. These flatter stellar profiles might be due to too violent stellar feedback in the central stellar-dominated region of these seemingly `expanded' ETGs, since the effect of the total density profile flattening due to AGN feedback is sub-dominant in these low-mass systems according to their low AGN feedback energy rate given by the IllustrisTNG AGN model~\citep{2017MNRAS.465.3291W}. In these cases, the single aperture measurement of $\gamma^{\prime}_{\mathrm{mw}}$ at $R_{\mathrm{eff}}$ is subject to galaxy mass-dependent systematic bias. We propose that future observational studies of ETG mass and dynamical structure can benefit from a multi-aperture measurement for the $\gamma^{\prime}_{\mathrm{rm}}$ profile.

Overall, most IllustrisTNG ETGs have contracted dark matter halos, and their contraction level is mass dependent. This is in line with the non-universal halo response driven by gas inflows and outflows found in zoom-in simulations~\citep{2016MNRAS.461.2658D}. The contraction level of the halo is correlated with the integrated star formation efficiency $\epsilon_{\mathrm{SF}}$ (see Fig. 5 and Fig. 10 in ~\citealt{2016MNRAS.461.2658D}), with mild contraction at the high $\epsilon_{\mathrm{SF}}$ end, and no contraction or even expansion at the low $\epsilon_{\mathrm{SF}}$ end for the $\epsilon_{\mathrm{SF}}$ range covered by our ETG sample as shown in Fig.~\ref{fig:esf}. Since $\epsilon_{\mathrm{SF}}$ also anti-correlates with the halo mass $M_{200}$ and the stellar mass $M_{\ast}$, the mass dependence of the halo contraction level is accounted for by the mass dependence of the $\epsilon_{\mathrm{SF}}$. Combining this with the fact that the central dark matter fraction $f_{\mathrm{DM}}$($\lesssim R_{\mathrm{1/2}}$) increases with increasing stellar and halo mass in our ETG sample mass range (see Fig. 6 and Fig. 9 in \citealt{2018MNRAS.481.1950L}), we conclude that baryonic impact is more significant in the lower mass IllustrisTNG ETGs, contracting the dark matter halo, where there is a lower central dark matter fraction. Baryonic impact acts vice versa at the higher mass end, resulting in a standard or even expanded NFW dark matter halo. This is also consistent with the discussion in the last paragraph of Section~\ref{sec:4.3} that the variations in the total density profile maybe partially due to the different halo responses to the presence of baryons and baryonic feedback processes.

\section{Conclusions and discussion}
\label{sec:6}

In this work, we present the first systematic analysis of the total density profiles of early-type galaxies in the IllustrisTNG Simulations~\citep{2018MNRAS.480.5113M,2018MNRAS.477.1206N,2018MNRAS.475..624N,2018MNRAS.475..648P,2018MNRAS.475..676S}. In particular we have studied the statistical properties of various galaxy correlations with the total density profile, the effects of different feedback models on the total density profile, and how baryons affect the dark matter halo and hence alter the total density profile. We focus on the methodology of analyzing the density profiles and present the first attempt to compare IllustrisTNG ETGs' total density profiles with observations and other simulations. Through these comparisons along with the various numerical experiments of model variations presented above, we have understood that IllustrisTNG ETGs broadly agree with previous results from observations and other simulations, despite some discrepancies are still present. Most importantly, we have gained new insights on how different feedback physics are connected to the formation of near-isothermal total density profiles in ETGs. 

Our selection strategy which employed single and double component luminosity profile fitting of the rest frame SDSS $r$-band radial luminosity distribution of galaxies in the TNG100 run resulted in a sample of 559 (720) well-resolved `central' ETGs in the stellar mass range of $10^{10.7}\mathrm{M}_{\astrosun} \leqslant M_{\ast} \leqslant 10^{11.9}\mathrm{M}_{\astrosun}$ at $z=0$ ($z=0.2$). We measured the total power-law density slopes of all the IllustrisTNG ETGs within four different radial ranges, and demonstrated the correlations of the total density slope with other global galactic properties including stellar mass, effective radius, stellar surface density, stellar kinematics, central dark matter fraction and in-situ-formed stellar mass ratio. Also presented is the redshift evolution of the total power-law density slopes. All of these have been compared with diverse datasets from local ETGs through stellar kinematic modeling, higher-redshift ETGs from strong lensing surveys and other numerical simulations. The major findings of our analysis are summarized as follows:

$\bullet$ We calculated the total power-law density slope of each selected IllustrisTNG ETG by performing a linear fit to the radial distribution of the 3D density in logarithmic scale within four radial ranges, with the inner radius set to $0.4\,R_{\mathrm{1/2}}$ ($R_{1/2}$ stands for the stellar half mass radius), and the outer radius set to $R_{\mathrm{1/2}}$, $2\,R_{\mathrm{1/2}}$, $3\,R_{\mathrm{1/2}}$, and $4\,R_{\mathrm{1/2}}$. The total density slopes were found to be close to (slightly steeper than) isothermal across these radial ranges, and the intrinsic scatter of the total power-law density slope mildly decreased with increasing outer radial range (see Fig.~\ref{fig:denprof} and Table~\ref{tab:denprof}). 

$\bullet$ The total power-law density slope (measured over $0.4\,R_{\mathrm{1/2}}-4\,R_{\mathrm{1/2}}$ for all correlations) of the IllustrisTNG ETGs shows a mild anti-correlation with their total stellar mass, an anti-correlation with their effective radius, and a positive correlation with their stellar surface density. These trends are in broad agreement with observations and other simulations (see Fig.~\ref{fig:gam_mstar}, Fig.~\ref{fig:gam_reff} and Fig.~\ref{fig:gam_sstar}), except that the effective radius of the IllustrisTNG ETGs are larger by $\approx0.1\,\mathrm{dex}$ compared to observations~\citep{2018MNRAS.474.3976G}. Similar discrepancy compared to observations occur on the lower end of $\Sigma_{\ast}$ as a consequence. The $\gamma^{\prime}-M_{\ast}$ relation shows a steeper trend compared with the Magneticum ETGs~\citep{2017MNRAS.464.3742R}, which could be a result of the black hole kinetic winds implementation~(see middle left panel of Fig.~\ref{fig:vary} and Section~\ref{sec:4.2}).

$\bullet$ The total power-law density slope of the IllustrisTNG ETGs shows a mild anti-correlation with their central velocity dispersion, in tension with the observational datasets (see Fig.~\ref{fig:gam_sv}). The apparent mismatch could be accounted for by the problematic morphology-size relation in IllustrisTNG compared with observations~\citep{2019MNRAS.483.4140R}. Also, a new finding with respect to previous numerical experiments~(e.g. \citealt{2017MNRAS.469.1824X}) is that the adoption of a mass-weighted slope (Equation~\ref{equ:gmw}) consistent with the observations~\citep{2014MNRAS.445..115T} cannot fully reconcile the discrepancy with the observed trend. Nevertheless, the slopes derived for the observational sample may suffer from systematic biases under the isotropic velocity dispersion assumption, such that the more-massive galaxies which are radially anisotropic tend to have overestimated total density slopes (see Fig.~\ref{fig:gam_beta} and also Fig.~17 of \citealt{2017MNRAS.469.1824X}). The apparent underestimation of $\sigma_{\mathrm{e/2}}$ in comparison with observations suggests limitations in the feedback models of IllustrisTNG.  

$\bullet$ The total power-law density slope of the IllustrisTNG ETGs exhibits an anti-correlation with the central dark matter fraction, and the trend marginally agrees with the comparison datasets at $z = 0$ (see Fig.~\ref{fig:gam_fdm}).  The agreement with the strong lensing dataset at $z>0$ would be better had we chosen a Salpeter IMF instead of a Chabrier IMF (chosen for stellar mass consistency), due to the $f_{\mathrm{DM}}-\mathrm{IMF}$ degeneracy. However, the inclusion of a black hole kinetic wind feedback channel, that leads to an increase of the dark matter fractions (top right panel of Fig.~\ref{fig:vary}), improves the agreement with lensing observations compared to other numerical experiments such as Magneticum.

$\bullet$ The power-law density slope of the IllustrisTNG ETGs also shows a clear positive correlation with the in-situ-formed stellar mass ratio, which indicates that gas-poor galaxy mergers may have played a dominant role in evolving the total density profile shallower with time. Such an effect is more significant in higher-mass ETGs compared to their lower-mass counterparts (see Fig.~\ref{fig:gam_f_insitu}). 

$\bullet$ The total power-law density slope is nearly independent of redshift below $z=1$, mildly decreasing with time. The trend shows some tension with the observed slope redshift dependence (see Fig.~\ref{fig:z_obs}).

$\bullet$ AGN feedback and stellar feedback both affect the distribution and the correlations of $\gamma^{\prime}$ with various galaxy properties (see Fig.~\ref{fig:vary} and Table~\ref{tab:vary}). Reducing AGN or stellar feedback results in higher stellar masses, central velocity dispersions and stellar surface surface densities, as well as lower dark matter fractions and effective radii, hence altering the various galaxy correlations with $\gamma^{\prime}$. The net effect of steepening $\gamma^{\prime}$ by reducing feedback can be seen as a combination of compacting galaxy sizes and enhancing the proportion of the stellar component which has a stellar density profile in the galactic central regions. In general, AGN feedback dominates over stellar feedback in altering $\gamma^{\prime}$. A new finding is that the low accretion rate black hole kinetic wind feedback mechanism is a crucial component that efficiently reduces $\gamma^{\prime}$ and forms realistic correlations with galaxy properties that matches observations. This indicates that $\gamma^{\prime}$ is sensitive to the AGN feedback model, especially the low-accretion rate feedback mode, which could provide important constraints for future calibrations of simulation models.

$\bullet$ Another genuinely new finding of the TNG model variations analysis in comparison with previous works on stellar feedback is that the effect of stellar feedback on $\gamma^{\prime}$ cannot be simply assumed to be independent of AGN feedback~(see Section~\ref{sec:4.3}). In the absence of AGN feedback, stronger stellar winds enhance in-situ formed stellar populations and increase the amount of dissipation during mergers that eventually lead to steeper $\gamma^{\prime}$~\citep{2013MNRAS.436.2929H,2013ApJ...766...71R,2017MNRAS.464.3742R}. However, for the TNG model variations, stronger stellar feedback increases the amount of hot gas in the galaxy, which acts as the receiving phase of AGN feedback that expands the galaxy by the expulsion of baryons from the central regions, thereby working in concordance with AGN feedback enhancing accreted stellar populations and lowering $\gamma^{\prime}$. Thus we find that stellar wind feedback reduces (increases) $\gamma^{\prime}$ with (without) the presence of AGN feedback.

$\bullet$ We calculated the slopes of the dark matter component of the IllustrisTNG ETGs and their corresponding dark matter only counterparts over the same radial ranges used for the measurements of their total slopes. The slopes of the dark matter only halos were in general shallower than their full physics counterparts and possessed larger intrinsic scatter in comparison (see Fig.~\ref{fig:denprof_DMO} and Table~\ref{tab:denprof_DMO}). The inner slopes of the best fit gNFW profile were shallower than the total power-law density slopes, and the FP inner slopes were much steeper than the standard NFW profile, indicative of halo contraction.

$\bullet$ The dark matter profile inner slope is anti-correlated with the halo mass in the FP run, and shows almost no correlation with the halo mass in the DMO run. The trends were in good agreement with the observed and simulated datasets (see Fig.~\ref{fig:gam_m200_fp}). The total density slope $\gamma^{\prime}$ is positively correlated with the FP inner slope $\Gamma_{\mathrm{FP}}^{\prime}$ and anti-correlated with the halo mass (see Fig.~\ref{fig:Gg}). We suggest that the prior of the dark matter profile inner slope of the gNFW profile should be broadened to [$0$, $2$] in dynamic modeling studies.

$\bullet$ The dark matter inner slope is anti-correlated with the halo concentration parameter (see Fig.~\ref{fig:gam_m200_DMO}), indicating non-universal representation of dark matter halos by the gNFW profile, possibly due to different environment and merger history issues (see Fig.~\ref{fig:CM} for an illustration). Comparison of the $\gamma_{\mathrm{mw}}^{\prime}-f_{\mathrm{DM}}$ with semi-analytical models and observations revealed that the dark matter halos of the IllustrisTNG ETGs are indeed contracted. The IllustrisTNG ETGs with smaller stellar mass and central dark matter fraction have more significant contraction, while the ETGs with larger stellar mass and central dark matter fraction are consistent with no contraction or even slight expansion (see Fig.~\ref{fig:AC}). The halo contraction level of the IllustrisTNG ETGs is mass- and baryon fraction- dependent (Fig.~\ref{fig:esf},\ref{fig:AC}). It is consistent with the indications of the IllustrisTNG galaxy dark matter fractions presented in \citep{2018MNRAS.481.1950L}, and indicates that the presence of baryons and their related feedback processes could back-react on the galaxy's halo to further alter the total density profile through halo response.

$\bullet$ There are also a small number of low mass ETGs that show fiducial halo expansion, due to a flattened core in the central stellar density profile (Fig.~\ref{fig:AC}). The mass weighted total density slope $\gamma^{\prime}_{\mathrm{mw}}$ varies with radius in these galaxies, and we suggest making multi-aperture measurements for obtaining the $\gamma^{\prime}_{\mathrm{mw}}$ of ETGs in future observations.

Our selected sample of IllustrisTNG ETGs reproduced reasonably well the observed statistical property of near-isothermal density profiles with little intrinsic scatter. The sample also demonstrated its fidelity through the broad agreement of the correlations between the total density slope and the global galactic properties in comparison with observations and other simulations. Since the IllustrisTNG model was not intentionally tailored to match the observed statistical properties and correlation trends of the total density profiles, the overall successful realization of a realistic ETG sample indicates that the main processes which shape the total density profiles are implemented effectively within the simulation prescription. Nonetheless, certain discrepancies with observations and systematic biases still stand. The slight overestimation of IllustrisTNG galaxy sizes~\citep{2018MNRAS.474.3976G,2018MNRAS.481.1950L} affect $R_{\mathrm{eff}}$, $\Sigma_{\ast}$, $f_{\mathrm{DM}}$, and $\gamma^{\prime}_{\mathrm{mw}}$. The $\gamma^{\prime}-\sigma_{\mathrm{e/2}}$ discrepancy with observations (Fig.~\ref{fig:gam_sv}), along with the fiducial `expansion' for low mass ETGs caused by a flattened central stellar density profile (Fig.~\ref{fig:AC}), suggest limitations in the baryonic models of IllustrisTNG.  Baryons contract dark matter halos at different levels in ETGs with different masses (Fig.~\ref{fig:gam_m200_fp}) and baryon fractions (Fig.~\ref{fig:esf}), leading to diverse and non-NFW dark matter profiles (Fig.~\ref{fig:beta_gam},\ref{fig:Gg},\ref{fig:CM}). These discrepancies may serve as valuable perspectives for future improvements of the subgrid models in cosmological hydrodynamic simulations. One important aspect of improvement enlightened by the TNG model variations analysis suggests further refinement of the black hole kinetic winds in the AGN feedback model (Section~\ref{sec:4.3}).

With the reasonably realistic sample of IllustrisTNG ETGs we present in this paper, we could trace these ETGs to high redshift and further study the key physical processes (including merger events, AGN feedback, star formation activities etc.) relevant for the formation and evolution of the isothermal density profiles. One key evidence for AGN feedback and galaxy mergers regulating the evolution of $\gamma^{\prime}$ is its correlation with the in-situ-formed stellar mass ratio (Fig.~\ref{fig:gam_beta}). Comparison with the Oser and Wind ETGs also demonstrate the necessity of AGN feedback to produce realistic $\gamma^{\prime}$ values (e.g. Fig.~\ref{fig:gam_reff}). Furthermore, a more detailed stellar mass-dependent redshift evolution comparison of $\gamma^{\prime}$ of the statistical sample and the main branch progenitor sample will further constrain the sample biases in the slope evolutionary trend. These analyses will be presented in a follow-up paper (Paper II, Wang et al.  in prep.).

\section*{Acknowledgements}
\label{sec:ack}

We thank Daniel Eisenstein, Dylan Nelson, Luca Ciotti, Michele Cappellari, Peter Schneider, Rhea-Silvia Remus, Ryan McKinnon, Sebastien Peirani, Stephanie O'Neil, and Yiping Shu for helpful discussions and support during the preparation of this paper. We thank the anonymous referee for constructive comments that helped to improve this paper. YW would like to thank the Tsinghua Xuetang Talents Program for funding his research at MIT. MV acknowledges support through an MIT RSC award, a Kavli Research Investment Fund, NASA ATP grant NNX17AG29G, and NSF grants AST-1814053 and AST-1814259. DX would like to thank the supercomputing facilities at the Heidelberg Institute for Advanced Studies and the Klaus Tschira Foundation. This work is partly supported by the National Key Basic Research and Development Program of China (No. 2018YFA0404501 to SM), and by the National Science Foundation of China (Grant No. 11333003, 11390372 and 11761131004 to SM). 




\bibliographystyle{mnras}
\bibliography{isothermal}




\appendix

\section{Comparison datasets of the power-law profiles}
\label{sec:AA}

\begin{table*}
	\begin{center}
	    \begin{tabular}{p{0.12\textwidth} P{0.08\textwidth} P{0.08\textwidth} P{0.10\textwidth} P{0.08\textwidth} P{0.12\textwidth} P{0.08\textwidth} P{0.065\textwidth} P{0.06\textwidth}}
	    \hline
	    Paper & Survey & z & $\gamma^{\prime}$ definition & IMF & Stellar kinematics & $f_{\mathrm{DM}}$ aperture & $\langle\gamma^{\prime}\rangle$ & $\sigma_{\gamma^{\prime}}$ \\
	    \hline
            \citet{2007MNRAS.382..657T} & $-$ & $\approx 0$ & Logarithmic potential, best-fit within $R_{\mathrm{eff}}$ & Constant $M/L$ & Luminosity weighted stellar orbits with anisotropy & $r\leqslant R_{\mathrm{eff}}$ & 1.95 & 0.26 \\
            \citet{2014MNRAS.445..115T} & SPIDER+ $\mathrm{ALTAS^{3D}}$ & $< 0.1$ & Mass-weighted slope at $R_{\mathrm{eff}}$ & Variable & Two-component model for velocity dispersion, $\beta$ = 0.1 or 0.2 & $r\leqslant R_{\mathrm{eff}}$ & $-$ & $-$ \\
            \citet{2015ApJ...804L..21C} & SPIDER+ $\mathrm{ALTAS^{3D}}$ & $<0.006$ & Power-law slope on [$0.1\,R_{\mathrm{eff}}$, $4\,R_{\mathrm{eff}}$] & JAM modeling $M/L_{\mathrm{r}}^{\mathrm{JAM}}$ & SAURON IFU, Keck/DEIMOS spectrograph, $\sigma_{\mathrm{e}}$ within $R_{\mathrm{eff}}$ & $r\leqslant R_{\mathrm{eff}}$ & $2.19\pm0.03$ & 0.11\\
            \citet{2016MNRAS.460.1382S} & $\mathrm{ALTAS^{3D}}$ &  $<0.006$ &  \textsc{Hi} circular velocity averaged slope within $R_{\textsc{Hi}}$~($4\,R_{\mathrm{eff}}-16\,R_{\mathrm{eff}}$) & JAM modeling $M/L_{\mathrm{r}}^{\mathrm{JAM}}$ & Velocity dispersion projected within $R_{\mathrm{eff}}$ & $-$ & $2.18 \pm 0.03$ & 0.11 \\
            \citet{2017MNRAS.467.1397P} & $\mathrm{ALTAS^{3D}}$ & $< 0.01$ & Mean power-law slope on [$0.1\,R_{\mathrm{eff}}$, $R_{\mathrm{eff}}$] (constrained data) & Salpeter & IFU, JAM modeling with variable anisotropy, velocity dispersion within $R_{\mathrm{eff}}$ & $r\leqslant R_{\mathrm{eff}}$ & $2.193 \pm 0.016$ & $0.168 \pm 0.015$ \\
            \citet{2018MNRAS.476.4543B} & SLUGGS & $< 0.005$ & Power-law slope on [$0.1\,R_{\mathrm{eff}}$, $4\,R_{\mathrm{eff}}$] & Constant $M/L$ & JAM modeling with variable anisotropy, velocity dispersion within $R_{\mathrm{eff}}$ & $r\leqslant R_{\mathrm{eff}}$ (\citealt{2017MNRAS.467.1397P} model III) & $2.12 \pm 0.05$ & $-$ \\
            \hline
            \citet{2004ApJ...611..739T} & LSD & [0.5, 1.0] & Power-law slope within $R_{\mathrm{Ein}}$ & Joint lensing and dynamics $M/L$ & $\sigma_{\mathrm{SIE}}$, Osipkov-Merritt anistropy/ constant $\beta (r)$ & $r\leqslant R_{\mathrm{Ein}}$ & $1.75 \pm 0.10$ & 0.2  \\
            \citet{2006ApJ...649..599K} & SLACS & [0.06, 0.33] & Power-law slope within $R_{\mathrm{Ein}}$ & Constant $M/L_{\mathrm{B}}$ & LOS velocity dispersion, variable anisotropy & $r\leqslant R_{\mathrm{Ein}}$ & $2.01_{-0.03}^{+0.02}$ & 0.12 \\
            \citet{2010ApJ...724..511A} & SLACS & [0.24, 0.78] & Power-law slope  & Chabrier/ Salpeter & Velocity dispersion within $R_{\mathrm{eff}}/2$, $\beta = 0$ & $r\leqslant R_{\mathrm{eff}}/2$ & $2.078 \pm 0.027$ & $0.16 \pm 0.02$ \\
            \citet{2011MNRAS.415.2215B} & SLACS & [0.08, 0.33] & Axisymmetric power-law slope & Chabrier/ Salpeter & Axisymmetric model, two-integral Schwarzschild model & $r\leqslant R_{\mathrm{eff}}$ & $2.074_{-0.041}^{+0.043}$ & $0.144_{-0.014}^{+0.055}$ \\
            \citet{2011ApJ...727...96R} & SL2S & [0.24, 0.77] & Power-law slope within $R_{\mathrm{Ein}}$ & Salpeter & Velocity dispersion within $R_{\mathrm{Ein}}$, $\beta = 0$ & $r\leqslant R_{\mathrm{eff}}/2$ (projected 2D) & $2.16_{-0.16}^{+0.16}$ & $0.25_{-0.07}^{+0.10}$ \\
            \citet{2013ApJ...777...98S} & SL2S & [0.2, 0.8] & Power-law slope within $R_{\mathrm{Ein}}$ & Salpeter & Velocity dispersion within $R_{\mathrm{Ein}}$, $\beta = 0$ & $-$ & $2.05_{-0.06}^{+0.06}$ & $0.14_{-0.03}^{+0.04}$ \\
            \hline
            \end{tabular}
        \end{center}
		\caption[2.\columnwidth]{The observational datasets of ETGs analyzed by dynamic modeling and strong lensing surveys utilized for comparison in Section 3. The dynamic modeling dataset of local ETGs is given in the upper half of the table, while the strong lensing dataset of higher redshift ETGs is given in the lower half of the table. For each study, its parent survey, the sample redshift $z$, the definition of the total density slope $\gamma^{\prime}$, the assumed initial mass function (IMF), the specifications of stellar kinematics, the aperture size in which the central dark matter fraction $f_{\mathrm{DM}}$ is calculated, the mean of the slope $\langle\gamma^{\prime}\rangle$ and the scatter of the slope $\sigma_{\gamma^{\prime}}$ are given in the table. We note that the total density slope for the strong lensing dataset is inferred from joint lensing and dynamics analysis, constrained by the total mass enclosed within the lens Einstein radius $M_{\mathrm{Ein}}$, the aperture velocity dispersion $\sigma_{\mathrm{apt}}$, and the de Vaucouleurs fit to the lens light profile. A `$-$' is assigned to any field that is not applicable.}
		\label{tab:comp_1}
\end{table*}

\begin{table*}
	\begin{center}
	    \begin{tabular}{p{0.14\textwidth} P{0.2\textwidth} P{0.1\textwidth} P{0.08\textwidth} P{0.03\textwidth} P{0.08\textwidth} P{0.08\textwidth} P{0.04\textwidth} P{0.035\textwidth}}
	    \hline
	    Paper & Simulation & z & $\gamma^{\prime}$ definition & IMF & Stellar kinematics & $f_{\mathrm{DM}}$ aperture & $\langle\gamma^{\prime}\rangle$ & $\sigma_{\gamma^{\prime}}$ \\
	    \hline
            \citet{2017MNRAS.464.3742R} & Magneticum (cosmological, with AGN, weak wind) & [0, 0.5, 1, 2] & Power-law slope & $-$ & $-$ & $r\leqslant R_{\mathrm{eff}}$ & 2.05 & 0.13 \\
            \citet{2017MNRAS.464.3742R} & Oser (zoom-in, no AGN, no wind) & [0, 0.5, 1, 2] & Power-law slope & $-$ & $-$ & $r\leqslant R_{\mathrm{eff}}$ & 2.30 & 0.28 \\
            \citet{2017MNRAS.464.3742R} & Wind (zoom-in, no AGN, strong wind) & [0, 0.5, 1, 2] & Power-law slope & $-$ & $-$ & $r\leqslant R_{\mathrm{eff}}$ & 2.56 & 0.03 \\
            \hline
	    \end{tabular}
        \end{center}
		\caption[2.\columnwidth]{The simulation dataset of ETGs utilized for comparison in Section 3. For each paper, its parent simulation, the sample redshift $z$, the definition of the total density slope $\gamma^{\prime}$, the assumed initial mass function (IMF), the specifications of stellar kinematics, the aperture size in which the central dark matter fraction $f_{\mathrm{DM}}$ is calculated, the mean of the slope $\langle\gamma^{\prime}\rangle$ and the scatter of the slope $\sigma_{\gamma^{\prime}}$ are given in the table. A `$-$' is assigned to any field that is not applicable.}
		\label{tab:comp_2}
\end{table*}

In this section we give a more detailed documentation of the datasets we used for the comparison of the different slope correlation trends with the IllustrisTNG ETGs in Section 3. The dynamic modeling dataset of local ETGs is given in the upper half of Table~\ref{tab:comp_1}, while the strong lensing dataset of higher redshift ETGs is given in the lower half of Table~\ref{tab:comp_1}. The simulation dataset is given in Table~\ref{tab:comp_2}. For each study, its parent survey or simulation, the sample redshift $z$, the definition of the total density slope $\gamma^{\prime}$, the assumed initial mass function (IMF), the specifications of stellar kinematics, the aperture size in which the central dark matter fraction $f_{\mathrm{DM}}$ is calculated, the mean of the slope $\langle\gamma^{\prime}\rangle$ and the scatter of the slope $\sigma_{\gamma^{\prime}}$ are given in the tables. We note that the total density slope for the strong lensing dataset is inferred from joint lensing and dynamics analysis, constrained by the total mass enclosed within the lens Einstein radius $M_{\mathrm{Ein}}$, the aperture velocity dispersion $\sigma_{\mathrm{apt}}$, and the de Vaucouleurs fit to the lens light profile.

\section{Comparison datasets of the dark matter inner slopes}
\label{sec:AB}

\begin{table*}
		\begin{center}
		\begin{tabular}{p{0.2\textwidth}P{0.15\textwidth}P{0.08\textwidth}P{0.15\textwidth}P{0.12\textwidth}P{0.07\textwidth}P{0.065\textwidth}}
			\hline
			Paper & Survey/Simulation & z & ~~~~~$\Gamma^{\prime}$ definition~~~~~ & $\mathrm{log}\,M_{200}\,$[$M_{\astrosun}$] & ~~~$\langle\Gamma^{\prime}\rangle$~~~ & ~~~$\sigma_{\Gamma^{\prime}}$~~~ \\
			\hline
            \citet{2015ApJ...814...26N} & $-$ & 0.208 & $\gamma_{\mathrm{DM}}^{\prime}$~($r\leqslant R_{\mathrm{eff}}$) & [13.65, 14.45] & 1.12 & 0.22 \\
            \citet{2015ApJ...800...94S} & SL2S+SLACS & [$0.24$, $0.88$] & gNFW inner slope & $13.44_{-0.16}^{+0.16}$ & $0.80_{-0.22}^{+0.18}$ & $-$ \\
            \citet{2018MNRAS.476..133O} & $-$ & $0.185$ & gNFW inner slope & [$11.04$, $12.46$] & $1.62$ & $0.61$ \\
            \citet{2018ApJ...863..130W} & $-$ & 0.185 & gNFW inner slope & $13.28_{-0.28}^{+0.47}$ & $0.96_{-0.41}^{+0.24}$ & $-$ \\
            \hline
            \citet{2015MNRAS.452..343S} & EAGLE & 0.0 & gNFW inner slope & [$13.40$, $14.11$] & $1.36$ & $0.11$ \\
            \hline
		\end{tabular}
        \end{center}
		\caption[2.\columnwidth]{The observational and simulation datasets of ETG dark matter density profiles utilized for comparison in Section 4. For each study, its parent survey or simulation, the sample redshift $z$, the definition of the dark matter profile slope $\Gamma^{\prime}$, the sample halo mass $M_{200}$, the mean of the slope $\langle\Gamma^{\prime}\rangle$ and the scatter of the slope $\sigma_{\Gamma^{\prime}}$ are given in the table. A `$-$' is assigned to any field that is not applicable.}
		\label{tab:comp_3}
\end{table*}

In this section we give a more detailed documentation of the datasets we used for the comparison of the dark matter profiles with the IllustrisTNG ETGs in Section 4. The observation and simulation datasets are given together in Table~\ref{tab:comp_3}. For each study, its parent survey or simulation, the sample redshift $z$, the definition of the dark matter profile slope $\Gamma^{\prime}$, the sample halo mass $M_{200}$, the mean of the slope $\langle\Gamma^{\prime}\rangle$ and the scatter of the slope $\sigma_{\Gamma^{\prime}}$ are given in the table.


\bsp	
\label{lastpage}
\end{document}